\documentclass{aa}

\usepackage{graphicx}
\usepackage{xcolor,colortbl}
\usepackage{txfonts}
\usepackage{soul}
\usepackage[mathscr]{euscript}

\DeclareRobustCommand{\VAN}[3]{#2}
\let\VANthebibliography\thebibliography
\def\thebibliography{\DeclareRobustCommand{\VAN}[3]{##3}\VANthebibliography}

\newcounter{count}

\begin{document}

   \title{The stellar initial mass function \\of nearby young moving groups}

   \author{Rafael Bertolotto-Stefanelli\inst{1,2}
          \and
          Juan Jos\'e Downes\inst{1}
          \and
          Genaro Su\'arez\inst{3}
          \and
          Cecilia Mateu\inst{1}
          \and
          Jonathan Gagn\'e\inst{4,5}
          \and
          Carlos Rom\'an Z\'u\~niga\inst{6}
          }

   \institute{Departamento de Astronom\'ia, Facultad de Ciencias, Universidad de la     Rep\'ublica, Igu\'a 4225, Montevideo, CP 11400, Uruguay
    \email{rbertolotto@arizona.edu}
    \and
    Steward Observatory and Department of Astronomy, University of Arizona, 933 N. Cherry Avenue, Tucson, AZ 85721, USA
    \and
    American Museum of Natural History, 200 Central Park W, New York, NY 10024, USA
    \and
    Plan\'etarium de Montr\'eal, Montr\'eal, Qu\'ebec, Canada
    \and
    Trottier Institute for Research on Exoplanets, Universit\'e de Montr\'eal, D\'epartement de Physique, Montr\'eal, Qu\'ebec, Canada
    \and
    Universidad Nacional Aut\'onoma de M\'exico, Instituto de Astronom\'ia, AP 106, Ensenada 22800, BC, M\'exico
    }
   \date{Received October 30, 2025; accepted December 29, 2025}

  \abstract
   {The solar neighbourhood is populated by nearby ($< 200~\text{pc}$), young ($< 100~\text{Myr}$) moving groups (NYMGs) of stars, whose origins are still debated. A possible explanation is that they are remnants of individual stellar clusters and associations, currently dispersing in the galactic disc.}
   {We aim to derive the initial mass function (IMF) of a large sample of NYMGs.}
   {We developed and applied an algorithm that uses photometric and astrometric data from Gaia DR3 to detect NYMGs as over-densities in a kinematic space, whose members distribute along young isochrones. We inferred individual masses from the photometry of both the detected and the previously known candidates. We estimated the IMFs for 33 groups, 30 of them for the first time, in an average mass range $0.1<m/M_\odot<5$ with some groups going as low as $0.02~M_{\odot}$ and as high as $10~M_{\odot}$. We parameterized these IMFs using a log-normal for $m<1~M_\odot$ and a power-law for $m>1~M_\odot$.}
   {We detected 4166 source candidate members of 44 known groups, including 2545 new candidates. We recovered 44-54\% of the literature candidates and estimated a contamination rate from old field stars of 16-24\%. The candidates of the detected groups distribute along young isochrones, which suggests that they are potential members of NYMGs. Parameterizations of both the average of the 33 IMFs based on our detections ($m_c=0.25\pm0.17~M_{\odot}$, $\sigma_c=0.45\pm0.17$, and $\alpha=-2.26\pm0.09$) and the one based on the known candidates from the literature ($m_c=0.22\pm0.14~M_{\odot}$, $\sigma_c=0.45\pm0.17$, and $\alpha=-2.45\pm0.06$) are in agreement with the IMF parameterization of the solar neighbourhood and young stellar associations.}
   {Our parameterization of the average IMF together with the distribution of the detected group members along young isochrones provide strong evidence suggesting that the NYMGs are remnants of individual stellar associations and clusters and that there are no systematic biases in our detection and in the literature in the range $0.1<m/M_{\odot}<10$.}


   \maketitle

\section{Introduction}
\label{sec:introduction}

The Initial Mass Function (IMF) is the distribution of 
stellar and sub-stellar masses resulting from a single star formation event \citep[e.g.,][]{hopkins2018}.
Despite extensive research since its first observational determination
\citep{salpeter1955}, the central question about its universality, 
meaning whether the IMF of all stellar populations originates from 
the same distribution, remains a subject of intense research. Reviews 
from \cite{bastian2010}, \cite{offner2014}, and \cite{hopkins2018}, 
show that while the IMF in the mass range $0.1 < M/M_\odot < 10$ is 
roughly consistent across different nearby populations in the 
Milky Way, there are exceptions, such as Upper Sco \citep{slesnick2008}, 
and large uncertainties do not allow for constraining the IMF, 
particularly in the sub-stellar mass domain. Furthermore, \cite{sami_2023} 
found a correlation between the slope of the high-mass end of the IMF 
and the surface density of young clusters, suggesting a dependence of 
the massive star formation on the environment. Additionally, the extent 
to which observational and methodological biases have influenced our 
understanding of the IMF remains unclear \citep{hopkins2018}, and more 
statistically robust observational studies of the IMF are needed to 
explore its universality.

Numerous stellar comoving groups have been found in the solar 
neighbourhood \citep[e.g.,][]{Gagne2018} with ages varying from 
a few Myr to a Gyr. Among these groups are the Nearby Young Moving 
Groups (NYMGs), which we consider here as those younger than $100$ Myr 
at distances shorter than $200$ pc. Currently, $68$ groups 
meet these characteristics according to the Montreal Open Clusters and Associations \citep[MOCA;][]{gagne2024} database, with a total of a few hundred known 
stellar and sub-stellar members extensively studied 
e.g. \cite{gagne_2021} and references therein. The most widely accepted hypothesis regarding the origin 
of these groups suggests that they are remnants of stellar associations and disrupted clusters that are now 
being scattered across the galactic disc \citep{gagne_2021} 
as a consequence of their stars having formed gravitationally unbound within their 
parental molecular cloud, as expected from most stellar populations 
\citep{lada2003}. Based on this idea, several efforts were made to integrate the orbits of young comoving groups located as far as 1 kpc \citep[e.g.][]{fernandez_2008,quillen_2020,swiggum_2024}. Although these works present different spatial origin to these groups, they all suggest they were form from giant molecular clouds structures within or related to the passage of spiral arms.

The young ages and proximity of NYMGs make them natural targets 
for determining the IMF because they should have lost only 
their most massive stars as a consequence of stellar evolution, and 
their closeness allows for both the observation of the least massive members
and the detection of some binary systems.
However, there are some significant challenges in identifying the members 
of each NYMG: $(i)$ their proximity make the 
NYMGs project onto extensive regions of the sky; $(ii)$ most 
NYMGs are loose associations with a small number of members, 
producing subtle over-densities in phase-space; and $(iii)$ if these groups are indeed 
scattering through the galactic disc, some members may have 
completely lost their kinematic memory, making it impossible 
to determine their membership.

The IMF of NYMGs is relevant in the description of the solar 
neighbourhood. Despite progress in identifying and characterizing
NYMGs and their members, there exist to our knowledge only 
three studies of the IMF of NYMGs. The first corresponds to \cite{gagne_2017}, who focused on the TW-Hya association. 
They inferred the IMF from 55 candidate members 
($0.01<m/M_{\odot}<2$), finding a log-normal distribution 
with a characteristic mass $m_c = 0.21^{+0.11}_{-0.06}M_\odot$, 
consistent with previous results for the solar neighbourhood
 \citep[$m_c = 0.20\pm0.05M_\odot$;][]{chabrier2005b}, and a width
$\sigma = 0.8^{+0.2}_{-0.1}$ dex, which they claim is significantly 
larger than the $0.3-0.55$ dex usually found in young associations. 
However, the authors stress that their results could be biased by 
an unknown observational or physical effect, especially in the 
low-mass regime, which could explain the large $\sigma$. The second study corresponds to \cite{gagne_2018_volans_carina}, who inferred the IMF of the Volans-Carina group in the mass range $0.15<m/M_{\odot}<3$. In this work, they were able to identify 58 candidate members with parallaxes from Gaia data release 2, from which they inferred their mass distribution. They concluded that the mass distribution found is well described by a fiducial log-normal IMF with $m_c=0.25~M_{\odot}$ and $\sigma_c=0.5$ in the range $m>0.2~M_{\odot}$, suggesting their census of this group is almost complete. Finally, \cite{kraus_2014} inferred the IMF of Tucana-Horologium. They found that when compared to the typical solar neighbourhood IMF \citep{chabrier2005b,salpeter1955}, there appear to be signs of incompleteness in the mass range $0.2<m/M_{\odot}<0.7$. However, they claim that it is not clear if this is indeed the product of a true incompleteness in the data or the result of an error in the stellar models they used that could have pulled stars of spectral types M0-M1 to spectral types K$7.5$.

In this work, we present the development and application of a 
method to identify members of 
NYMGs and infer 
their IMFs in the mass range $0.02<m/M_{\odot}<10$ based on photometric and astrometric data from 
Gaia Data Release 3 \citep{gaia2023} and the 
DBSCAN algorithm \citep{dbscan_original}. We only focus on 44 of the 68 known NYMGs from the literature, which we were able to detect. In Section \ref{sec:data}, we describe how we selected and completed the sample of all GDR3 sources within $200$ pc from the Sun and the compilation of known NYMG members and bona fide candidates from the literature. In Section \ref{sec:search} we explain the detection algorithm to find NYMGs and the significance of the results. 
In Section \ref{sec:imf_of_NYMGs}, we apply Bayesian inference 
to estimate masses from photometry and parallaxes, and present 
the resulting IMFs characterized in terms of their purity, completeness, and parameterization. Finally, in Section \ref{sec:conclusions} we summarize our results and conclude.

\section{The samples}
\label{sec:data}

We conducted our search using the GDR3 catalog, which offers 
the best combination to date of spatial and photometric 
completeness, as well as precision in photometric and astrometric 
data for the solar neighbourhood. As a validation sample, we 
compiled a list of known confirmed members and bona 
fide candidates of NYMGs from MOCA.

\subsection{The Gaia sample in the solar neighbourhood}
\label{subsec:GDR3}

Our starting sample corresponds to the 5,244,458 GDR3 sources with 
parallaxes $\varpi \geqslant 5$ mas, after applying the zero-point 
correction for $\varpi$ provided by the $gaiadr3\_zeropoint$ Python 
package based on \cite{Lindegren_2021_zpt}. The resulting sample 
corresponds to distances $d \leqslant 200$ pc, if we assume 
$d=1/\varpi$. As shown in Appendix \ref{appendix_distance_from_parallax}, 
distances estimated in this way 
show good agreement (with different distance estimates for less than $\sim5\%$ of the sample, almost all fainter than $G=17$ mag) with the photo-geometric distances estimated by 
\cite{Bailer_Jones_2021} if we impose a threshold of $<0.1$ on parallax fractional error ($\sigma_{\varpi}/\varpi$). Additionally, we estimated that only $\sim1.3\%$ of the stars that truly belong to the solar neighbourhood are discarded with this method. Only 43\% of the starting sample sources fulfill the condition $\sigma_{\varpi}/\varpi<0.1$, which corresponds to almost all sources from GDR3 with photo-geometric distances smaller than 200pc. 

We define two subsets from the starting sample of GDR3 sources with $\varpi\geqslant5$ mas: $S$ and $S_{\text{NYMG}}$. The first sample is defined to select all the solar neighbourhood sources with precise magnitudes and parallaxes that could be potential NYMG members. The sample $S_{\text{NYMG}}$ is the result of applying additional photometric and astrometric cuts to the sources from $S$.

\begin{figure}
  \resizebox{\hsize}{!}{\includegraphics{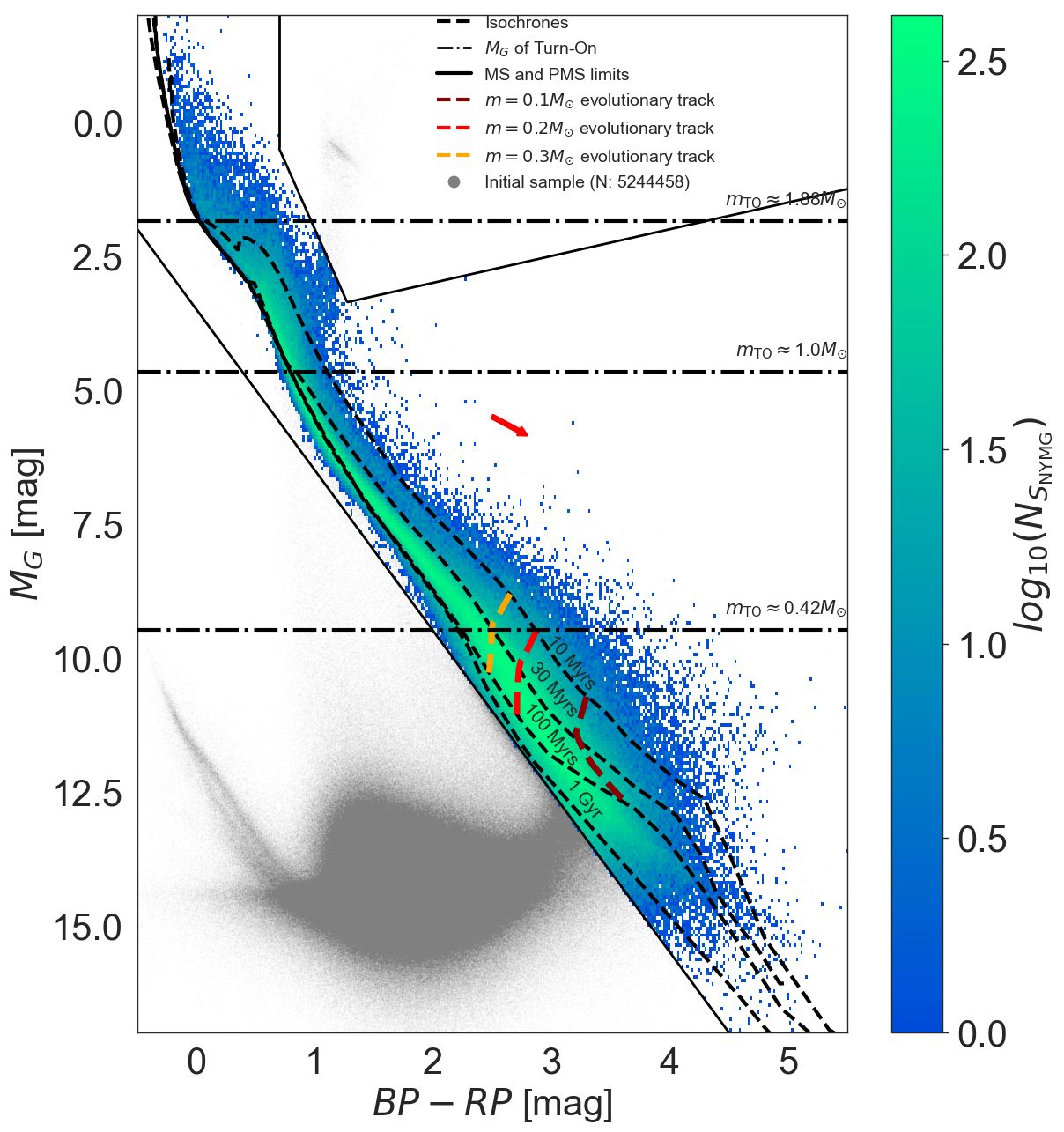}}
  \caption{Color-absolute magnitude diagram of the starting sample (grey points) and the $S_{\text{NYMG}}$ sample (coloured scale indicating the density of points). The solid lines indicate the limits of the MS and PMS locus, the black dashed curves show the isochrones from \cite{Phillips_2020}, \cite{baraffe2015}, and \cite{marigo2017} stellar evolutionary models for $10$ Myr, $30$ Myr, $100$ Myr, and $1$ Gyr, and the horizontal black dash-dotted lines indicate the $M_G$ values corresponding to the Turn-On (see Section \ref{subsec:nymg_candidates}) for the labelled masses. The red, orange, and blue dashed lines indicate the $0.1$, $0.2$, and $0.3~M_{\odot}$ evolutionary tracks respectively. The red arrow indicates two times the extinction vector for an M4 dwarf star with $A_V=0.22$~mag.}
  \label{fig:CMD_starting_sample}
\end{figure}

The sample $S$ was obtained by applying the following conditions to the GDR3 sources with $\varpi\geqslant5$ mas:

\begin{enumerate}

    \item Precise parallaxes: We discarded sources with $\sigma_{\varpi}/\varpi > 0.1$ to ensure that $1/\varpi$ is a good estimate of $d$. 

    \item Main sequence (MS) and pre-main sequence (PMS) stellar 
    and substellar objects: We discarded all sources outside the MS and 
    PMS locus in the $M_G~vs.~BP-RP$ colour absolute-magnitude diagram 
    (CMD), as shown in Figure \ref{fig:CMD_starting_sample}. 
    
    \item Precise $G$ magnitudes: We discarded all sources with 
    signal-to-noise ratio (SNR) in the G passband of 
    $\text{log}_{10} (F_G/\sigma_{F_G}) < 2.2$, where $F_G$ and $\sigma_{F_G}$ 
    are the flux and the corresponding uncertainty, respectively. Most of the 
    discarded sources appear in a non-physical position on the CMD for stars 
    (overdensity around $BP-RP=2$~mag and $M_G=14$~mag in 
    Figure~\ref{fig:CMD_starting_sample}), probably due to unreliable parallaxes as a result of crowding for being near the galactic plane \citep{gaia_nearby_stars_2021}. Additionally, many of these sources show proper motions very close to zero, suggesting they could also be
    extragalactic sources.

    \item Coherent tangential velocities: We discarded sources with 
    tangential velocities $[v_{\alpha_{LSR}}^2+v_{\delta_{LSR}}^2]^{1/2} > 20$ km s$^{-1}$ 
    where $v_{\alpha_{LSR}}$ and $v_{\delta_{LSR}}$ are the 
    components of the tangential velocity in the local standard 
    of rest ($LSR$) computed using $d$, the equatorial coordinates 
    $(\alpha,\delta)$,  and the proper motions $(\mu_{\alpha},\mu_{\delta})$.
    The $20$ km s$^{-1}$ corresponds to the maximum value for the NYMG members
    as shown in Appendix \ref{appendix_section_lsr}.

\end{enumerate}

Although the conditions on $\sigma_{\varpi}/\varpi$ and the photometric 
SNR may primarily exclude some faint objects and thus affect the lower 
end of the IMF, in Appendix \ref{table:parallax_photogeo} we estimate 
that this corresponds to only $\sim 1.3\%$ of the objects in the sample 
$S$.

The sample $S_{\text{NYMG}}$ was obtained by applying the following 
quality condition to the $S$ sample:

\begin{enumerate}

    \item Uncontaminated photometry: We discarded sources whose photometry is affected by contamination from other sources and other effects by applying
    the condition suggested by \cite{gaiadr3_galacticAnticenter_2021}:
    $\log_{10}(\text{phot\_bp\_rp\_excess\_factor}) \leqslant 0.001+0.039(G_{BP}-G_{RP})$ or $0.012+0.039(G_{BP}-G_{RP}) \leqslant \log_{10}(\text{phot\_bp\_rp\_excess\_factor})$,  
    where $G_{BP}$ and $G_{RP}$ are magnitudes in the Gaia passbands 
    and $phot\_bp\_rp\_excess\_factor=(F_{BP}+F_{RP})/F_G$ is 
    the flux excess.
    
    \item Reliable astrometry: We discarded sources with bad astrometry, often caused by multiplicity. To do so, we followed the prescription from \cite{gdr3_validation_2021} and applied the condition  
    $ruwe\geqslant 1.4$, where $ruwe$ is the re-normalized unit weighted 
    error \citep{Lindegren2018}. This condition affects the entire sample 
    and removes potential multiple systems that will not be included in the IMFs. We show in Section~\ref{subsec:completing_sample} how we correct the sample for these missing sources.
    
\end{enumerate}

Figure \ref{fig:CMD_starting_sample} shows the CMD distribution of the starting 
sample as well as the $S$ and $S_{\text{NYMG}}$ samples together with the 
corresponding number of sources. The isochrones in the figure were obtained from a combination of three sets of 
different stellar and substellar evolutionary models:  
1) the ATMO 2020 models by \cite{Phillips_2020} for masses 
$0.01 \leqslant M/M_{\odot} \leqslant 0.015$ and ages younger than 20~Myr, as well
as masses $0.01 \leqslant M/M_{\odot} \leqslant 0.03$ and ages older than 
20~Myr, following the conclusions of Section 5.5 from \cite{Phillips_2020}, 
2) the \cite{baraffe2015} models for ages and masses 
greater than the previous limits and up to 
$0.75~M_\odot$, and 3) the \cite{marigo2017} models for masses $0.75 < M/M_\odot < 10$.

Figure \ref{fig:CMD_starting_sample} also shows a reference extinction vector 
for an M4 dwarf star with $A_V=0.22$~mag, which corresponds to the average extinction for the 
known NYMG candidate members (Section~\ref{subsec:known_nymgs}) according to the 3D extinction maps from \cite{Gontcharov2017}. As Gaia's photometric BP, RP, and $G$ bands are very wide (effective widths $\gtrsim 300$~nm), the colour dependence on extinction laws in those bands cannot be ignored, as demonstrated in Appendix A from \cite{ramos_2020}\footnote{See also \cite{danielski_2019}.}, who shows that Gaia's extinction law in the G band drastically decreases with BP-RP from $k_G\sim1$ for BP-RP$\sim0$ to $k_G\sim0.55$ mag for BP-RP$\sim4$ mag. As the expressions $E(RP-J)/E(J-H)\simeq2.4$, $E(G-RP)/E(J-H)\simeq0.6$, and $E(BP-RP)/E(J-H)\simeq3$ were estimated by \cite{Luhman_2022} for the overall associations of Upper Sco and Ophiuchus, they can be used to estimate an average extinction across spectral types. Using these equations together with the values $k_J=0.282$, $k_H=0.175$ and $k_K=0.112$ from \cite{rieke_1985}, we were able to estimate that on average for an M4 dwarf in an NYMG $A_{\text{BP}}-A_{\text{RP}}\approx0.11$~mag and $A_G\approx0.13$~mag. Since the value of $A_G$ is very small and should be even smaller for fainter stars because of the previously mentioned colour dependence, only the faintest stars can be dimmed beyond the completeness photometric limit $(G\sim17$~mag) as a consequence of extinction. The reddening component $A_{\text{BP}}-A_{\text{RP}}$ of the extinction vector, however, is non-negligible and must be taken into account in the inference of mass based on photometry. Finally, we note that $S_{\text{NYMG}}$ may lack some of the brightest stars of the solar neighbourhood since sources with $G<3$ mag are not included in Gaia \citep{gaia_nearby_stars_2021}. Although this may affect the high-mass end of the IMF, if these groups contain low numbers of members as suggested by the literature (see section \ref{subsec:known_nymgs}) and assuming that their IMF follows some power-law like any other stellar population, then they should contain very low numbers of high-mass stars, which we expect to be comparable with Poisson's noise. This means that although we may be missing some very massive stars, their absence should not significantly affect the statistical behaviour of the measured IMFs.

\subsection{Towards a complete Gaia sample of the solar neighbourhood}
\label{subsec:completing_sample}

It is expected that $S_{\text{NYMG}}$ will suffer 
from some level of incompleteness due to the combination
of the intrinsic biases of the GDR3 survey and potential 
biases introduced by the conditions we used to build it. Such incompleteness may compromise the correct identification of some of the groups or some of their sub-structures and ultimately affect our measurement of their IMFs. 
We use the GaiaUnlimited Selection Functions (GSFs; 
\citealt{gaiaunlimitedSubsample_2023}) Python library and the prescription from Appendix \ref{appendix_completeness} to 
estimate the real completeness of $S_{\text{NYMG}}$, which we define as the number of stars that we are effectively studying, divided by the real number of stars that we wish to study.

The GSFs give us the tools to compute the general selection 
function of GDR3. This corresponds to a continuous map $\mathscr{F}(q)$ defined in a four dimensional space of observables $q=(\alpha,\delta,G-RP,G)$ that returns, for a given $q$, the probability $\mathscr{F}(q)=P_q(\text{GDR3})$ that a source with $q\in \Delta q = (q-dq,q+dq)$ has made it into GDR3. As explained by \citet{Rix_2021}, for each $q$, $\mathscr{F}(q)$ is a dimensionless probability with values between 0 and 1, which is why $\mathscr{F}(q)$ is not a probability distribution over all $q$. The GSFs also contain tools to directly estimate the selection function of a sub-sample $C$ based on some arbitrary conditions applied to the GDR3 catalogue. In other words, it allows us to estimate the probability $\mathscr{F}_C(q)=P_q(C|\text{GDR3})$ that 
a source with $q\in (q-dq,q+dq)$ belongs to $C$, knowing that 
the source is in GDR3. Then, from Appendix \ref{appendix_completeness}, the fraction of real completeness $f_q$ 
for $q\in (q-dq,q+dq)$ can be estimated as:
\begin{equation}\label{eq_completeness_5D}
     f_q\sim P_q(S_{\text{NYMG}}|S_{\text{real}}) = \frac{P_q(S_{\text{NYMG}}|\text{GDR3})}{P_q(S|\text{GDR3})}P_q(\text{GDR3})
\end{equation}
where $S_{\text{real}}$ stands for the real sample of MS and PMS stars 
in the solar neighbourhood, or in other words the sample that we wish to study. The term $P_q(\text{GDR3})$ represents the probability that a star with observables $q$ is in GDR3 while the term $\frac{P_q(S_{\text{NYMG}}|\text{GDR3})}{P_q(S|\text{GDR3})}$ approximates the probability that this star ends up in the final sample $S_{\text{NYMG}}$, assuming that it belongs to ancillary sample $S$ defined in Section~\ref{subsec:GDR3}. If these two probabilities are independent, their product should approximate the value of $f_q$.

As discussed in Section~\ref{subsec:algorithm_description}, our work focuses on the 5-dimensional kinematic space (X,Y,Z,$v_{\alpha_{LSR}}$,$v_{\delta_{LSR}}$) that we call the restricted phase-space (RPS) and the CMD (BP-RP,$M_G$). The RPS is defined as the union of the galactic Cartesian positions (X,Y,Z) and the equatorial tangential velocities in the LSR ($v_{\alpha_{LSR}}$,$v_{\delta_{LSR}}$). We only need to increase the completeness of $S_\text{NYMG}$ in the RPS and the CMD. To do so, we first divide $S_\text{NYMG}$ into 
$\Delta q$ bins of $1$ mag in $G$, $0.1$ mag in $G-RP$, and a HEALPix level of 4 for $(\alpha,\delta)$. Then, we computed $f_q$ for each $\Delta q$ using Equation \ref{eq_completeness_5D}, counted the number of stars $N_{\Delta q}$ in $\Delta q$, and stochastically sampled $N_{\Delta q}/f_q - N_{\Delta q}$ stars from a model of the 7D kinematic and photometric distribution (X,Y,Z,$v_{\alpha_{LSR}}$,$v_{\delta_{LSR}}$,G,BP-RP). Assuming that the dependence between the RPS and CMD distribution is negligible, we sample separately these two spaces, using the full $S_{\text{NYMG}}$ for the RPS and only the sources from this sample within $\Delta q$ to model the CMD. We used a kernel density estimator (KDE) with a Gaussian kernel to model the RPS distribution. The KDE is built by multiplying all the velocities by 10~pc~km$^{-1}$~s so the scale is the same for positions and velocities. In this rescaled space, we used a bandwidth equal to the 90th percentile of the kinematic errors of the $S_{\text{NYMG}}$ sample in that scaled RPS, which is $\approx$9.72~pc. For the CMD, we modelled the $S_{\text{NYMG}}$ sources in $\Delta q$ by sampling the cumulative distributions of $M_G$ within BP-RP bins of $0.1$ mag across the entire BP-RP range of $-0.5$ to $5$ mag.

The resulting catalogue $S_c$ includes the $S_\text{NYMG}$ 
sample plus $N_{\Delta q}/f_q-N_{\Delta q}$ synthetic sources. We created 
ten realizations of $S_c$ to achieve more 
statistically robust results. 
As expected, the number of added synthetic sources increases 
with $G$ and $G-RP$ because the completeness decreases for 
fainter sources.

As explained in Appendix \ref{appendix_completeness}, 
Equation \ref{eq_completeness_5D} and the GSFs do not account 
for potential incompleteness in the $S$ sample due to the 
quality conditions $\log_{10}$(SNR)>2.2 
and $\sigma_{\varpi}/\varpi<0.1$. 
However, as explained in Appendix 
\ref{appendix_distance_from_parallax}, 
we estimated that these conditions only excluded $\sim7000$ objects ($\sim1.3\%$ of final sample) that would have been part of $S$ and that could belong to the solar neighbourhood, according to \cite{Bailer_Jones_2021}, which would have corresponded to only $\sim1\%$ of $S$.

\subsection{The catalogue of known members and candidates}
\label{subsec:known_nymgs}

In this study, we say a source is a candidate member 
of a specific known NYMG if its photometry and kinematics from GDR3 
are consistent with those of previously known members of the 
NYMG. We used the MOCA database \citep{gagne_2024} to compile most of the objects reported in the literature as members of the NYMGs following the prescription from Appendix \ref{appendix_nymgs_from_moca}. This catalogue includes relevant information from most, if not all of the known groups in the literature. We only kept in the resulting sample candidate members from MOCA with a membership probability estimated by BANYAN in April 2025 $p>0.95$. 

The resulting sample consists of a total of 5926 sources distributed 
between 68 different NYMGs. In this work and as discussed in Section \ref{subsec:detecting_nymgs}, we were able to re-detect 46 of the 68 known groups, whose general statistics are summarized in Table \ref{table:knowngroups}. We refer to the subset of candidate members from the literature that belong to the mentioned 46 groups as the literature candidate sample ($S_{\text{Lit}}$), which includes a total of 4104 sources ($\sim 69\%$ of the original 5926 sources).

\section{Detecting NYMG member candidates}
\label{sec:search}

In this section, we present the procedure and algorithm designed 
to search for NYMGs based on GDR3 photometric and astrometric data. 
We explain how we evaluated the statistical significance of our 
detections and estimate the contamination and completeness of both 
the detected groups in this work and the recovered groups from $S_{\text{Lit}}$. From now on, we use the term field to refer to the stars that populate the solar neighbourhood but 
do not belong to any NYMG. The stars from the field can be classified into two categories: $(i)$ members of stellar associations that are not NYMGs, such as massive and older moving groups or open clusters, or $(ii)$ old stars that do not belong to any stellar association. We refer to the latter subset of field stars as the Besan\c{c}on field. We chose this name because our model of the RPS of the field is based on the Besan\c{c}on galactic models, which correspond to an axisymmetric model of the Milky Way.

\subsection{DBSCAN: description and caveats}
\label{subsec:dbscan}

The procedure is grounded in the Density-Based Spatial Clustering 
of Applications with Noise (DBSCAN) algorithm \citep{dbscan_original}, 
a clustering algorithm designed to detect point over-densities 
in an N-dimensional space. It operates based on two essential parameters:
$\varepsilon$, which sets the maximum distance between two points 
to be considered neighbours, and $N_{\text{min}}$, which defines the minimum 
number of neighbours required for a point to be labeled as a core 
point. DBSCAN identifies clusters as sets of neighbouring core 
points and their associated neighbours.
Once $\varepsilon$ and $N_{\text{min}}$ are chosen, the DBSCAN 
algorithm will only detect clusters denser than the threshold
$\rho_{\varepsilon,N_{\text{min}}} = N_{\text{min}}/V_\varepsilon$, 
where $V_\varepsilon$ is the volume of a hyper-sphere of dimension 
equal to the number of dimensions in which DBSCAN is being used. 
This makes the algorithm a density-based clustering algorithm 
\citep{Malzer2020, ratzenbock_2022}. 

In order to detect a set of groups with different densities in phase space without significantly increasing the contamination from the field, we need different combinations of $(\varepsilon, N_{\text{min}})$ for each group. For this reason, it is necessary to explore the hyper-parameter space 
of $(\varepsilon, N_{\text{min}})$ to ensure that all the clusters are 
detected. One possible way to do so is to use Hierarchical 
DBSCAN \citep[HDBSCAN;][]{hdbscan_original}, which builds 
a dendrogram of clusters detected with DBSCAN at different 
density levels and uses different statistical methods to select 
the final clusters. Although HDBSCAN has been used to search for co-moving groups \citep{kerr2021}, it is computationally more costly than DBSCAN for large sets of data. Regarding sensitivity, it is not clear if DBSCAN is better or worse than HDBSCAN as \cite{ratzenbock_2022} found higher recovery rates when using DBSCAN instead of HDBSCAN in a study of substructures in the Sco-Cen association, while \cite{hunt_reffert_2021} found opposite results on a sample of open clusters. In terms of detecting real members however, \cite{hunt_reffert_2021} also found that DBSCAN presented more precise results (i.e. less false positive detections) than HDBSCAN. Since our study focuses on the statistical behaviour of the IMF of the NYMGs rather than on the detection of all their members, and because of the cheaper computational cost of DBSCAN, we decided to use this algorithm rather than HDBSCAN for our study.

\subsection{The core of the detection algorithm}
\label{subsec:algorithm_description}

It is expected that a NYMG forms an over-density in the 
6-dimensional phase-space of positions $(X,Y,Z)$ and 
velocities $(U,V,W)$ in the Cartesian galactic heliocentric 
coordinate system. This over-density might overlap the characteristic 
distribution of the field stars, and the members of the NYMGs also 
are distributed along their corresponding isochrone in the CMD. 
However, since only $44\%$ of the sources in the $S_{\text{NYMG}}$ 
sample have measured radial velocities, we cannot search for 
NYMGs in the 6-dimensional phase-space without significantly 
reducing the completeness of the sample. Therefore, following
\cite{ratzenbock_2022}, we restrict our kinematic search to 
the RPS (X,Y,Z,$v_{\alpha_{LSR}}$,$v_{\delta_{LSR}}$). In Appendix
\ref{appendix_section_lsr}, we show how the tangential 
velocities reduce the scattering that these groups have 
in the proper-motions space and how working in the LSR 
helps us avoid the Sun's motion from causing scattering between groups or
joining different groups together.

As shown by \cite{dbscan_original}, the performance of DBSCAN at both maximizing the number of detected real members and minimizing the number of detection of non-members increases 
as the difference between the RPS density of the cluster stars and the field 
increases. This is why, to build the $S_{\text{NYMG}}$ sample, we excluded all
stars lying outside the MS and PMS loci and with $v_{tan}>20$~km s$^{-1}$, which lowers 
the field density without affecting the NYMG densities. Following this 
reasoning and based on the idea that if the NYMGs are remnants of stellar associations and clusters 
then each NYMG should follow an isochrone on the CMD, we divided each of 
the $S_c$ samples into 10 sub-samples according to the isochrones in the 
$M_G$ vs. $G_{BP} - G_{RP}$ CMD. The first sub-sample, called 
$S_{c,10}$, includes all sources that fall above the curve 
which is 0.5 mag fainter than the 10 Myr isochrone. This way, the sample 
includes all sources younger than 10 Myr and accounts 
for photometric uncertainties. The selection did not 
include a brighter limit because this could bias the 
sample by excluding sources affected by extinction.
We repeated the same procedure for nine additional isochrones 
with ages from 20 Myr up to 100 Myr at intervals of 10 Myr, 
and produced the corresponding sub-samples $S_{c,\mathcal{A}}$, 
where $\mathcal{A}$ is the age of the isochrone used for 
the selection of each sample.

In order to run DBSCAN, all coordinates must be scaled and 
made dimensionless. Thus, we define the scaled restricted phase-space 
(SRPS) as 
$(X/\varepsilon_r,Y/\varepsilon_r,Z/\varepsilon_r,v_{\alpha,\text{LSR}}/\varepsilon_v,v_{\delta,\text{LSR}}/\varepsilon_v)$,
where $\varepsilon_r$ and $\varepsilon_v$ are the scaling factors 
for the position and velocity spaces, respectively. Then, the core of the detection algorithm simply consists of running the DBSCAN algorithm in the SRPS using $\varepsilon=0.847\sqrt{2}$ based on the analysis in Appendix \ref{appendix_dbscan_hyper_parameters} where we show how this approach
approximates the separate use of DBSCAN in the position and velocity spaces. For each detected over-density, a hypothesis test is performed at a $99\%$ significance level, where our 
null hypothesis $\mathcal{H}_0$ is this cluster is  
not a NYMG, but a stochastic fluctuation in the density of 
Besan\c{c}on field stars. Finally, we keep only the clusters that are not ruled out by 
$\mathcal{H}_0$. The implementation of the hypothesis test 
is presented in Section \ref{subsec:significance_and_synthetic_field}. 

The hyper-parameters of the core of the detection algorithm 
are: the age $\mathcal{A}$ of the isochrone used for the photometric 
selection, the scale parameters $\varepsilon_r$ and 
$\varepsilon_v$, and the minimum number of neighbours 
$N_{\text{min}}$. To explore the detection of NYMGs 
with different ages and densities in the SRPS, we ran
the core of the detection algorithm over one hundred sub-samples $S_{c,\mathcal{A}}$
with a set of combinations for the hyper-parameters.
Specifically, we use $\varepsilon_r$ ranging from 
5 pc up to 30 pc in steps of 2.5 pc; $\varepsilon_v$ 
ranging from 0.25 km~s$^{-1}$ to 5 km~s$^{-1}$ in steps of 0.25 km~s$^{-1}$; 
and $N_{\text{min}}$ ranging from 2 to 30 in steps of 2.

The detection algorithm runs independently over each 
sub-sample $S_{c,\mathcal{A}}$, which means that if a cluster is 
detected in a certain sample, it should also be detected 
in all the older ones. The advantage of this procedure 
is twofold: it confirms the detection of a given cluster 
from different sub-samples and it allows us to find the sub-sample $S_{c,\mathcal{A}}$ with the oldest age $\mathcal{A}$ that minimizes contamination and maximizes completeness.

\subsection{Spurious over-densities from field fluctuations}
\label{subsec:significance_and_synthetic_field}

To evaluate the statistical significance of the detections, 
we estimate the probability of identifying over-densities 
resulting from stochastic density fluctuations in the Besan\c{c}on field of the solar neighbourhood. We run the algorithm on a set of ten synthetic catalogues generated 
from a numeric model that mimics the distribution of the Besan\c{c}on field in both the RPS and the $M_G$ vs. $BP-RP$ CMD. 

We modelled the RPS and CMD distribution of the Besan\c{c}on field using the same method and bandwidth values for the kernel widths and bin size used in Section~\ref{subsec:completing_sample} to model the RPS and CMD of $S_{\text{NYMG}}$ but using different samples for the RPS and CMD. In the case of the RPS, we utilized the Gaia Object Generator 
(GOG) synthetic catalogue \citep{gog}, which is generated by 
applying the GDR3 error models to the galactic stellar 
populations predicted by the Besan\c{c}on models \citep{gums}. 
We chose GOG for constructing a model of the distribution of Besan\c{c}on field stars because it does not incorporate stellar clusters and NYMGs and includes GDR3 errors.
We applied the conditions outlined in Section 
\ref{subsec:GDR3} to GOG and completed it by using 
the method described in Section \ref{subsec:completing_sample} (using this sample from GOG instead of $S_{\text{NYMG}}$). In the case of the CMD, we used the ten combined $S_c$ samples. 

As stated in Section~\ref{subsec:completing_sample}, this method assumes that the dependence between the CMD and RPS distributions is negligible. Additionally, for this particular case (to model the Besan\c{c}on field), we are assuming that the vast majority of stars in the $S_c$ samples are field stars whose CMD distribution does not significantly change due to the presence of NYMGs or other young massive clusters. With this model, we can create a synthetic Besan\c{c}on field for the $S_{\text{NYMG}}$ sample by simply sampling stochastically a number of sources from this model equal to the number of sources in any $S_c$.

\begin{figure}
  \resizebox{\hsize}{!}{\includegraphics{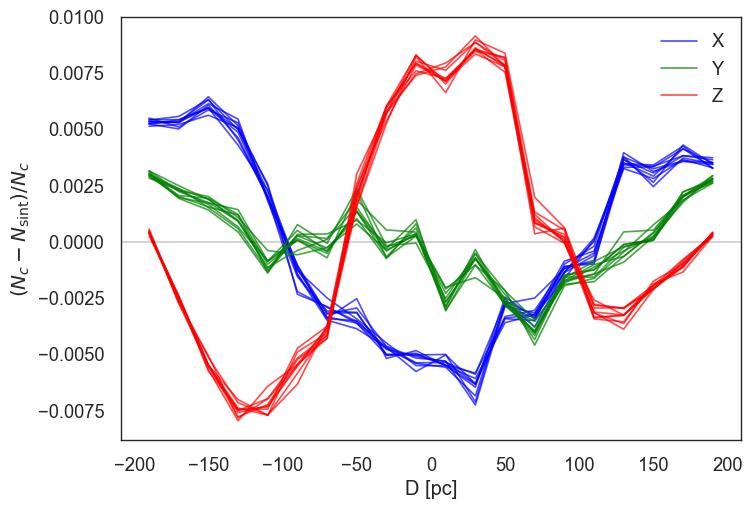}}
  \resizebox{\hsize}{!}{\includegraphics{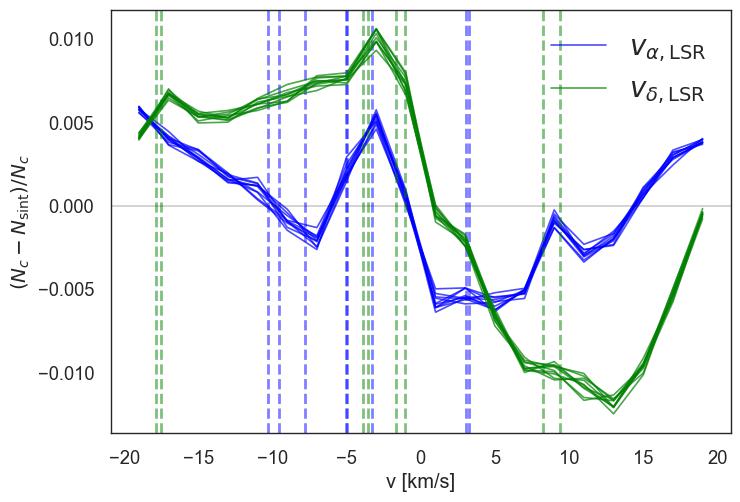}}
  \caption{Distribution of residuals between ten synthetic fields 
and $S_{\text{NYMG}}$ samples across the position space (top panel) and 
the LSR tangential velocity space (bottom panel). Dashed lines indicate 
the mean velocities of known open clusters according to the 
MOCA database.}
  \label{fig:gog_vs_gaia}
\end{figure}

Figure \ref{fig:gog_vs_gaia} shows the fractional residuals between 
ten synthetic fields and one of the $S_c$ 
samples, which do not exceed 1\% in both spaces. Notably, we found 
that the differences in the velocity space very likely correspond to certain known open clusters 
present in the $S_{\text{NYMG}}$ sample, according to the MOCA database. 
The remarkably small residuals underscore the effectiveness 
of GOG in accurately reproducing the GDR3 Besan\c{c}on field, even within 
the limited volume of our study.

We generated ten synthetic fields and applied our detection algorithm 
to each of them for every hyper-parameter combination $(\mathcal{A}, N_{\text{min}}, 
\varepsilon_r, \varepsilon_v)$. For each field, we obtained the 
number $K$ of members detected in each over-density. Then, for 
each hyper-parameter combination and each synthetic field, we 
derived the distribution of $K$ and identified $K_{min,\alpha}$, 
which is defined as the $\alpha$-quantile of this distribution. 
Finally, we computed the mean value of $K_{min,\alpha}$ across 
the ten synthetic fields for each hyper-parameter combination. 
This mean value represents the minimum number of members required 
for an over-density to be classified as a cluster with a 
significance level of $\alpha$. In this work, we 
used $\alpha = 0.99$.

\begin{figure}
  \resizebox{\hsize}{!}{\includegraphics{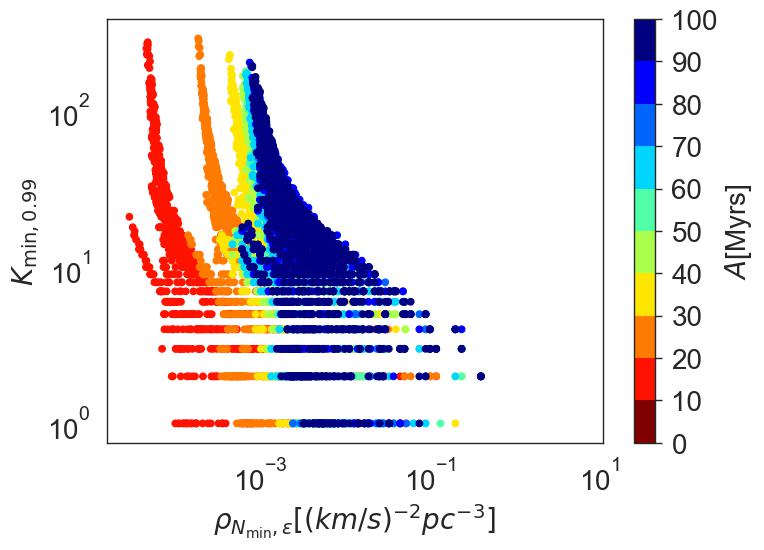}}
  \resizebox{\hsize}{!}{\includegraphics{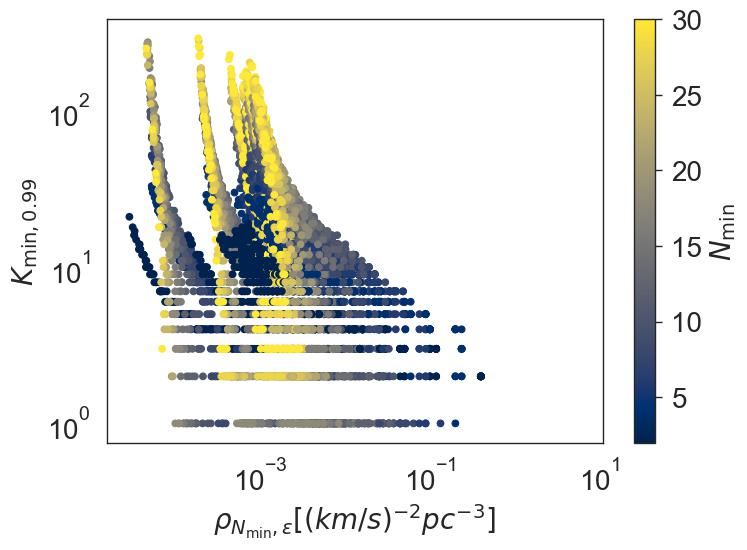}}
  \caption{Resulting $K_{min,0.99}$ for all the hyper-parameter combinations as a function of the RPS density threshold $\rho_{\varepsilon_{r,v},N_{\text{min}}}$ corresponding to the hyper-parameter combination. The color-map for the top and bottom plots indicates the age $\mathcal{A}$ and the value of $N_{\text{min}}$ respectively.}
  \label{fig:K_vs_density}
\end{figure}

Figure \ref{fig:K_vs_density} shows the distribution 
of all final $K_{min,0.99}$ as a function of the DBSCAN 
density threshold 
$\rho_{\varepsilon_{r,v},N_{\text{min}}} = N_{\text{min}}/(\varepsilon_r^3\varepsilon_v^2)$ for all the hyper-parameter combinations. As expected, $K_{min,0.99}$ decreases as $\rho_{\varepsilon_{r,v},N_{\text{min}}}$ increases. Additionally, we note that the values of $\rho_{\varepsilon_{r,v},N_{\text{min}}}$ increase with the isochrone age $\mathcal{A}$. This is due to the fact that the older the isochrone, the more sources are included in the sample and hence the higher the density of the field, which means higher density thresholds are needed in order to ensure statistical significance. Finally, we also note that for a given age $\mathcal{A}$, we need to use high values of $N_{\text{min}}$ in order to use low values of $\rho_{\varepsilon_{r,v},N_{\text{min}}}$. Overall, the trends in Figure \ref{fig:K_vs_density} indicate that, to confidently detect lower-number and/or older associations, we either need to increase the density threshold fixed by the hyper-parameters or $N_{\text{min}}$, as expected.

It is important to notice that the statistical significance evaluated with this method is relative to the Besan\c{c}on field only. This means that a statistically significant detection does not necessarily mean that it is real or not highly contaminated by stars from an older or more massive moving group or open cluster. This is discussed in Section \ref{subsec:detecting_nymgs}.

\subsection{Identifying NYMG candidates in real over-densities}
\label{subsec:detecting_nymgs}

We ran the detection algorithm on the ten $S_c$ samples, considering 
over-densities with $\alpha \geqslant 0.99$ as significant detections, as 
explained in Section \ref{subsec:significance_and_synthetic_field}. The 
results consist of a list of statistically significant over-densities for each hyper-parameter combination. Some over-densities may be detected in several combinations of hyper-parameters, so we developed in this section a criterion to choose among the different detections.

We used the Bayesian Information Criterion \citep[BIC;][]{schwarz1978} to fit a set of multivariate Gaussians to each over-density in the RPS. This allows us to identify different substructures of a single NYMG in the RPS and reduce the level of field star contamination in the groups. Each star for the original over-density is assigned to the Gaussian it most likely belongs to. Then, we compute for each member of the group the ratio $r_c =f(\vec{r},\vec{v})/f(\vec{r}_{\text{mean}},\vec{v}_{\text{mean}})$, where $f(\vec{r},\vec{v})$ corresponds to the probability density function (PDF) of the fitted Gaussian evaluated on the RPS position of the member and $f(\vec{r}_{\text{mean}},\vec{v}_{\text{mean}})$ to the PDF evaluated on the mean RPS position of the Gaussian. Although this ratio is not a probability, it tells us how likely it is that a star belongs to the distribution based on its RPS position relative to the mean position of the Gaussian, which has by definition the highest likelihood of belonging to the Gaussian. We then compute for each group the score function $C$:
\begin{equation}\label{eq:groupScore}
    \mathcal{C} = N\displaystyle\sum_{i=0}^Nr_{c,i}=N\displaystyle\sum_{i=0}^N\frac{f(\vec{r}_i,\vec{v}_i)}{f(\vec{r}_{\text{mean}},\vec{v}_{\text{mean}})}
\end{equation}
Where $N$ is the number of members in the group. The more tightly grouped the members of a group are and the more members a group has, the higher its score. We consider all the 
different groups detected in all the hyper-parameter combinations 
and identify the one with the highest $\mathcal{C}$. Then, we discard all the
other groups that contain at least one of the members of the group with 
maximum $\mathcal{C}$. We iterate this process over the remaining groups until 
all the stars are either assigned to a group or discarded. Finally, we identify 
from our final list of groups the ones that contain at least ten members of 
one of the known NYMGs. We follow all the previous steps for each of the ten 
$S_c$ samples and merge the results into one final list of detected NYMGs.

\subsection{NYMG candidates from this work}
\label{subsec:nymg_candidates}

Following the procedure presented in Section \ref{subsec:detecting_nymgs}, we detected a total of 1705 over-densities. As discussed in Section \ref{subsec:significance_and_synthetic_field}, over-densities that are statistically significant when compared to the Besan\c{c}on field may still not be significant relative to the full field, a result of the real non-axisymmetric potential of the Milky Way, or may simply be substructures of other stellar associations different from the NYMGs. As shown in Appendix \ref{appendix_overdensities_rejections} and following the approach from \cite{ratzenbock_2022}, we fitted a double Gaussian to the uni-variate distribution of over-density sizes from which we estimated that only 473 ($\sim28\%$) of the detected over-densities are more likely to be real. 

We found that 35 of the 1705 detected over-densities include members of 47 of the 68 groups from $S_{\text{Lit}}$. Only 2 of these 35 over-densities do not belong to the likely real 473 over-densities. The first of those over-densities corresponds to the group ETAC, while the other corresponds to a fragment of RATZSCO12. We detected however another over-density with members from RATZSCO12 that does belong to the likely real 473 over-densities. By discarding the 2 known over-densities corresponding to ETAC and the fragment of RATZSCO12, we end up with 33 over-densities corresponding to 44 known NYMGs ($\sim65\%$ of the 68 known groups), whose general statistics are shown in Table \ref{table:knowngroups}. Although the list of the remaining 440 over-densities is interesting, it is very likely that many are associated with sub-structures of massive old moving groups or star-forming regions. A more detailed analysis of these over-densities will be presented in a separate publication. In this work, we focus on studying only the 33 detected over-densities corresponding to the 44 recovered group candidates. As pointed out in the MOCA database, in numerous cases, it is not clear if NYMGs with similar photometric and kinematic distribution are indeed separate associations or just sub-structures of a common association. This could be the reason why in many cases our detection algorithm classified different NYMGs from MOCA as part of a single group, leading to this shorter list of 33 group candidates instead of a list of 44 groups.

\begin{table*}
\begin{flushleft}
\scriptsize
\caption{Relevant statistics for the NYMGs analyzed in this study.}
\begin{center}
\hspace*{-0.5cm}
\begin{tabular}{c c c c c c c c c c c c c c c c}
\hline
\hline
Name & $A_{\text{lit}}$ & $N$ & $N_{RV}$ & $d_{\text{mean}}$ & $N_{\text{det}}$ & $N_{\text{det,0.1}}$ & $f_r$ & $f_{r,0.1}$ & $f_p$ & $f_{p,0.1}$ & $m_c$ & $\sigma_c$ & $\alpha$ & Mass & Mass ($S_{\text{lit}}$)\\
     & [Myr] & & & [pc] & & & & & & & [$M_\odot$] & & & [$M_\odot$] & [$M_\odot$] \\
\hline
BPMG & 26 & 170 & 85 & 57 $\pm$ 9 & 92 & 51 & 0.26 & 0.33 & 0.81 & 1.00 & 0.50 $\pm$ 0.16 & 0.60 $\pm$ 0.16 & -2.36 $\pm$ 0.13 & 29.07 & 66.64 \\
CAR & 45 & 161 & 73 & 87 $\pm$ 12 & 127 & 73 & 0.25 & 0.33 & 0.92 & 1.00 & 0.28 $\pm$ 0.15 & 0.35 $\pm$ 0.17 & -1.73 $\pm$ 0.16 & 38.37 & 55.10 \\
COL & 42 & 249 & 110 & 91 $\pm$ 22 & 365 & 202 & 0.35 & 0.45 & 0.74 & 0.96 & 0.32 $\pm$ 0.17 & 0.50 $\pm$ 0.16 & -3.20 $\pm$ 0.15 & 101.94 & 87.55 \\
GRSCOS16 & 18 & 49 & 20 & 132 $\pm$ 4 & 69 & 39 & 0.73 & 0.97 & 1.00 & 0.80 & 0.28 $\pm$ 0.18 & 0.50 $\pm$ 0.17 & -2.16 $\pm$ 0.12 & 19.34 & 22.63 \\
GRSCOS2 & 43 & 85 & 38 & 129 $\pm$ 14 & 442 & 276 & 0.14 & 0.14 & 0.94 & 0.58 & 0.28 $\pm$ 0.17 & 0.40 $\pm$ 0.17 & -2.43 $\pm$ 0.06 & 121.58 & 37.09 \\
GRSCOS27D & 23 & 182 & 94 & 111 $\pm$ 4 & 225 & 117 & 0.15 & 0.23 & 1.00 & 0.89 & 0.32 $\pm$ 0.16 & 0.35 $\pm$ 0.18 & -1.14 $\pm$ 0.12 & 51.78 & 77.49 \\
HSC1373 & 44 & 41 & 11 & 145 $\pm$ 9 & 81 & 41 & 0.87 & 1.00 & 1.00 & 0.62 & 0.32 $\pm$ 0.16 & 0.50 $\pm$ 0.17 & -3.05 $\pm$ 0.16 & 21.25 & 9.74 \\
HSC1733 & 64 & 53 & 12 & 120 $\pm$ 25 & 254 & 144 & 0.41 & 0.53 & 0.63 & 0.86 & 0.28 $\pm$ 0.16 & 0.40 $\pm$ 0.17 & -2.08 $\pm$ 0.09 & 67.74 & 20.35 \\
HSC1923 & 85 & 98 & 20 & 144 $\pm$ 20 & 146 & 78 & 0.65 & 0.84 & 0.86 & 0.79 & 0.28 $\pm$ 0.15 & 0.40 $\pm$ 0.18 & -2.60 $\pm$ 0.09 & 36.44 & 31.79 \\
HSC1964 & 88 & 282 & 98 & 138 $\pm$ 30 & 800 & 442 & 0.81 & 0.91 & 0.72 & 0.55 & 0.32 $\pm$ 0.17 & 0.40 $\pm$ 0.17 & -2.79 $\pm$ 0.16 & 243.68 & 114.03 \\
HSC2597 & 35 & 79 & 22 & 159 $\pm$ 19 & 296 & 166 & 0.67 & 0.82 & 0.89 & 0.77 & 0.25 $\pm$ 0.17 & 0.45 $\pm$ 0.17 & -2.89 $\pm$ 0.08 & 78.54 & 26.75 \\
HSC958 & 41 & 140 & 29 & 175 $\pm$ 8 & 148 & 89 & 0.66 & 0.95 & 0.65 & 0.76 & 0.25 $\pm$ 0.16 & 0.45 $\pm$ 0.17 & -3.60 $\pm$ 0.13 & 40.02 & 46.84 \\
OCSN99 & 93 & 51 & 15 & 142 $\pm$ 13 & 242 & 139 & 0.85 & 0.91 & 0.72 & 0.60 & 0.32 $\pm$ 0.17 & 0.40 $\pm$ 0.17 & -1.80 $\pm$ 0.13 & 77.68 & 17.99 \\
OCT & 35 & 187 & 56 & 168 $\pm$ 19 & 445 & 243 & 0.34 & 0.52 & 0.86 & 0.96 & 0.28 $\pm$ 0.17 & 0.45 $\pm$ 0.17 & -3.12 $\pm$ 0.17 & 121.28 & 62.14 \\
RATZSCO12 & 19 & 83 & 21 & 160 $\pm$ 3 & 41 & 24 & 0.28 & 0.43 & 1.00 & 0.86 & 0.22 $\pm$ 0.14 & 0.45 $\pm$ 0.18 & -1.38 $\pm$ 0.11 & 11.88 & 25.07 \\
RATZSCO14 & 15 & 279 & 104 & 139 $\pm$ 6 & 429 & 216 & 0.16 & 0.27 & 1.00 & 0.95 & 0.25 $\pm$ 0.15 & 0.35 $\pm$ 0.16 & -2.08 $\pm$ 0.14 & 93.60 & 110.32 \\
RATZSCO17 & 21 & 201 & 57 & 153 $\pm$ 11 & 237 & 135 & 0.14 & 0.20 & 1.00 & 0.88 & 0.28 $\pm$ 0.16 & 0.40 $\pm$ 0.17 & -1.50 $\pm$ 0.11 & 63.54 & 63.27 \\
RATZSCO19 & 14 & 112 & 42 & 123 $\pm$ 4 & 113 & 61 & 0.13 & 0.23 & 1.00 & 1.00 & 0.18 $\pm$ 0.13 & 0.40 $\pm$ 0.17 & -2.10 $\pm$ 0.12 & 30.09 & 46.18 \\
RATZSCO31 & 14 & 62 & 18 & 132 $\pm$ 5 & 67 & 33 & 0.50 & 0.52 & 1.00 & 0.63 & 0.25 $\pm$ 0.16 & 0.45 $\pm$ 0.17 & -0.65 $\pm$ 0.24 & 13.08 & 15.56 \\
TAUMGLIU13 & 33 & 14 & 4 & 123 $\pm$ 9 & 338 & 196 & 1.00 & 1.00 & 0.92 & 1.00 & 0.28 $\pm$ 0.18 & 0.50 $\pm$ 0.17 & -2.24 $\pm$ 0.10 & 112.79 & 5.42 \\
THA & 40 & 284 & 153 & 51 $\pm$ 10 & 146 & 88 & 0.33 & 0.45 & 0.98 & 0.93 & 0.18 $\pm$ 0.15 & 0.45 $\pm$ 0.17 & -2.33 $\pm$ 0.10 & 33.62 & 95.13 \\
TWA & 10 & 122 & 62 & 71 $\pm$ 7 & 78 & 35 & 0.17 & 0.19 & 0.98 & 1.00 & 0.22 $\pm$ 0.20 & 0.80 $\pm$ 0.12 & -2.41 $\pm$ 0.07 & 15.57 & 33.55 \\
UTAU & 50 & 30 & 5 & 183 $\pm$ 8 & 79 & 46 & 0.87 & 1.00 & 0.76 & 0.69 & 0.28 $\pm$ 0.17 & 0.45 $\pm$ 0.17 & -1.17 $\pm$ 0.23 & 25.38 & 17.80 \\
VCA & 89 & 51 & 26 & 101 $\pm$ 16 & 187 & 100 & 0.88 & 0.96 & 0.68 & 0.60 & 0.28 $\pm$ 0.16 & 0.40 $\pm$ 0.17 & -2.58 $\pm$ 0.15 & 46.67 & 21.29 \\
ETAU$^{1}$ & 50 & 57 & 12 & 151 $\pm$ 14 & 395 & 227 & 0.91 & 1.00 & 0.93 & 0.56 & 0.22 $\pm$ 0.14 & 0.40 $\pm$ 0.17 & -1.76 $\pm$ 0.16 & 131.88 & 25.27 \\
MUTAU$^{1}$ & 61 & 232 & 39 & 151 $\pm$ 14 & 395 & 227 & 0.26 & 0.44 & 0.93 & 0.87 & 0.22 $\pm$ 0.14 & 0.40 $\pm$ 0.17 & -1.76 $\pm$ 0.16 & 131.88 & 81.23 \\
TAUMGLIU7$^{2}$ & 3 & 15 & 3 & 158 $\pm$ 6 & 159 & 78 & 0.87 & 0.87 & 0.74 & 1.00 & 0.18 $\pm$ 0.14 & 0.45 $\pm$ 0.16 & -1.62 $\pm$ 0.24 & 36.70 & 3.15 \\
TAUMGLIU8$^{2}$ & 4 & 24 & 9 & 158 $\pm$ 6 & 159 & 78 & 0.91 & 1.00 & 0.74 & 0.71 & 0.18 $\pm$ 0.14 & 0.45 $\pm$ 0.16 & -1.62 $\pm$ 0.24 & 36.70 & 15.31 \\
THOR$^{3}$ & 25 & 60 & 24 & 103 $\pm$ 12 & 290 & 146 & 0.82 & 1.00 & 0.51 & 0.63 & 0.18 $\pm$ 0.13 & 0.45 $\pm$ 0.16 & -2.54 $\pm$ 0.17 & 59.85 & 14.75 \\
118TAU$^{3}$ & 10 & 48 & 22 & 103 $\pm$ 12 & 290 & 146 & 0.98 & 1.00 & 0.51 & 0.58 & 0.18 $\pm$ 0.13 & 0.45 $\pm$ 0.16 & -2.54 $\pm$ 0.17 & 59.85 & 17.62 \\
GRTAUS9B$^{4}$ & 12 & 33 & 14 & 171 $\pm$ 9 & 306 & 160 & 0.97 & 1.00 & 0.57 & 0.66 & 0.22 $\pm$ 0.17 & 0.60 $\pm$ 0.15 & -2.55 $\pm$ 0.07 & 58.63 & 10.75 \\
TAUMGLIU9$^{4}$ & 8 & 61 & 17 & 171 $\pm$ 9 & 306 & 160 & 0.85 & 1.00 & 0.57 & 0.76 & 0.22 $\pm$ 0.17 & 0.60 $\pm$ 0.15 & -2.55 $\pm$ 0.07 & 58.63 & 18.62 \\
GRSCOS27C$^{5}$ & 15 & 111 & 60 & 106 $\pm$ 4 & 294 & 150 & 0.17 & 0.27 & 1.00 & 0.94 & 0.32 $\pm$ 0.17 & 0.45 $\pm$ 0.17 & -2.15 $\pm$ 0.15 & 64.23 & 49.15 \\
RATZSCO23$^{5}$ & 10 & 25 & 8 & 106 $\pm$ 4 & 294 & 150 & 0.33 & 0.62 & 1.00 & 1.00 & 0.32 $\pm$ 0.17 & 0.45 $\pm$ 0.17 & -2.15 $\pm$ 0.15 & 64.23 & 12.40 \\
GRSCOS10$^{6}$ & 20 & 33 & 15 & 149 $\pm$ 3 & 64 & 41 & 0.76 & 1.00 & 1.00 & 0.84 & 0.40 $\pm$ 0.16 & 0.55 $\pm$ 0.16 & -2.88 $\pm$ 0.17 & 22.29 & 18.03 \\
RATZSCO28$^{6}$ & 14 & 29 & 14 & 149 $\pm$ 3 & 64 & 41 & 0.48 & 0.76 & 1.00 & 0.93 & 0.40 $\pm$ 0.16 & 0.55 $\pm$ 0.16 & -2.88 $\pm$ 0.17 & 22.29 & 18.72 \\
HSC1368$^{7}$ & 47 & 87 & 18 & 183 $\pm$ 8 & 134 & 75 & 0.49 & 0.83 & 1.00 & 0.88 & 0.22 $\pm$ 0.12 & 0.35 $\pm$ 0.17 & -4.32 $\pm$ 0.10 & 25.04 & 30.24 \\
TAUMGLIU18$^{7}$ & 39 & 29 & 1 & 183 $\pm$ 8 & 134 & 75 & 0.52 & 0.92 & 1.00 & 1.00 & 0.22 $\pm$ 0.12 & 0.35 $\pm$ 0.17 & -4.32 $\pm$ 0.10 & 25.04 & 5.40 \\
TAUMGLIU19$^{7}$ & 40 & 18 & 2 & 183 $\pm$ 8 & 134 & 75 & 0.78 & 0.78 & 1.00 & 1.00 & 0.22 $\pm$ 0.12 & 0.35 $\pm$ 0.17 & -4.32 $\pm$ 0.10 & 25.04 & 4.88 \\
GRTAUS3A$^{8}$ & 19 & 61 & 12 & 163 $\pm$ 8 & 66 & 39 & 0.44 & 0.80 & 0.88 & 0.84 & 0.25 $\pm$ 0.16 & 0.55 $\pm$ 0.16 & -2.21 $\pm$ 0.13 & 14.50 & 10.93 \\
GRTAUS3B$^{8}$ & 14 & 10 & 2 & 163 $\pm$ 8 & 66 & 39 & 1.00 & 1.00 & 0.88 & 1.00 & 0.25 $\pm$ 0.16 & 0.55 $\pm$ 0.16 & -2.21 $\pm$ 0.13 & 14.50 & 2.95 \\
TAUMGLIU15$^{9}$ & 29 & 28 & 4 & 189 $\pm$ 8 & 375 & 216 & 0.78 & 0.85 & 0.80 & 0.61 & 0.22 $\pm$ 0.14 & 0.40 $\pm$ 0.17 & -1.51 $\pm$ 0.18 & 116.36 & 13.85 \\
TAUMGLIU20$^{9}$ & 42 & 42 & 13 & 189 $\pm$ 8 & 375 & 216 & 0.93 & 1.00 & 0.80 & 0.55 & 0.22 $\pm$ 0.14 & 0.40 $\pm$ 0.17 & -1.51 $\pm$ 0.18 & 116.36 & 15.98 \\
TAUMGLIU21$^{9}$ & 40 & 36 & 10 & 189 $\pm$ 8 & 375 & 216 & 0.97 & 1.00 & 0.80 & 0.47 & 0.22 $\pm$ 0.14 & 0.40 $\pm$ 0.17 & -1.51 $\pm$ 0.18 & 116.36 & 13.08 \\
\hline
Mean & 34 & 93 & 34 & 140 $\pm$ 11 & 228 & 126 & 0.59 & 0.71 & 0.86 & 0.81 & 0.27 $\pm$ 0.06 & 0.45 $\pm$ 0.09 & -2.27 $\pm$ 0.75 & 61.35 & 33.91 \\
All & - & 4104 & 1474 & - & 7530 & 4166 & 0.44 & 0.54 & 0.84 & 0.76 & 0.25 $\pm$ 0.16 & 0.45 $\pm$ 0.17 & -2.26 $\pm$ 0.09 & 2024.41 & 1492.03 \\
\hline
\hline
\end{tabular}
\end{center}
\label{table:knowngroups}
\par \textbf{Description of the columns.} 
Name: names of the groups from $S_{\text{lit}}$ recovered using the detection technique presented here. Identical superscripts denote literature groups that were detected as a single group, which we named as indicated by the subscript. 
$A_{\text{lit}}$: ages from the MOCA catalogue \citep{gagne_2024}. 
$N$: total number of sources in $S_{\text{lit}}$. 
$N_{\text{RV}}$: sources in $S_{\text{lit}}$ with RVs from GDR3. 
$d_{\text{mean}}$: mean distance of the detected group. 
$f_r$ and $f_{r,0.1}$: recovery relative to the $S_{\text{lit}}$ sources with $r_c>0$ and $r_c>0.1$, respectively (see Section \ref{subsec:purity_completeness_nymgs}). 
$f_p$ and $f_{p,0.1}$: purity estimated from the purity curves (see Section \ref{subsec:purity_completeness_nymgs}) and the $S_{\text{lit}}$ sources with $r_c>0$ and $r_c>0.1$, respectively. 
$N_{\text{det}}$ and $N_{\text{det,0.1}}$: number of candidate members with $r_c>0$ and $r_c>0.1$, respectively. 
$m_c$, $\sigma_c$ and $\alpha$: characteristic mass, characteristic width, and slope of the parametrization used to represent the IMFs of the detected groups (see Section \ref{subsec:imf_parametrization}). The row All for these columns shows the values for the mean normalized IMF. 
Mass and Mass ($S_{\text{lit}}$): mass of the group estimated from integrating the IMF based on our detection and the $S_{\text{lit}}$ members respectively.
 $^1$EMUTAU group, $^2$TAUMGLIU-A group, $^3$THOR-118TAU group, $^4$TAUMGLIU-GRTAUS group, and $^5$GRSCOS-RATZSCO-A group, $^6$GRSCOS-RATZSCO-B group, $^7$TAUMGLIU-HSC group, $^8$GRTAUS3 group and $^9$TAUMGLIU-B group. \\
\end{flushleft}
\end{table*}

Our detection algorithm classified the known groups TAUMGLIU9 and GRTAUS9B from $S_{\text{lit}}$ as one group which we call TAUMGLIU-GRTAUS. This means that the sources classified by our algorithm as candidate members share a common region of the RPS. Similarly, our detection algorithm detected the following group candidates that merge different sets of known NYMGs from MOCA: RATZSCO23 and GRSCOS27C as GRSCOS-RATZSCO-A; ETAU and MUTAU as EMUTAU; TAUMGLIU7 and TAUMGLIU8 as TAUMGLIU-A; THOR and 118TAU as THOR-118TAU; GRTAUS9B and TAUMGLIU9 as TAUMGLIU-GRTAUS; RATZSCO-28 and GRSCOS10 as GRSCOS-RATZSCO-B; TAUMGLIU18, TAUMGLIU19 and HSC1368 as TAUMGLIU-HSC; GRTAUS3A and GRTAUS3B as GRTAUS3; and TAUMGLIU15, TAUMGLIU20 and TAUMGLIU21 as TAUMGLIU-B. In all of these cases, we find that the $S_{\text{lit}}$ members of the MOCA groups of each detected group share a common sequence in the CMD, likely corresponding to young isochrones. 

In the specific case of EMUTAU, we notice that the $S_{\text{lit}}$ members of the MOCA group MUTAU are more scattered in the CMD than the members of EMUTAU. We observe a similar behaviour for TAUMGLIU9 within TAUMGLIU-GRTAUS, HSC1368 within TAUMGLIU-HSC, and GRTAUS3A within GRTAUS3. In each one of these cases, our detection algorithm alone does not provide the necessary tools to conclude if the different MOCA groups within each detected group are indeed separate physical associations or just sub-structures of the same association. In the case of RATZSCO23 and GRSCOS27C for example, it has been shown that these are likely to be separate groups \citep{kerr2021,ratzenbock_2022}, however their similar ages and proximity in phase-space make it difficult for our algorithm to differentiate them. Additionally, as a consequence of our algorithm prioritizing purity over recovery by design, the spread of these two groups in the CMD along their respective potential isochrones may have been diminished, leading to an even more important degeneracy for our method.

We detected a total of 7530 candidate members across the identified groups. In order to lower the level of contamination from field stars, we discarded the source candidates with $r_c<0.1$, leaving only 4166 (55\% of all detected candidates) source candidates in the final set of recovered NYMG candidates. The value of the threshold on $r_c$ was chosen by testing which value would maximize the amount of discarded MS stars that are photometrically older than 100 Myr and minimize the amount of discarded stars from the PMS. To identify stars older than a certain age $\mathcal{A}$ on the MS, we simply selected stars from the MS that are below the Turn-on locus on the CMD where the isochrone of age $\mathcal{A}$ transitions from the PMS into the MS. Figure \ref{fig:CMD_starting_sample} shows the mass and $M_G$ of the Turn-on of different isochrones as an example. The initial list of the 7530 candidate members including the final list of 4166 candidate members as well as the general statistics and information of the groups from Table \ref{table:knowngroups} is available at the CDS. 

\begin{figure}
  \resizebox{\hsize}{!}{\includegraphics{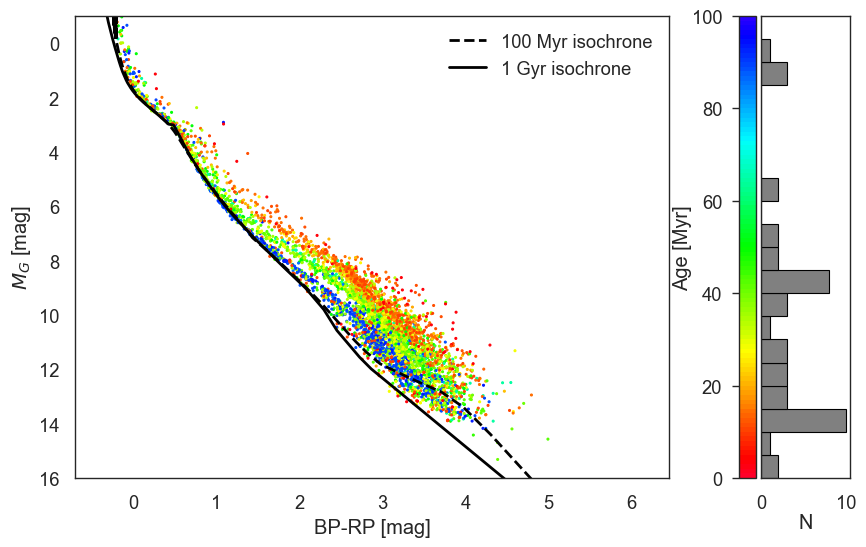}}
  \caption{CMD distribution of all the 4166 candidate members detected in this work. The colour map shows the age of the identified groups according to MOCA. The black solid and dashed curve show the 1 Gyr and 100 Myr isochrones respectively. The rotated histogram shows the distribution of the detected NYMG ages according to MOCA. The age-axis of the histogram is the same as the colour bar axis.}
  \label{fig:allCmds}
\end{figure}

Figure \ref{fig:allCmds} shows a CMD distribution of all the 4166 detected candidate members from this work. We can see that the ages from MOCA correlate well with sequences that are very likely to be isochrones. This means that each of the detected NYMG candidates appears to follow an isochrone on the CMD, which strongly suggests that stars of a same NYMG were born together. We also notice that the group candidates are mostly distributed into three age categories: the youngest groups with ages around $\mathcal{A}\sim10$ Myr, the middle-aged groups with $\mathcal{A}\sim40$ Myr, and the old groups with $\mathcal{A}\sim90$ Myr.

\subsection{Estimating the purity and recovery rates of the NYMGs}
\label{subsec:purity_completeness_nymgs}

\begin{figure}
  \resizebox{\hsize}{!}{\includegraphics{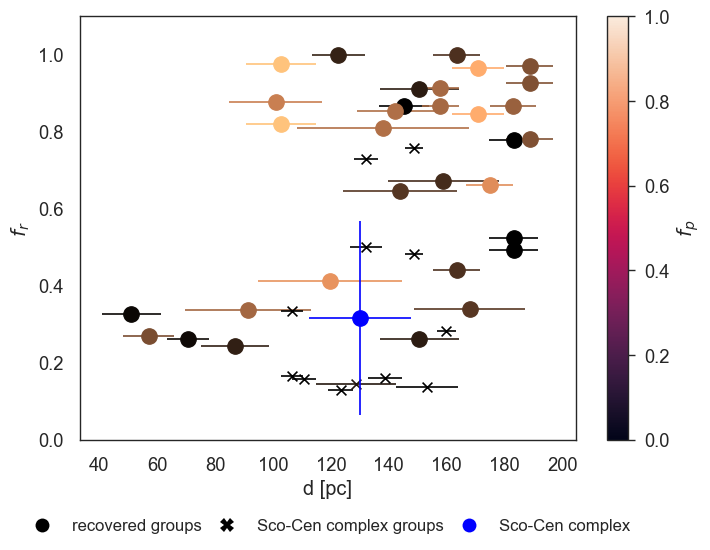}}
  \caption{Recovery rates $f_r$ as a function of the mean distance of each detected group. The colour indicates the purity rate $f_p$. The dots correspond to all the recovered groups that are not part of the Sco-Cen complex while the crosses correspond to the ones that do belong. The blue dot indicates the mean recovery of the Sco-Cen complex groups.}
  \label{fig:recovery_vs_d}
\end{figure}

We estimated the recovery rate $f_r$ of each group candidate as the fraction $n_{\text{det},\text{lit}}/n_{\text{lit}}$ where $n_{\text{det},\text{lit}}$ corresponds to the number of members of the group from $S_{\text{lit}}$ detected as members of the group with $r_c>0.1$ and $n_{\text{lit}}$ corresponds to the total number of members of the group from $S_{\text{lit}}$. We found a total $f_r$ relative to the 44 recovered groups of 44\% and an average $f_r$ per group of 59\%. Figure \ref{fig:recovery_vs_d} shows the $f_r$ of each group as a function of distance. If we ignore the groups that are part of the Sco-Cen complex, we see that on average, $f_r$ appears to increase with heliocentric distance. This can be explained partially by the fact that groups that are closer to the Sun appear more scattered in the sky, leading to a higher dispersion in sky tangential velocities, which makes the detection of their members more challenging. Although the values of $f_r$ may seem low for the closest recovered NYMGs, these values are based on the members from the literature, so the presence of any contamination in the literature members that would have survived the selection made in Section \ref{subsec:known_nymgs} will lower the real recovery rates. Regarding the groups from the Sco-Cen complex, we notice that they cover a large range of recovery rates ($\Delta f_r\sim0.6$) for a short range of distances ($\Delta d\sim60$~pc). However, we can notice that the mean $f_r$ of the groups from the Sco-Cen complex does follow the trend of the other groups.

\begin{figure*}
\includegraphics[width=180mm]{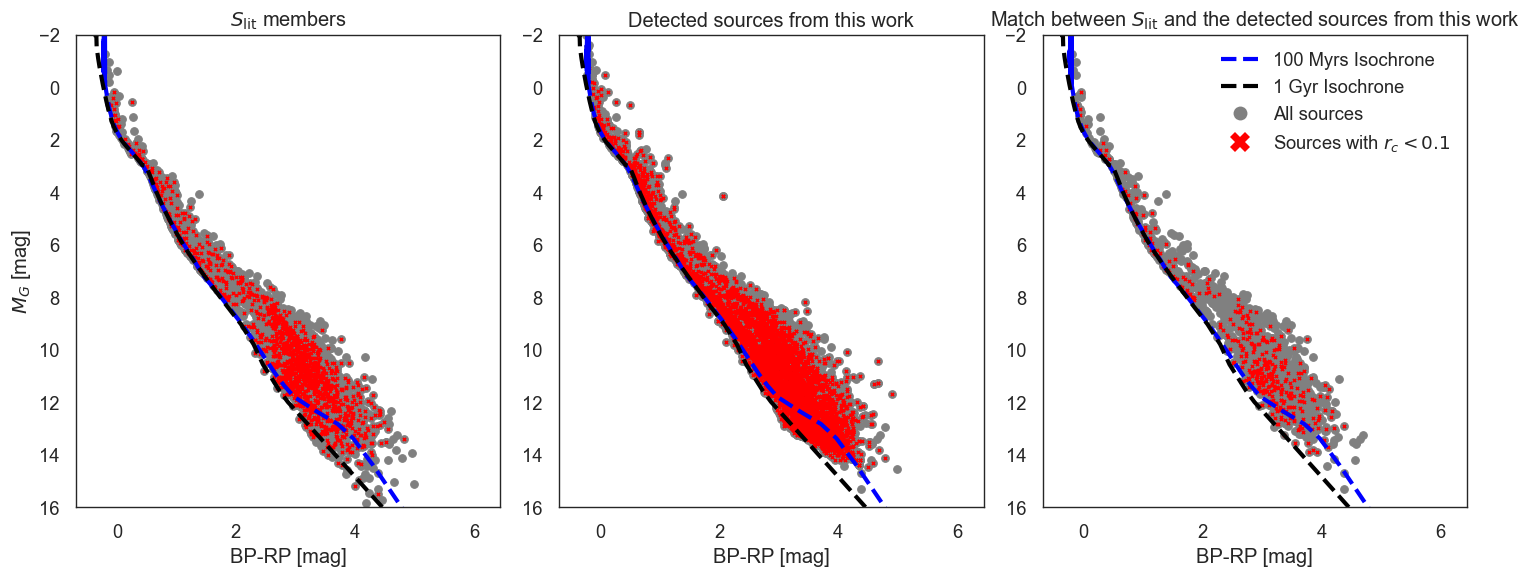}
\caption{$M_G$ vs. BP-RP CMD of the candidate members of the NYMG candidates before applying the $r_c>0.1$ cut (grey dots), highlighting the sources that are discarded with this cut (red crosses). The dashed lines indicate the 100 Myr (blue line) and 1 Gyr (black line) isochrones. Top, middle, and bottom plots correspond, respectively, to the members according to the candidates of $S_{\text{lit}}$, the candidates of the detected groups from this work, and the exact match between the two latter samples.}
\label{fig:detections_and_contaminants}
\end{figure*}

Figure \ref{fig:detections_and_contaminants} shows the CMD of the candidates from
$S_{\text{lit}}$, the 7530 original source candidates from this work with $r_c>0$, and the recovered source candidates. We can see that many of 
the sources from $S_{\text{lit}}$ and the original 7530 candidates from our 
work are located in the MS locus, some of which are located below the Turn-on of 100 Myr. 
Some of those stars might very likely correspond to field star contaminants. To 
test this possibility, we fitted a Gaussian to each NYMG from $S_{\text{lit}}$ 
in the RPS and selected the sources with $r_c>0.1$. In the case of the $S_{\text{lit}}$ members, we find that $\sim51$\% of the MS stars have $r_c<0.1$ while only $\sim38$\% of the PMS stars have $r_c<0.1$. Similarly, in the case of the original 7530 detected candidates from this work, we find that $\sim76$\% of the MS stars fulfil the condition $r_c<0.1$ and $\sim55$\% of the PMS stars have $r_c<0.1$. The latter statistic tells us that for both the $S_{\text{lit}}$ sample and the detections from this work, the $r_c$ parameter suggests that stars from the MS tend to be more scattered than the ones from the PMS. A first potential explanation could be that older groups tend to be more kinematically scattered than younger groups, but we do not notice from Table \ref{table:knowngroups} any correlation between the age and the purity rate $f_p$ of the groups, which quantifies the fraction of candidate members that are true members. We present the method to estimate $f_p$ in Section \ref{subsec:purity_completeness_nymgs}. A second more likely explanation could be that these statistics are simply a consequence of the fact that the fraction of field contaminants in the MS is way higher than in the PMS. If this is the case, it means that the $r_c>0.1$ threshold is a good criterion to lower the level of contaminants within the detected groups. 
 
Based on the previous analysis, we computed a likely less contaminated recovery rate $f_{r,0.1}=n_{\text{det},\text{lit},0.1}/n_{\text{lit},0.1}$ where the subindex $0.1$ indicates the threshold $r_c>0.1$ on the $S_{\text{lit}}$ members. With this new recovery rate, we estimated a total $f_{r,0.1}$ of 71\% and an average $f_{r,0.1}$ of 54\% per cluster. Additionally, we can estimate the level of purity of each group according to the literature as $f_{p,0.1}=n_{\text{det},\text{lit},0.1}/n_{\text{det},\text{lit}}$. This purity parameter measures the fraction of recovered sources with $r_c>0.1$ over the number of recovered sources. We obtained a total $f_{p,0.1}$ of 76\% and a mean $f_{p,0.1}$ of 81\% per group.

\section{IMF of selected NYMGs}
\label{sec:imf_of_NYMGs}
In this section we infer the individual IMFs of the 33 detected NYMGs from the $S_c$ samples as well as the individual IMFs of the 44 recovered groups based on their members from the $S_{\text{lit}}$ sample. This involves the estimation of masses for the member candidates and the stars added for completeness using the GSFs. We first show how to infer masses of the individual sources from their photometry and parallaxes. Then, we present the resulting IMFs of each group after correcting them for contamination due to field stars in each mass bin and compare them with the IMFs from the $S_{\text{lit}}$ members. Finally, we obtain the mean normalized IMFs of the detected and recovered groups and compare them to each other and to results from the literature.

\subsection{Inferring individual masses and ages}
\label{subsec:infering_masses}

For a source with apparent magnitude $G$, colour $BR = BP-RP$, measured parallax $\varpi$, assumed extinction vector $\vec{A}=(A_{\text{BR}},A_G)$, and assumed age $\mathcal{A}$, we can infer its true parallax $\varpi_0$, true extinction $\vec{A}_0$ and mass $m$ by following Bayes' rule:
{\small
\begin{equation}\label{eq_imf_bayes_0}
    P_{\mathcal{A}}(m,\varpi_0,\vec{A}_0|G,BR,\varpi,\vec{A}) \propto 
    \mathcal{L}_{\mathcal{A}}(G,BR,\varpi,\vec{A}|m,\varpi_0,\vec{A}_0)\, 
    \mathcal{P}(m,\varpi_0,\vec{A}_0)
\end{equation}
}
Assuming an agnostic prior on $\varpi_0$ and $\vec{A}_0$ we have that $\mathcal{P}(m,\varpi_0,\vec{A}_0)=\mathcal{P}(m)$, which is by definition our a priori assumption of the IMF. We can then infer the mass alone by marginalizing on $\varpi_0$ and $\vec{A}_0$:
{\small
\begin{equation}\label{eq_imf_bayes_1}
    P_{\mathcal{A}}(m|G,BR,\varpi,\vec{A})\propto\displaystyle\int\mathcal{L}_\mathcal{A}(G,BR,\varpi,\vec{A}|m,\varpi_0,\vec{A}_0)\,d\varpi_0\,dA_G\,dA_{\text{BR}}~\mathcal{P}(m)
\end{equation}
}
Assuming that $G$, $BP$, $RP$, and $\varpi$ have Gaussian errors, and that for a given NYMG the distribution of $A_G$ and $A_{\text{BR}}$ are Gaussian, we can then approximate the likelihood as a product of normal distributions:
\begin{equation}\label{eq_likelihood}
\mathcal{L}_{\mathcal{A}}(G,BR,\varpi,\vec{A}|m,\varpi_0,\vec{A}_0)=\displaystyle\prod_{i=1}^5\frac{1}{\sqrt{2\pi}\sigma_{o_i}}e^{-}\frac{(o_i-o_{i,0})^2}{2\sigma_{o_i}^2}
\end{equation}
where $o_i\in\{G,\text{BR},\varpi,A_G,A_{\text{BR}}\}$. In this notation, $o_i$ corresponds to the measured or assumed parameter, while $o_{i,0}$ corresponds to the value being inferred or integrated. In the particular cases of $G_0$ and $\text{BR}_0$, these are functions of $m$, $\varpi_0$, $\vec{A}_0$, and the age $\mathcal{A}$ computed as
\[
G_0(m,\varpi_0,A_{G,0},\mathcal{A})=M_{G,\mathcal{A}}(m)-5+5\log_{10}(1/\varpi_0)+A_{G,0}
\]
and
\[
\text{BR}_0(m,A_{\text{BR},0},\mathcal{A})=M_{\text{BP},\mathcal{A}}(m)-M_{\text{RP},\mathcal{A}}(m)+A_{\text{BR},0},
\]
where $M_{G,\mathcal{A}}(m)$, $M_{\text{BP},\mathcal{A}}(m)$, and $M_{\text{RP},\mathcal{A}}(m)$ are the mass–magnitude relations for Gaia passbands interpolated from the isochrone of age $\mathcal{A}$ resulting from the combination of the same models used for the isochrones presented in Section~\ref{subsec:GDR3} \citep{baraffe2015, marigo2017, Phillips_2020}. 

To infer mass from observables $\{G,\text{BR},\varpi,A_G,A_{\text{BR}}\}$, we used Equations \eqref{eq_imf_bayes_1} and \eqref{eq_likelihood} assuming for the prior the \cite{chabrier2005b} system IMF, 
described by a log-normal distribution with a characteristic mass 
$m_c=0.2M_\odot$ and $\sigma=0.6$ for masses lower than $1~M_\odot$, 
and a \cite{salpeter1955} power-law for higher masses. We consider this model because it is representative of disc stellar populations with unresolved stellar systems \citep{henebelle_2024}. For the extinctions, we used the mean extinction $A_V$ and its standard deviation estimated for each NYMG from the 3D maps from \cite{Gontcharov2017}, and used it to estimate the corresponding values in Gaia bands following the same procedure presented in Section~\ref{subsec:GDR3}.

With this method, if we know the age $\mathcal{A}$ of a star, we can simply find the mass $m$ that maximizes the posterior. In this work, we used the ages from MOCA reported in Table \ref{table:knowngroups} to infer the masses.

\subsection{Accounting for contamination}
\label{subsec:contamination}

The $S_c$ sample corresponds to $S_{\text{NYMG}}$ corrected for incompleteness 
due to most observational biases. Then, the resulting IMFs are already corrected for these incompletenesses. 
We present here the correction of the IMFs for contamination from 
the overlap of Besan\c{c}on field stars and NYMGs in the RPS. 
To estimate the fraction of contaminants per mass bin for each group, 
we start by identifying the source candidate members of the group that were 
classified by the DBSCAN as core-points using the hyper-parameter combination 
used for the detection of the group. Then, we generate a synthetic field from the 
model built in Section \ref{subsec:significance_and_synthetic_field} and select 
the synthetic sources that are classified by the DBSCAN as neighbours 
of the core points of the group. We infer the masses of these synthetic stars 
using the same isochrones used to infer the masses of the 
source candidates of the NYMG and measure in each bin of mass $\Delta m$ the 
purity fraction as $f_p=n_{\text{group},\Delta m}/(n_{\text{field},\Delta m}+n_{\text{group},\Delta m})$, where $n_{\text{group},\Delta m}$ 
and $n_{\text{field},\Delta m}$ are the number of source 
candidates and synthetic sources, respectively. We interpolate 
the purity as a function of mass and repeat the whole process for 200 
random synthetic fields. Finally, we compute the average purity function 
$f_p(m)$ of the 200 fields and smooth it using a Gaussian Kernel at 1-$\sigma$. 
As shown in Table \ref{table:knowngroups}, we estimated an average purity 
per NYMG of $\langle f_p\rangle$ of 86\% and a total purity of 84\%, which are 
similar values to the values of $f_{p,0.1}$ estimated from the $S_{\text{lit}}$ 
recovered members.

\begin{figure}
  \resizebox{\hsize}{!}{\includegraphics{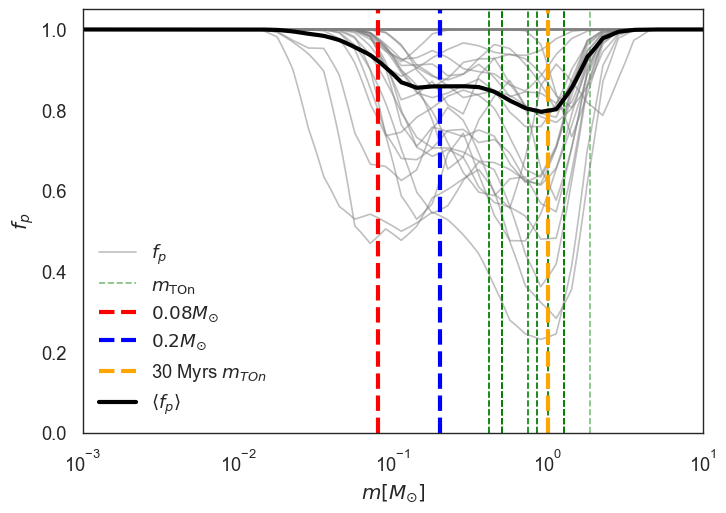}}
  \caption{Purity functions $f_p(m)$ estimated from the simulated fields for all 
the detected NYMGs (grey curves) and the mean purity $\langle f_p(m)\rangle$ 
(black curve). The red, blue, and green vertical dashed lines indicate respectively 
the $0.08~M_{\odot}$, $0.2~M_{\odot}$, and the masses of the TO of the different 
ages used to photometrically select the NYMGs.}
  \label{fig:all_purities}
\end{figure}

Figure \ref{fig:all_purities} shows the purity curves $f_p(m)$ for each of 
the 33 detected NYMG candidates as well as the mean $\langle f_p\rangle(m)$,
which, in most cases, shows between one and three minima. More particularly, 
$\langle f_p\rangle$ shows two local minima. 
For a given group, we expect to encounter higher contamination in mass ranges 
corresponding to the most populated loci of the CMD that survive the photometric 
isochrone selection of the detection algorithm. We observe in Figure
\ref{fig:CMD_starting_sample} that apart from the MS, there are two other loci 
with a high density of stars. The first is located in the locus between the 
$0.1~M_{\odot}$ and $0.3~M_{\odot}$ evolutionary tracks, which corresponds to the characteristic mass range of the typical IMF of field stars in the solar neighbourhood. 
This peak in density can be explained by the fact that since the magnitude $M$ of a star is proportional to the logarithm of its luminosity $L$ and 
if we assume that its luminosity is proportional to a power of 
the mass, then we have $M\propto \log(m)$, which implies that 
if the typical IMF of the solar neighbourhood field stars
shows a peak around the characteristic mass, we should 
also see a peak in the CMD around the locus associated with the evolutionary tracks 
of such mass. The second region of high density corresponds to the locus between 
the $\sim30$ Myr Turn-on and the 1 Gyr Turn-Off where many isochrones older than 
1 Gyr intersect the isochrones younger than $\sim30$ Myr. These two loci of high 
density in the CMD could explain why one of the minima of $\langle f_p\rangle(m)$ 
is located around the peak of the field IMF and the other one right after the 30 Myr isochrone Turn-on. Additionally, for a given isochrone, we 
should also expect contamination around its Turn-on as it is the closest MS locus 
of that isochrone to the peak of the field IMF. Depending on the age of the isochrone 
and, thus, on how close its Turn-on is to the peak of the IMF and the locus between the 
$\sim30$ Myr Turn-on and the 1 Gyr Turn-Off, the contamination caused by the Turn-on
may blend with the $f_p$ minimum associated with one of the other two highly dense loci 
or simply produce a new $f_p$ minimum.

We notice that some $f_p(m)$ curves predict very high rates of contamination with minima as low as $\sim0.2$ in certain mass ranges. Although such contamination may be true, 
it could also be explained by the fact that the synthetic fields used to estimate 
$f_p(m)$ do not include the presence of RPS over-densities produced by massive 
moving groups and open clusters but have the same number of objects as in $S_c$, 
as discussed in Section \ref{subsec:significance_and_synthetic_field}. This means 
that in regions of the RPS where there are no massive stellar associations, our 
model of the field will over-estimate the density of stars. Similarly, $f_p(m)$ 
will tend to underestimate the contamination in regions of the RPS that are 
close to some massive stellar association. The impact of under-or overestimating $f_p(m)$ is hard to predict as it is case-dependent, but we discuss this in Sections \ref{subsec:inferred_imf} and \ref{subsec:imf_parametrization}.

Finally, contamination can also be produced by extinction as some stars from the MS that share kinematics with a younger group could be reddened and dimmed in the CMD towards the isochrone of the group. However, since our model of the Besan\c{c}on field of stars is based on GOG, whose distribution in the CMD and phase space is based on Gaia photometric errors, the effect of extinction is encoded in our model of the field. This means that our estimation of purity already considers the effect of extinction.

\subsection{Resulting IMFs}
\label{subsec:inferred_imf}

We estimated the IMF distribution by computing the KDE of the mass 
distribution of each group using a Gaussian Kernel and a 
bandwidth of $d\text{log}_{10}(m/M_{\odot})=0.1$. We consider this bandwidth because of a combination of factors: it corresponds to the 
typical mass bin size used in the literature to represent mass functions with histograms, and 95\% of the errors on the inferred masses 
of all the source candidates are smaller than this value. We then estimated the IMF of
each group as $\Phi(m)=\Phi^\prime(m)f_p(m)$, where $\Phi^\prime(m)$ 
corresponds to the KDE of the original mass distribution. The final reported IMF is the one corrected by the purity curves: $\Phi(m)$. Finally, we used the 
mean square error to fit a log-normal 
$\Phi(m)\propto \exp\!\left(\frac{(\text{log}_{10}(m)-\text{log}_{10}(m_c))^2}{2\sigma_c^2}\right)$ to the KDE $\Phi(m)$ in the mass range $m_{\text{min}}<m<1~M_{\odot}$ and a power-law $\phi(m)\propto m^{\alpha}$
in the mass range $1~M_{\odot}<m<m_{\text{max}}$. Here, $m_{\text{min}}$ and $m_{\text{max}}$ are defined as the points where $\Phi(m)$ falls below $\sim0.54$. 
This threshold comes from the fact that, with a Gaussian KDE bandwidth of $0.1$, a single object produces 
a normal distribution whose PDF at $\mu \pm 2\sigma$ is $\sim0.54$. 
Continuity was 
not imposed between the two parametrizations. Table \ref{table:knowngroups} shows the fitted 
parameters $m_c$, $\sigma_c$, and $\alpha$ of the inferred IMFs of the 33 detected groups.

\begin{figure*}
\includegraphics[width=60mm]{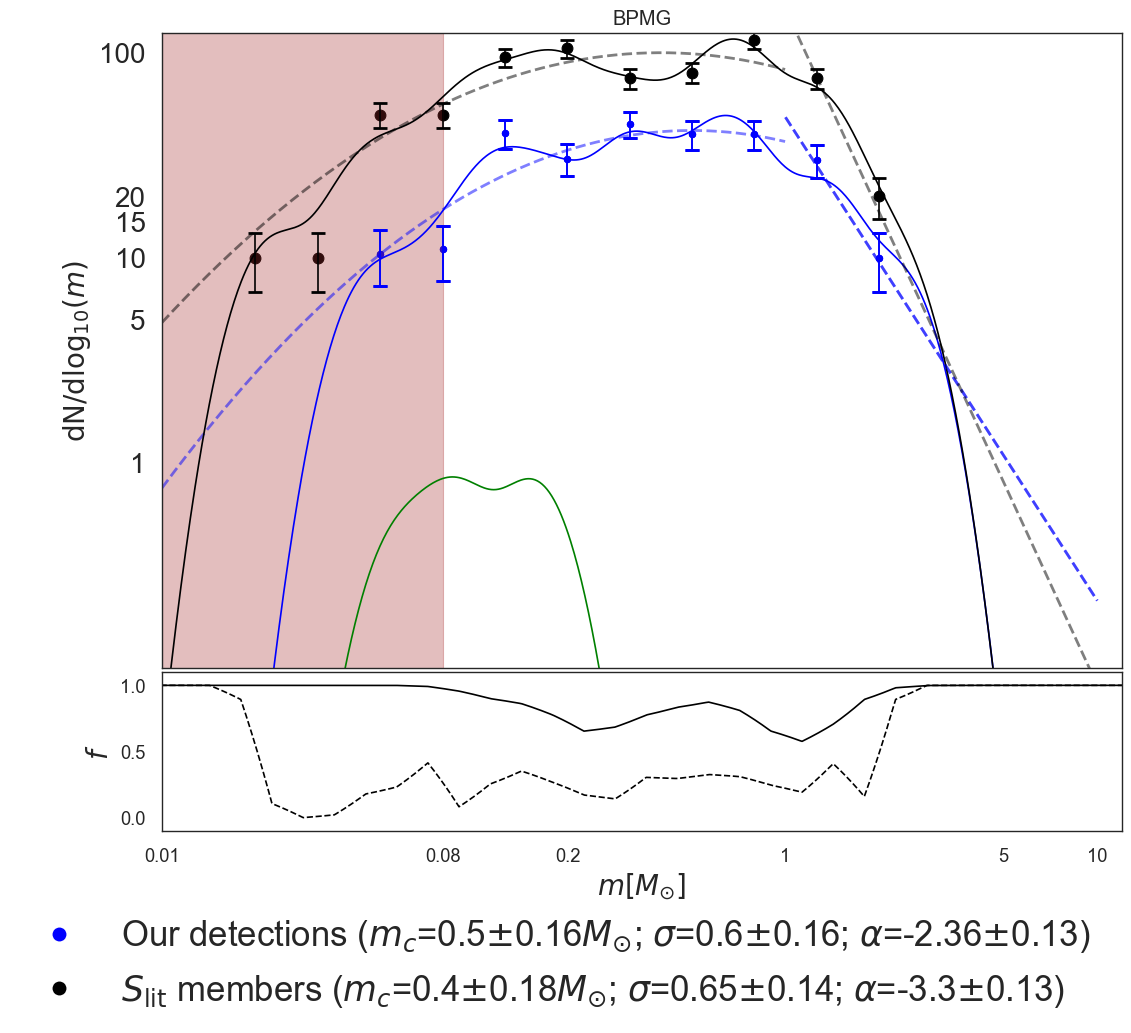}
\includegraphics[width=60mm]{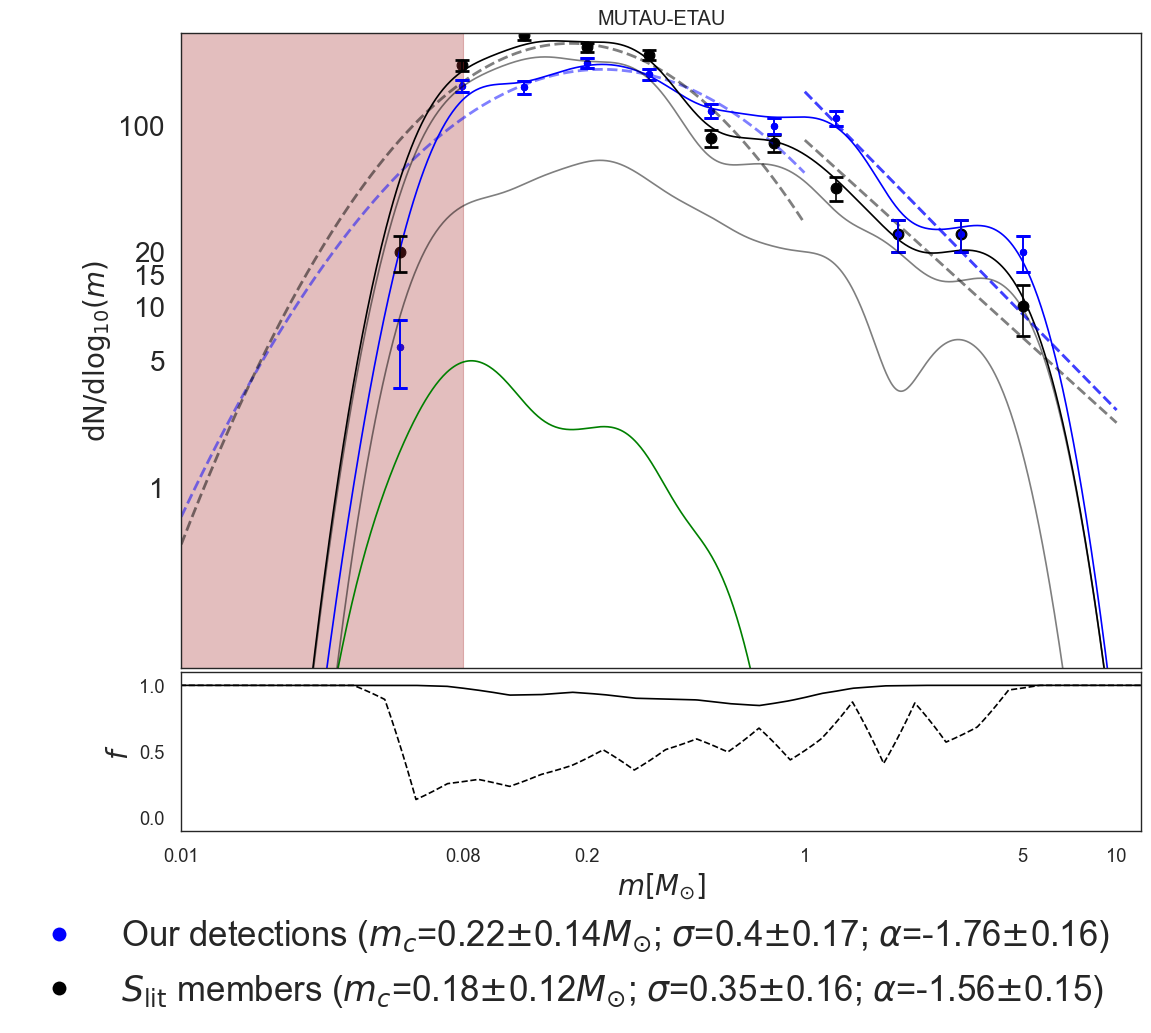}
\includegraphics[width=60mm]{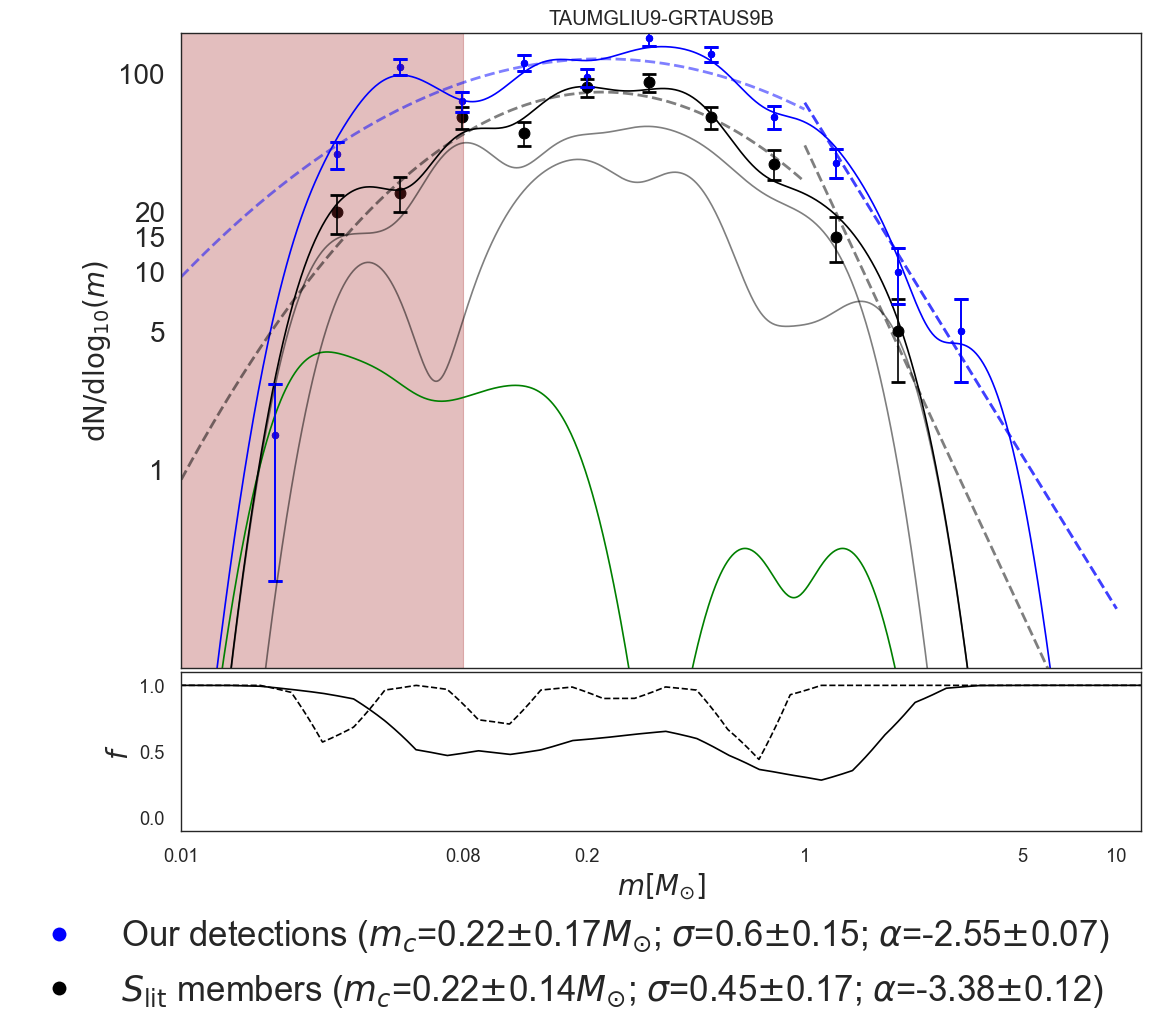}
\caption{
Characteristic inferred IMFs which illustrate the different cases found among the 33 groups: $\beta$-Pictoris (left panel), EMUTAU (middle panel), and TAUMGLIU-GRTAUS (right panel). Solid and dashed lines in the bottom panels show respectively the purity and recovery functions. The dots with error bars correspond to the histogram of the distributions with Poisson error bars using $\text{log}_{10}(0.1~M_{\odot})$ bins. The smooth solid curves correspond to the KDEs of the masses using a Gaussian Kernel and bandwidth of $\text{log}_{10}(0.1~M_{\odot})$. The areas of both the histogram and the KDE are normalized to the total number of members of the group. The green solid line correspond to the IMF of the synthetic sources of $S_c$ detected as members of the groups. The grey solid curves correspond to the IMFs of the the individual MOCA groups for the cases of EMUTAU and TAUMGLIU-GRTAUS. The dashed lines in the top panels represent the log-normal and power-law parametrization of the KDE. The colours black and blue correspond to the results based on the members from $S_{\text{lit}}$ and our detections respectively. The brown region represents the mass range of brown dwarfs ($m\lesssim 0.08~M_{\odot}$).The value of the fitted parameters are indicated in the legend. In the cases where more than one NYMG from MOCA was detected, we show their individual IMFs in grey while the total IMF is in black.}
\label{fig:single_imf}
\end{figure*}

Figure \ref{fig:single_imf} shows the inferred IMFs for the particular cases of groups $\beta$-Pictoris (BPMG), EMUTAU, and TAUMGLIU-GRTAUS with their corresponding purity functions $f_p(m)$. As a comparison, we also show the IMFs we inferred considering only the members from $S_{\text{lit}}$.
We computed the KDEs of these mass distributions and parametrized them in the same way as we did for our detections. For each group, we show their respective purity function $f_p(m)$ as well as the recovery rates as a function of mass $f_r(m)$. These three individual cases illustrate some of the different cases found among the 33 groups. The same plots for the remaining 30 groups are shown in Appendix \ref{appendix_individual_imfs}.

For the specific case of BPMG, we detected only 56 candidate members, 36 of which are also reported in $S_{\text{lit}}$ as candidate members of BPMG. Meanwhile, this group according to $S_{\text{lit}}$ is made of 137 candidate members. Although we obtained similar parametrizations in the low-mass regime ($m<1~M_{\odot}$) using our detections and the members from $S_{\text{lit}}$, we notice that in the high-mass range ($m>1~M_{\odot}$) the fitted power-law is steeper for the IMF inferred from the $S_{\text{lit}}$ members than for our detections. Moreover, the slope ($\alpha=-2.36\pm0.13$) obtained from our detections is much closer to the typical \cite{salpeter1955} slope of $\alpha_{\text{Salpeter}}=-2.35$ than the one estimated from the $S_{\text{lit}}$ candidates ($\alpha=-3.30\pm0.13$). The discrepancy in the high-mass regime between the parametrizations based on our detections and the candidate members from $S_{\text{lit}}$ could be explained by the fact that the IMF based on our detection was corrected using $f_p(m)$. Indeed, the latter function presents a minimum around $1~M_{\odot}$, preventing contaminants from the Turn-ons of the isochrones with ages between 10 and 30 Myr mentioned in Section \ref{subsec:contamination} from increasing the number of sources in that mass range, which would have led to a steepening of the slope. This means that the steepening of the slope in the case of using the candidates from $S_{\text{lit}}$ could be explained by the presence of contaminants forming a bump on the IMF around the solar mass. From now on, we refer to this bump as the $1~M_{\odot}$ bump.

In the case of EMUTAU, we found similar parametrizations between our detections and $S_{\text{lit}}$ in the whole mass range with similar number of candidate members. Moreover, we notice for the $S_{\text{lit}}$ members that the shape of the combined distribution of e-Tau and $\mu$-Tau is representative of the distributions of the individual groups. Compared to the typical solar neighbourhood IMF ($\sigma_c=0.6$), we found narrow characteristic widths ($0.40\pm0.17$ and $0.35\pm0.16$ for our detections and the $S_{\text{lit}}$ members respectively) in the low mass regime. This is most likely due to a lack of brown dwarfs and other low mass objects. Such incompleteness is most likely due to the observational bias that we lack some of the faintest objects in $S_{\text{NYMG}}$ because of their low SNR leading for many of those sources to not make it through the selection process from Section \ref{subsec:GDR3}. Indeed, if we directly interpolate the $M_G$-mass relationship from the isochrone with the average age of e-Tau and $\mu$-Tau ($\sim55$ Myr) and use it to estimate the mass of a star with the minimum G magnitude in $S_{\text{NYMG}}$ ($\sim20$ mag) and at the distance of EMUTAU's closest member ($\sim120$ pc), we find that the minimum mass we could obtain this way for an EMUTAU member is $\sim0.05~M_{\odot}$. This explains why the IMF of this group based on both our detections and $S_{\text{lit}}$ drops very rapidly for masses $m<0.08~M_{\odot}$. In the high mass regime, we obtain shallow slopes ($-1.76\pm0.16$ and $-1.56\pm0.15$ for our detections and the $S_{\text{lit}}$ members respectively). This flattening can be explained by the presence of a higher amount of high mass stars than what is predicted by the typical $\alpha_{\text{Salpeter}}$. Although the $f_p(m)$ curve suggests that there is almost no contamination from the Besan\c{c}on field, this does not necessarily imply that the potential excess of high mass stars we detected is real as it could simply be the product of contaminants from other more massive clusters.

Finally, the case of TAUMGLIU-GRTAUS is a typical case in which we may have over-estimated the contamination rate over the whole range of mass, as we can see from Figure \ref{fig:single_imf} where $f_p(m)$ gets as low as $f_p(m)\sim0.4$. This potential over-estimation of contamination could explain why we obtained for our detections a steeper slope ($\alpha=-2.55\pm0.07$) than $\alpha_{\text{Salpeter}}$. We also notice the presence of a bump on the IMF around $m\sim0.05~M_{\odot}$. This may be explained by the fact that the purity curve starts to drop at that mass to $f_p(m)\sim0.5$, which means that this bump is probably not real, but rather a consequence of an over-estimation of the contamination rate for masses higher than $0.05~M_{\odot}$ or an under-estimation for lower masses. This potential over-estimation of contamination may be a consequence of the fact that this group covers a large region of the RPS, which makes it more likely to capture synthetic stars from our model of the Besan\c{c}on field during our estimation of $f_p(m)$.

\subsection{What the parametrized IMFs tell us}
\label{subsec:imf_parametrization}

\begin{figure}
  \resizebox{\hsize}{!}{\includegraphics{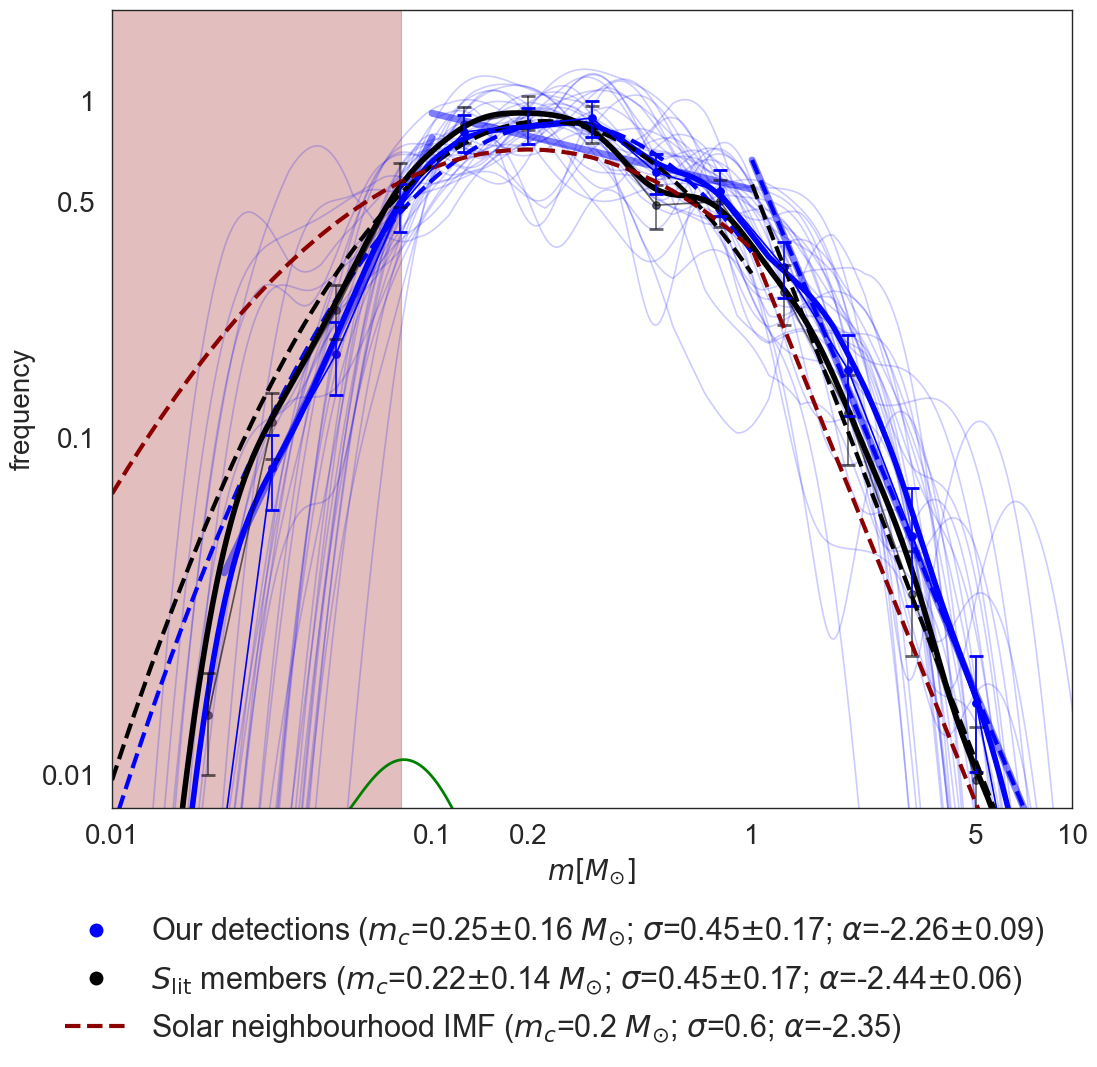}}
  \caption{Normalized individual IMF KDEs of the 33 groups (blue transparent curves), the mean normalized IMF from our detections (blue solid curve) and from the $S_{\text{lit}}$ groups (black). The dots with Poisson error bars show the mean normalized histogram of the masses of the groups and the dashed curves show the log-normal and power-law parametrizations of the mean normalized IMFs. The green curve shows the mean normalized IMF of the synthetic sources. The thick blue transparent lines show the triple power law fitted to the mean normalized IMF of our detections. The brown dashed line shows the typical IMF of the solar neighbourhood: \cite{chabrier2005b} log-normal for $m<1~M_{\odot}$ and \cite{salpeter1955} power-law for higher masses.}
  \label{fig:all_imfs}
\end{figure}

Although some maxima and minima in the inferred IMFs could be explained by potential contamination or incompleteness, others could simply be the result of Poisson noise since many groups present a low number of source candidates, which can highly affect the estimated parameters. Additionally, as shown with three different groups in Section \ref{subsec:inferred_imf}, each group has its own specific reasons to show potential contamination or incompleteness in different mass ranges. To decrease the impact of Poisson noise and identify systematic behaviours in the IMFs of the NYMGs, we normalized the 33 inferred IMFs based on our detections to their respective areas and computed the mean normalized IMF. The resulting IMF, together with its parametrization and the individual normalized IMFs are shown in Figure \ref{fig:all_imfs}. We also show the mean normalized IMF of the $S_{\text{lit}}$ members and as a visual guide the typical IMF of the solar neighbourhood as a \cite{chabrier2005b} log-normal for $m<1~M_{\odot}$ and a \cite{salpeter1955} power-law for higher masses. The latter was built based on the 44 individual MOCA groups that we recovered rather than the 33 combined groups based on our detections. The mean normalized IMF tells us on average what is the shape of the IMF of the NYMGs. We note that, given the ages and the typical number of members in the NYMGs, it is unlikely that they host massive stars which have already evolved off the main sequence. Therefore, the mean normalized IMF provides a representative average shape without significant bias from evolutionary depletion at the high-mass end. Additionally, since the mean normalized IMF of the NYMGs is well described by \cite{salpeter1955} slope in the high mass range, it is very likely that the number of high mass stars that are members of the NYMGs and that are not included in Gaia DR3 based on what was discussed in Section \ref{subsec:GDR3} is very low if not zero. Hence, our results suggest that these potentially missing O and B stars should not significantly affect our results.

As we can see in Figure \ref{fig:all_imfs}, the mean normalized IMF of our detections is very smooth in the whole mass range. This is also the case of the mean normalized IMF of the $S_{\text{lit}}$ members. However, we do notice in both cases a very slight bump between $\sim0.8$ and $\sim1~M_{\odot}$. This bump may correspond to the $1~M_{\odot}$ bump that is the product of contamination from some of the Turn-Ons of the NYMGs as discussed in the specific case of $\beta$-Pictoris in Section \ref{subsec:inferred_imf}. We notice that the bump for our detection is less significant than for the $S_{\text{lit}}$ members, which supports the idea that this bump is indeed the product of contamination since the IMFs based on our detections were corrected using the $f_p(m)$. This also suggests that our method to estimate contamination on average corrected well for the contamination from the Turn-Ons in our detections.

As in the cases of the individual IMFs, we fitted a log-normal and a power-law in the low ($m<1~M_{\odot}$) and high ($m>1~M_{\odot}$) mass ranges, respectively. We obtained parameter values $m_c=0.25\pm0.16~M_{\odot}$, $\sigma_c=0.45\pm0.17$, and $\alpha=-2.26\pm0.09$ for the mean normalized IMF based on our detections and $m_c=0.22\pm0.14~M_{\odot}$, $\sigma_c=0.45\pm0.17$, and $\alpha=-2.45\pm0.06$ for the one based on the $S_{\text{lit}}$ members. We notice that the high mass range slope is slightly steeper for the $S_{\text{lit}}$ members than for our detections. This difference may be produced by the potential contamination of the $1~M_{\odot}$ bump, whose effect was lowered by the $f_p(m)$ curves in the case of our detections, as discussed previously. Additionally, the slope based on our detection may be shallower than $\alpha_{\text{Salpeter}}$ because of cases similar to EMUTAU, as discussed in Section \ref{subsec:inferred_imf}. However, the difference between the parametrization of our detection and the one of the $S_{\text{lit}}$ members is still very small, and both cases are only one standard deviation away from $\alpha_{\text{Salpeter}}$. The combined IMF can also be decently described by a three-segment power-law function, analogous to Kroupa’s \citep{kroupa2002} representation of the field IMF. For this parametrization, we obtained slope values of $2.03\pm0.02$ in the low mass range $m<0.1~M_{\odot}$, $-0.22\pm0.05$ in the intermediate mass range $0.1<m/M_{\odot}<1$, and $-2.26\pm0.09$ in the high mass range $m>1~M_{\odot}$. This parametrization helps us uncover more accurately a potential incompleteness in the low mass range that is also visible from comparing the normalized typical solar neighbourhood IMF with the mean normalized IMFs. Indeed, we notice that the estimated slope in the low mass range is much steeper than the one ($-0.3$) found by \cite{kroupa2002}. The origin of this potential incompleteness may partially come from observational biases imposed by the selection process from Section \ref{subsec:GDR3}. One example of this is shown for the case of EMUTAU in Section \ref{subsec:inferred_imf} where we show that the completeness in the lowest mass range is greatly affected by the distance of the group which leads to very low values of SNR for the apparent magnitude of the less massive objects of the group. Additionally, this potential incompleteness could also be partially explained by strong chromospheric activity from low mass objects. As summarized by \cite{suarez2017} based on \cite{Lopez_morales_2007} and \cite{stassun_2014}, intense chromospheric activity of low mass stars and brown dwarfs could either lead to underestimated masses due to lowered effective temperatures or to overestimated masses due to increased luminosities. However, the authors conclude that the mass differences should be of only $\sim5\%$, which is why we conclude that such phenomena should not have a significant impact on our estimation of the IMF.

\begin{figure*}
\sidecaption
   \includegraphics[width=12cm]{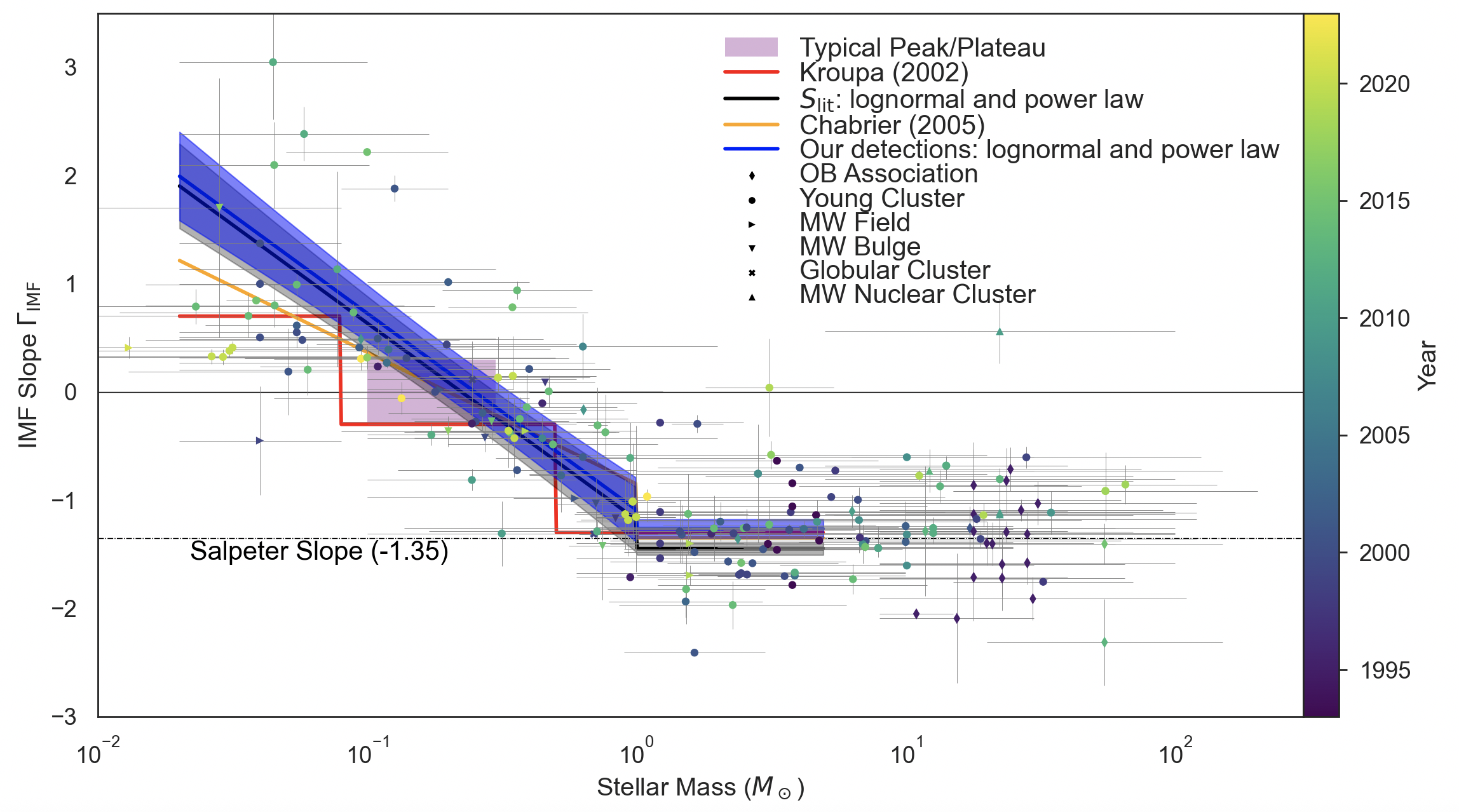}
     \caption{Figure taken from \cite{henebelle_2024} review on the IMF overlapped with the parametrization of the mean normalized IMF based on our detections (blue lines and shaded region) and the $S_{\text{lit}}$ members (black lines and shaded region). The plot shows the slope of different estimation of the IMFs of several populations in the Milky Way disc as a function of the stellar mass from the literature. The shadowed region corresponds to the propagated errors in the estimation of the parameters of the mean normalized IMF of the NYMGs.}
     \label{fig:imf_slopes_literature}
\end{figure*}

Figure \ref{fig:imf_slopes_literature} shows Figure 1 of the IMF review from 
\cite{henebelle_2024} on top of which we have added the parametrization of the mean normalized IMF of both our detections and the $S_{\text{lit}}$ groups. In this 
figure, $\Gamma_{\text{IMF}}=\alpha+1$. As discussed in \cite{henebelle_2024}, this plot shows that roughly seventy years of studying the IMF in the solar neighbourhood reveal a consistent trend. Although the inferred parametrization of individual young stellar clusters and associations can vary significantly due to intrinsic differences or observational biases, the overall IMF exhibits a certain regularity. On average, it is well described by the parametrization of the system IMF of field stars in the solar neighbourhood, as estimated by \cite{chabrier2005b} in the low-mass range and by \cite{salpeter1955} in the high-mass range. Whether this behaviour is a consequence of a locally invariant IMF for populations in the solar neighbourhood or even the Milky-Way disc is not known and partially depends on whether the differences between individual IMFs are dominantly produced by intrinsic physical differences or by observational biases. 

We notice in Figure \ref{fig:imf_slopes_literature}
that the mean normalized IMF of both our detections and the $S_{\text{lit}}$ members are essentially identical to the average IMF of the solar neighbourhood, especially for the values of $m_c$ and $\alpha$, while the parametrization of the individual IMFs shown in Table \ref{table:knowngroups} 
are consistent with the variety of estimations observed in Figure 
\ref{fig:imf_slopes_literature} for young clusters. While the standard deviation for both mean normalized IMFs ($\sigma_c=0.45\pm0.17$) is consistent with the typical value for the solar neighbourhood ($\sigma_c\sim0.6$), we note that the slightly narrower absolute value could be due to incompleteness in the very low mass range $0.02<m/M_{\odot}<0.1$ as it was previously discussed when comparing the fitted triple power-law with the parametrization from \cite{kroupa2002}. In the case of our detection, the potential over-estimation of contaminants when using the purity rates $f_p(m)$ for some of the groups may have also contributed to narrowing the IMF. Apart from this small difference in $\sigma_c$ and if we extrapolate the previously discussed statistical behaviour of the IMFs of young populations of the solar neighbourhood, we can conclude that the striking similarity between the parametrized mean normalized IMF of our detections and the IMF of \cite{chabrier2005b} and \cite{salpeter1955} is a strong argument in favour of the idea that, on average, there is no systematic bias in either our detections or the overall literature reflected by MOCA in the mass range $0.1<m/M_{\odot}<10$ with the exception of the $1~M_{\odot}$ bump due to potential contamination discussed previously.

\begin{figure*}
\sidecaption
  \includegraphics[width=6cm]{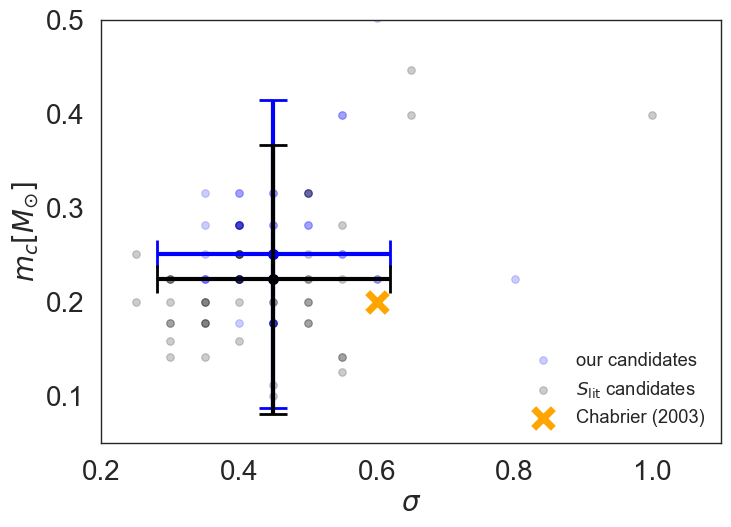}
  \includegraphics[width=6cm]{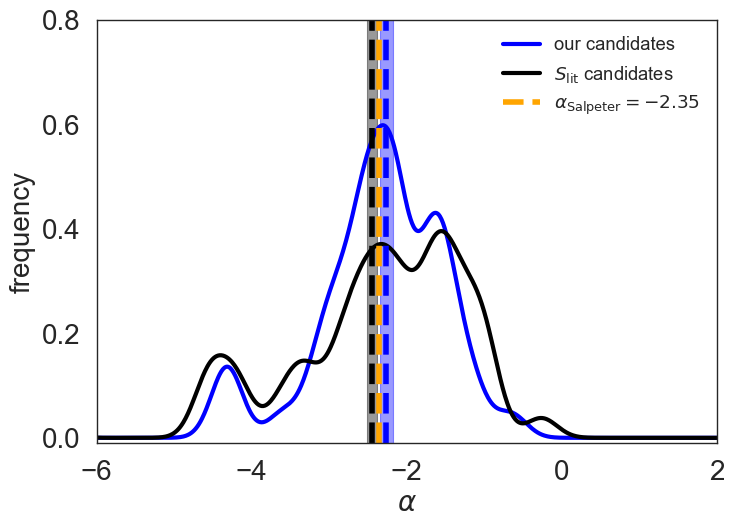}
     \caption{Estimated parameters of the inferred IMFs from our detections (blue) and the 
groups from $S_{\text{lit}}$ (black). The thick non transparent blue and black dots on the left panel and the dashed lines on the right panel indicate the values of the mean normalized IMFs. The left plot shows the estimated values (transparent dots with error bars) for the parameters of the 
log-normal with the values from the parametrization made by \cite{chabrier2005b}. The right plot shows the distribution of the power-law 
slopes estimated through a KDE with the value found by \cite{salpeter1955}.}
     \label{fig:imf_parameters}
\end{figure*}

Figure \ref{fig:imf_parameters} shows the distribution of the estimated parameters 
$m_c$, $\sigma_c$, and $\alpha$ based on our detections, the $S_{\text{lit}}$ groups 
and their respective mean normalized IMFs. As in the case of the mean normalized 
IMFs, we observe that the distribution of parameters in the low-mass range from our 
detections ($0.02<m/M_{\odot}<1$) is similar to the one 
based on the $S_{\text{lit}}$ groups. 

Regarding the high-mass range ($m>1~M_{\odot}$), we also notice similar distributions when using our detections and $S_{\text{lit}}$. Additionally, we notice that in both cases the distribution is bimodal with a greater maximum around \cite{salpeter1955} slope and a lower maximum around $\alpha\sim-1$. The maximum associated with shallower slopes is mainly the product of groups such as EMUTAU that show an excess of massive stars, which could be the product of contamination from other stellar clusters. We also notice an extended tail with a third smaller local maximum around $\alpha\sim-4.5$. As discussed in the case of $\beta$-Pictoris, these steep slopes may be the product of contamination from the $1~M_{\odot}$ bump or from a lack of massive stars. However, we notice fewer cases with steep slopes for the IMFs based on our detections than for the ones based on $S_{\text{lit}}$, which suggests that we have successfully reduced the contamination from the $1~M_{\odot}$ bump for the IMFs based on our detections.

\section{Summary, conclusions, and discussion}
\label{sec:conclusions}

This work presents the first inference of the IMFs for a large set 
of NYMGs, a crucial step toward understanding the origin of these groups. 
In this section we summarize and discuss our main conclusions, covering 
the tools and methods developed, the detected NYMG candidates, the inferred 
IMFs, and the implications for the origin of these groups.

\subsection{Tools and methods}
\label{subsec:conclusions_tools_methods}

We developed the following methods and tools:

\begin{enumerate}
    \item A general method to estimate the completeness of a sample within a given survey, based on the survey’s selection function and the criteria used to define the sample. We applied this method to enhance the completeness of our initial sample using the GaiaUnlimited Selection Functions. 

    \item A detection algorithm that combines the DBSCAN clustering method with GDR3 astrometric and photometric data to identify stellar co-moving groups, and that estimates the statistical significance of each detection through a comparison with the GOG synthetic catalogue. 
    
    \item A simple Bayesian method to infer the masses and ages of GDR3 sources classified as NYMG
    members, using their parallaxes, photometry, and extinction. Although the individual masses of the candidate members of the detected NYMGs were inferred assuming a mean extinction vector, we do not expect the inferred masses to differ significantly from those obtained using individual extinctions because the overall extinction is very small. Finally, we adopted stellar models typically employed in IMF studies. A systematic comparison with alternative models, while important, is beyond the scope of this work.
\end{enumerate}

\subsection{The detected NYMG candidates}
\label{subsec:conclusions_detected_nymgs}

We searched for NYMGs using our detection algorithm, and found the following results:

\begin{enumerate}
    \item Detected a total of 473 over-densities in the RPS, 33 of which correspond 
    to group candidates associated with at least one of the 44 of the 68 NYMGs listed in MOCA. Since this work focused only on the recovered known groups, the remaining over-densities will be studied in future works. The 33 NYMG candidates include a total of 4166 candidate members, of which 2545 are new members. In a few cases, different groups from the literature were detected as a single group candidate by our detection algorithm, and our method cannot distinguish whether they are separate populations with similar properties or substructures of the same association. However, we do notice shared ages and CMD distribution for groups that were merged by our detection algorithm. 
    
    \item The 24 remaining NYMG candidates from the $S_{\text{lit}}$ sample that we did not recover present $S_{\text{lit}}$ members that are either too scattered in the RPS or have too few members, resulting in a low RPS density that our method could not detect.

    \item Achieved a recovery rate relative to the literature between 44\% and 54\% with a mean recovery per group between 59\% and 71\%. We estimate that only 16\% to 24\% of our candidate members are contaminants. A conservative detection strategy was adopted by maximizing the $\mathcal{C}$ score function and applying a threshold of $r_c>0.1$, due to the RPS scattering introduced by the use of tangential velocities. 
    This likely reduced the recovery rates of the known groups but significantly minimized contamination. We prioritized purity over recovery with the aim of prioritizing statistical representation over completeness in number, because the goal of this work is to infer the IMF shape of the NYMGs rather than identifying all members of these groups. 

    \item Found that the members of the detected NYMG candidates follow sequences likely associated with young isochrones in the CMD, and that their ages are mostly distributed into three categories: young ($\sim10$ Myr), middle-aged ($\sim40$ Myr), and old ($\sim90$ Myr).
    
    \item Showed for all groups that for both our detections and the $S_{\text{lit}}$ members, MS candidate members tend to be more kinematically scattered than the PMS candidates. This is reflected in both cases by a higher percentage of stars with $r_c<0.1$ in the MS than in the PMS. This result could be a consequence of two non-excluding potential phenomena. The first is the possibility that older groups would tend to be more scattered than younger groups. The second is that older groups are more likely to have old star contaminants, as they are closer to the MS, and this contamination would increase the RPS dispersion of their host groups. Further information on radial velocities and chemistry is essential to confirm memberships of individual sources and hence to infer how the previously mentioned phenomena contribute to what we observe.
\end{enumerate}

\subsection{Inferred IMFs of NYMGs}
\label{subsec:conclusions_imfs}

For the 33 detected groups, we inferred the individual masses of their candidate members, and their corresponding IMFs in the mass range $0.02\lesssim m/M_\odot\lesssim10$. Our findings are as follows:

\begin{enumerate}
    \item We present the IMFs of 33 NYMG candidates (Table 1). Previous studies have constructed IMFs for only three of the 68 known NYMGs listed in MOCA: TW-Hya (TWA), Volans-Carina (VCA), and Tucana-Horologium (THA). In addition to estimating the IMFs of these three groups, we provide the first IMF estimates for the remaining 30 detected NYMGs.

    \item We parameterized the individual IMFs of the 33 detected groups, as well as the mean normalized IMF, using a log-normal function for $m<1~M_{\odot}$ and a power-law for $m>1~M_{\odot}$. The parameters of the mean normalized IMF from our detections ($m_c=0.25\pm0.17~M_{\odot}$, $\sigma_c=0.45\pm0.17$ and $\alpha=-2.26\pm0.09$) and the $S_{\text{lit}}$ candidate members ($m_c=0.22\pm0.14~M_{\odot}$, $\sigma_c=0.45\pm0.17$ and $\alpha=-2.45\pm0.06$) are both consistent with the typical system IMF of the solar neighbourhood estimated by \cite{chabrier2005b} and \cite{salpeter1955}. This result suggests that there is no significant systematic bias in our detections and the members from $S_{\text{lit}}$ based on our discussion of Figure \ref{fig:imf_slopes_literature}.

    \item Our method to estimate and correct for contamination significantly reduced the $1~M_{\odot}$ bump around the Turn-On of most of the NYMGs, which is likely associated with contamination as shown in Figure \ref{fig:all_purities}. Importantly, while this method helps to identify contamination rates from the Besan\c{c}on field within the inferred IMFs, it does not identify which specific candidate members are contaminants. This is why the presence of the bump is slightly more visible for the IMFs based on $S_{\text{lit}}$, for which no contamination correction was applied. This translates into a slightly steeper slope for the IMF based on $S_{\text{lit}}$ than for the one based on our detections. 
\end{enumerate}

\subsection{Hints on the origin of the NYMGs}
\label{subsec:conclusions_origin_nymgs}

Our results provide important insights into the origin of the NYMGs. The most 
significant conclusion is that the striking similarity between the mean normalized 
IMF of the NYMGs and the typical system IMF of stellar clusters and field stars 
in the Solar neighbourhood \citep[e.g.][]{bastian2010}, in the mass range 
$0.1 < m/M_{\odot} < 5$, together with the fact that most NYMGs follow sequences 
in the CMD likely corresponding to young isochrones, supports the hypothesis 
that NYMGs are remnants of disrupted clusters or associations potentially 
in the process of dispersing. 

The weak signatures of the NYMGs in the RPS and phase-space suggest 
that they could be unbound. To explore this possibility, we performed a simple 
estimation of the lower bounds $Q_{\text{min}}$ for the virial ratio $Q=T/|U|$ 
for each group, defined as $Q_{\text{min}} = T_{\text{min}}/|U|$. Here $U$ denotes 
the total potential energy of the group, while $T_{\text{min}}$ is the total kinetic 
energy computed without the radial velocities of individual sources (using only 
$v_{\alpha,\text{LSR}}$ and $v_{\delta,\text{LSR}}$). Since $T > T_{\text{min}}$, 
it follows that $Q_{\text{min}} < Q$. For the detected groups we find 
$1.3 < Q_{\text{min}} < 1.4 \times 10^3$, implying $Q > 1$ for all groups. This 
simple estimate of $Q_{\text{min}}$ therefore suggests that the studied NYMGs are 
not gravitationally bound. A similar analysis was performed by \cite{suarez2019} 
on the young ($5 < \text{age}/{\rm Myr} < 10$) 25-Ori group, showing that it 
is gravitationally unbound. If the NYMGs are indeed unbound, this raises the 
intriguing possibility that younger groups such as 25-Ori could represent 
NYMG-like progenitors. Nevertheless, the present results are not sufficient 
to draw a firm conclusion on this matter, which will require a deeper analysis
beyond the scope of this work. 

Furthermore, the IMF similarities indicate that if dispersion is occurring in the NYMGs, 
it does not preferentially affect a specific stellar-mass range within $0.1 < m/M_{\odot} < 5$. 
However, our results do not allow us to draw conclusions for $m < 0.1\,M_{\odot}$. If we assume 
that the IMF is invariant across the Galactic disc and resembles the system IMF of the Solar
neighbourhood, then—as discussed in Section \ref{subsec:imf_parametrization}—the estimated 
slope ($2.03 \pm 0.02$) in the mass range $m < 0.1\,M_{\odot}$ would imply a deficit of 
low-mass objects relative to the slope ($0.3$) from \cite{kroupa2003}. Based on the 
discussion in Section \ref{subsec:imf_parametrization}, however, a substantial fraction 
of this apparent deficit is very likely due to observational biases. If, on the other 
hand, such an incompleteness proved to be real and independent of observational biases, 
it would imply that the scattering process of the NYMGs may have undergone a rapid 
mass-segregation phase. The present results are not sufficient to distinguish between 
these two scenarios.

Finally, it is important to note that the idea of the NYMGs being remnants 
of stellar clusters and associations does not rule out the possibility that they are also 
shaped by dynamical resonances with the Milky Way's spiral arms or 
bar, like other more massive and/or older groups \citep{antoja_2010}. 
An intriguing scenario to explore is 
that different stellar associations and clusters could be brought together in phase-space 
through such resonances in a short period of time. If this is the case for 
two populations of similar ages, distinguishing them from each other in the 
CMD should be difficult, if not impossible. If their ages differ significantly, 
this could result in double sequences in the CMD. In such cases, the second 
sequence produced by the MS stars that we observed in some cases might in fact 
correspond to genuine NYMG members rather than contaminants. Photometry alone 
would not suffice to determine whether this is the case or not. 
Even chemical abundances may fail to break this degeneracy if the two 
populations are chemically similar to begin with. These considerations 
motivate future studies involving dynamical simulations of stellar clusters 
interacting with spiral arms and the bar, to investigate whether resonant 
processes could indeed bring together remnants of different young stellar 
clusters or associations and produce NYMG-like associations.

\section*{Data availability}
Table \ref{table:knowngroups} and the catalogue of detected members are only available in electronic form at the CDS via anonymous
ftp to cdsarc.u-strasbg.fr (130.79.128.5) or via
\url{http://cdsweb.u-strasbg.fr/cgi-bin/qcat?J/A+A/}.

\begin{acknowledgements} 
The author RB gratefully acknowledges the scholarship Becas de apoyo a docentes para estudios de posgrado en la Udelar, Maestr\'ia, 2022 from the Comisi\'on Acad\'emica de Posgrado (CAP) for financially supporting and making possible this work. Likewise, RB and JJD thank the Programa de Desarrollo de Ciencias B\'asicas (PEDECIBA) and the Apoyo de Movilidad Individual Acad\'emica (MIA) of the Universidad de la República (UdelaR, Uruguay) for their financial support. RB and GS also thank the Heising–Simons Foundation for its support, particularly; RB acknowledges grant \# 2022-3927 from the foundation. This work made use of the high-performance computing facilities of ClusterUY \citep{clusteruy}, for which the authors are thankful. The authors extend their deep gratitude to Anthony Brown for the enriching comments and advice provided through their early review of this manuscript. The authors thank Ronan Kerr as well as the referee of this article for cautiously reviewing this work and providing very kind and helpful comments. Finally, they warmly thank Luis Aguilar, Mauro Cabrera, Bruno Dom\'inguez, Pau Ramos, and Rodrigo Cabral for closely following this work from its early stages and for sharing their wisdom and insights, some of which were key to the development of this research.
\end{acknowledgements}

\bibliographystyle{aa}
\bibliography{references}

\appendix

\section{The distance from the inverse of parallax}
\label{appendix_distance_from_parallax}

Estimating distances using the inverse of parallax ($d_{\varpi}=1/\varpi$) is the simplest but 
least precise method \citep{bailerjones2015}. On the other hand, 
the photo-geometric distances $d_{pg}$ estimated by 
\citet{Bailer_Jones_2021} provide a more robust determination 
for the complete GDR3 catalog. We demonstrate in this section that these estimates show good 
agreement with $d_{\varpi}=1/\varpi$ when $\sigma_\varpi/\varpi < 0.1$, 
where $\sigma_\varpi$ is the uncertainty in $\varpi$.
Although smaller mean values of $\sigma_\varpi$ are expected for the 
solar neighbourhood compared to more distant regions due to the intrinsically larger $\varpi$ 
and higher mean signal-to-noise ratio observations for nearby objects, 
the effect of the 
sample selection according to $d_{\varpi}$, particularly for the 
fainter sources, needs to be addressed.

\begin{figure}[!t]
  \resizebox{\hsize}{!}{\includegraphics{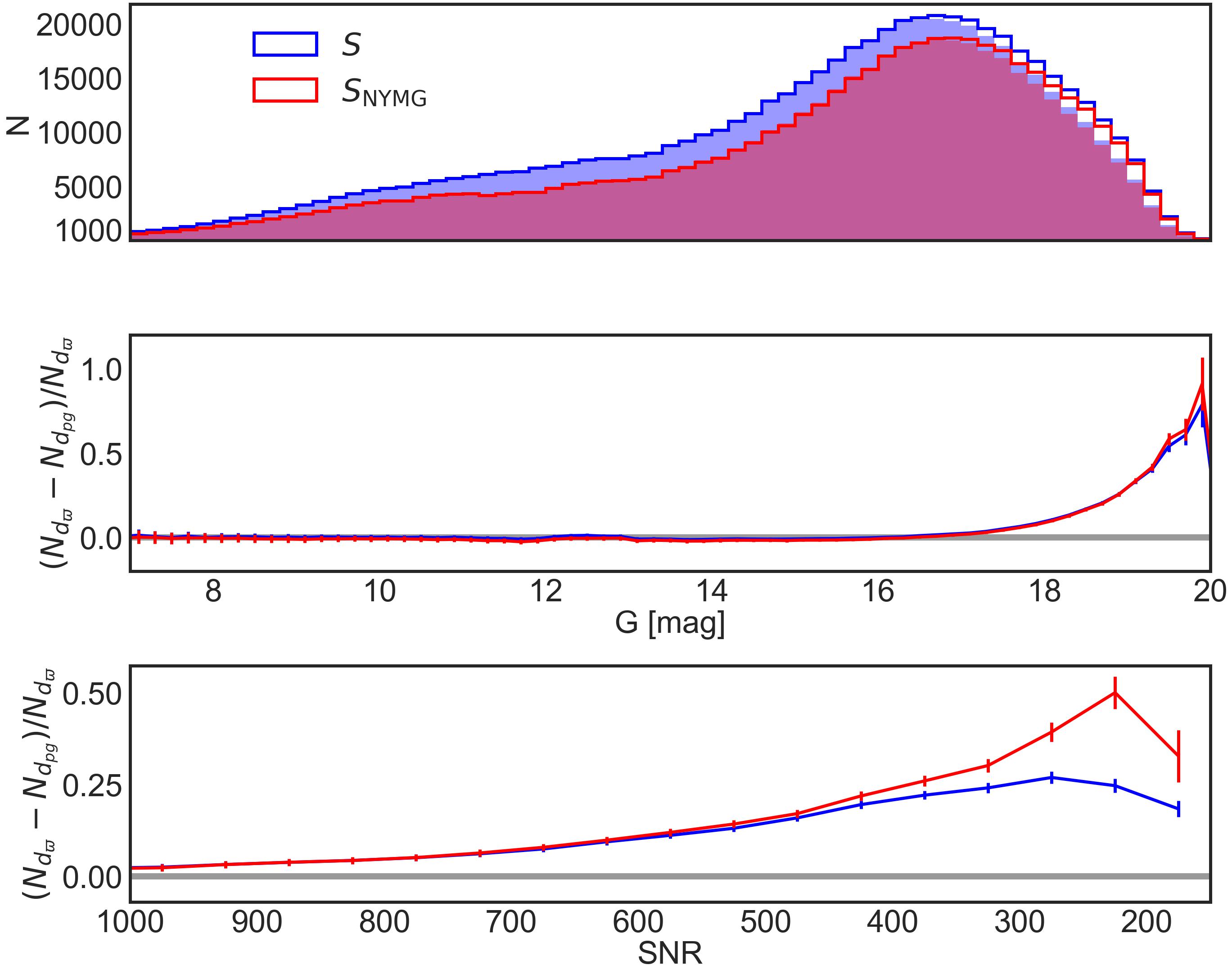}}
  \caption{\textbf{Top panel:} Distribution of $G$ magnitudes for the sample of 
stars with distances $d = 1/\varpi < 200$ pc 
(solid line histograms) and distances 
$d_{pg} < 200$ pc from \citet{Bailer_Jones_2021} (shaded 
histograms). The blue histograms represent the $S$ and 
$S_{pg}$ samples. The red histograms represent the 
$S_{\text{NYMG}}$ and $S_{NYMG_{pg}}$ samples, as explained in Section 
\ref{subsec:GDR3}. \textbf{Middle panel:} Residuals of the number of sources 
between $S_{\text{NYMG}}$ and $S_{NYMG_{pg}}$ (red) and between $S$ 
and $S_{pg}$ (blue) as a function of $G$. \textbf{Bottom panel:} Same residuals 
as the middle panel, shown as a function of the photometric $SNR$ of $G$. All error bars 
correspond to the propagated Poisson error.
}
  \label{fig:parallax_vs_photogeo}
\end{figure}

We compare two samples: the one we call starting sample, 
which includes all stars having $d_{\varpi} = 1/\varpi < 200$ pc, and a 
sample we call $GDR3_{d_{pg}}$, which includes all stars having 
$d_{pg} < 200$ pc. We applied the same procedures used to select 
the samples $S$ and $S_{\text{NYMG}}$ from the starting sample 
(Section \ref{subsec:GDR3}) to the $GDR3_{d_{pg}}$ sample, producing 
the samples $S_{pg}$ and $S_{NYMG_{pg}}$, respectively. The 
number of sources for each sample is shown in Table 
\ref{table:parallax_photogeo}. The difference in number between the samples 
considering different distance estimations is smaller than 5\%
and occurs for stars with $G > 17$~mag, as shown in Figure \ref{fig:parallax_vs_photogeo}. 
This difference increases with $G$ as 
the $SNR$ decreases.
We conclude that, in our case, it is equivalent to estimate 
distances with either method.

Finally, we can estimate how many stars that truly belong to the 
Solar neighbourhood MS and PMS with kinematics consistent with 
NYMGs are being discarded by the $\text{log}_{10}(\text{SNR}_G)>2.2$ 
and $\sigma_{\varpi}/\varpi<0.1$ conditions mentioned in Section \ref{subsec:GDR3}. To do so, we consider a new 
sample obtained by applying all the conditions used to build $S$ with the exception of the $\sigma_\varpi/\varpi\leqslant 0.1$ and $\text{log}_{10}(F_G/\sigma_{F_G})=2.2$ conditions to the match between the starting sample using $d_{\varpi}$ and the $GDR3_{d_{pg}}$ sample.
Let $N_{aux,match}^*$ be the number of sources in that 
sample and  $N_{aux,match}$ the number of sources in that sample 
after applying the SNR and $\sigma_{\varpi}/\varpi$ conditions. Then, we can 
roughly estimate the fraction of completeness after applying the SNR 
and $\sigma_{\varpi}/\varpi$ conditions as $N_{aux,match}/N_{aux,match}^*$. 
We found that this ratio is equal to 0.987, which means that we are 
discarding only 1.3\% of the sources of interest.

\begin{table}
\tiny{
\begin{center}
\begin{tabular}{l r r r}
\hline
 Subset                 & $N_{d_{\varpi}}$   & $N_{d_{pg}}$  & $N_{\text{match}}$ \\
\hline
\hline
\it{Starting sample}    & 5,244,458 & 2,328,436     & 2,320,981 (44\% of $N_{d}$) \\
$S$               & 594,297   & 577,003       & 568,703   (95\% of $N_{d}$) \\
$S_{\text{NYMG}}$              & 495,079   & 482,417       & 475,854   (96\% of $N_{d}$) \\
\hline
\end{tabular}
\end{center}
}
\caption{Number of solar neighbourhood sources from different samples selected from GDR3 following the prescription in Section \ref{subsec:GDR3} and using $d_{\varpi}$ and $d_{pg}$ as estimates of distances as explained in Appendix \ref{appendix_distance_from_parallax}. The column $N_{\text{match}}$ shows the number of sources after doing an exact match using GDR3 source ID between the two samples.}
\label{table:parallax_photogeo}
\end{table}

\section{Estimation of completeness}
\label{appendix_completeness}

We present a general method to estimate the real completeness of a sample of stars based on the knowledge of the selection function of that sample. 

Let us consider a real and arbitrary set of stars called the real sample of interest, which in our particular case corresponds to the real solar neighbourhood MS and PMS stars with kinematics similar to that of the known NYMGs. Although the real sample of interest corresponds to the sample that we wish to study, it does not necessarily correspond to the one we end up studying in practice. The procedure for selecting candidate members of the real sample of interest from a given survey catalogue is generally made through the application of conditions that can be classified into two categories: selection conditions, which define the real sample of interest within the catalogue, and quality conditions, which do not define the sample of interest in itself but ensure some degree of quality in the data. In our specific case, one selection condition is $\varpi~\geqslant~5$ mas and one quality condition is $ruwe~\leqslant~1.4$. 

If we assume that the number of contaminants in the sample resulting from applying all conditions is negligible, we can estimate the real completeness as $f=n_{s\cap q}/n_{\text{real}}$ with $n_{s\cap q}$ the number of sources selected using all the conditions and $n_{\text{real}}$ the number of sources in the real sample of interest. An example of a contaminant source in our case could be a source with true parallax $\varpi_{\text{true}}<5$ mas but with (poorly) measured parallax $\varpi_{\text{obs}}>5$ mas, since it would not be excluded by the condition $\varpi~\geqslant~5$. 

Although we do not know $n_{\text{real}}$ and hence $f$, we can use the GaiaUnlimited Selection Functions (GSFs; Section \ref{subsec:completing_sample}) to estimate the real completeness in a given bin of observables $\Delta q$, i.e. the fraction $f_{\Delta q}=n_{\Delta q}^{s\cap q}/n_{\Delta q}^{\text{real}}$. The real completeness $f_{\Delta q}$ can be estimated from the probability that a source within $\Delta q$ belongs to the studied sample resulting from applying selection and quality conditions, knowing that it belongs to the real sample of interest, by using Bayes:
\begin{equation}\label{eq_completeness_0}
    f_{\Delta q}\sim P_q(S_{s\cap q}|S_{\text{real}}) = \frac{P_q(S_{\text{real}}|S_{s\cap q})P_q(S_{s\cap q})}{P_q(S_{\text{real}})}
\end{equation}
Here, $S_{\text{real}}$ and $S_{s\cap q}$ represent the real sample of interest and the sample resulting from applying all conditions, respectively. In our case, $S_{s\cap q}$ corresponds to $S_{\text{NYMG}}$.

Since we assumed the amount of contaminants in the studied sample $S_{s\cap q}$ is negligible, then almost all sources in $S_{s\cap q}\bigcap S_{\text{real}}$ should belong to the real sample, and hence the likelihood $P_q(S_{\text{real}}|S_{s\cap q})\sim 1$. Regarding the prior $P_q(S_{s\cap q})$, all sources from the studied sample belong to $GDR3$, hence $P_q(S_{s\cap q})=P_q(S_{s\cap q}\cap \text{GDR3})=P_q(S_{s\cap q}|\text{GDR3})P_q(\text{GDR3})$. Finally, we can estimate the normalization probability $P_q(S_{\text{real}})$ as $n_{\Delta q}^{\text{real}}/n_{\Delta q}$, where $n_{\Delta q}$ is the real number of sources in the Universe (not only in the sample of interest) within $\Delta q$. We then notice that $n_{\Delta q}^{\text{real}}/n_{\Delta q}\sim n_{\Delta q}^{\text{sel}}/n_{\Delta q}^{\text{GDR3}}$, where $n_{\Delta q}^{\text{GDR3}}$ and $n_{\Delta q}^{\text{sel}}$ are the number of sources in $\Delta q$ that belong to GDR3 and to the sample resulting from only applying the selection conditions, respectively. The latter ratio can be estimated as the probability $P_q(S_{\text{sel}}|\text{GDR3})$ with $S_{\text{sel}}$ being the sample resulting from only applying selection conditions. In our case, $S_{\text{sel}}$ corresponds to the ancillary sample $S$ defined in Section \ref{subsec:GDR3}. We can then rewrite Equation \ref{eq_completeness_0} as:
\begin{equation}\label{eq_completeness_1}
    f_{\Delta q}\sim P_q(S_{s\cap q}|S_{\text{real}}) = \frac{P_q(S_{s\cap q}|\text{GDR3})}{P_q(S_{\text{sel}}|\text{GDR3})}P_q(\text{GDR3})
\end{equation}
All of the terms on the right side of Equation \ref{eq_completeness_1} 
can be computed using the GSFs. Although this reasoning was made with the GSFs, we notice that the method can be generalized to any survey with a known selection function.

In practice, most quality conditions also act as selection conditions because they might discard sources that would have been wrongly included in the studied sample if they were not applied. In our case, for example, a source with true parallax $\varpi_{\text{true}}<5$ mas but with (poorly) measured parallax $\varpi_{\text{obs}}>5$ mas would not be excluded by the condition $\varpi~\geqslant~5$ mas but would be excluded after applying the condition $\sigma_{\varpi}/\varpi < 0.1$. This means that the previously intended quality condition acted as a selection condition on this source in particular. We call conditions like this one mixed conditions. In the case of having mixed conditions, we cannot repeat the exact same reasoning because it would no longer be true that $n_{\Delta q}^{\text{real}}/n_{\Delta q}\sim n_{\Delta q}^{\text{sel}}/n_{\Delta q}^{\text{GDR3}}$. This is because we cannot know the number of contaminants that would have been included in the studied sample if we had only applied the selection conditions but that would have been discarded by a mixed condition. If we somehow know that this number is negligible, we can treat the mixed condition as a quality condition. If we cannot neglect such contaminants, we must treat the mixed condition as a selection condition and then 
use Equation \ref{eq_completeness_1}. In the latter case, this means we would not be accounting for the incompleteness produced by such mixed conditions.

\section{Selecting NYMGs from the MOCA database}
\label{appendix_nymgs_from_moca}

We used the MOCA database to easily and efficiently identify known NYMGs from the literature. We started by querying all the associations in MOCA classified as moving groups with the query:
\begin{quote}
\texttt{SELECT * FROM moca\_associations WHERE physical\_nature ='moving group'}
\end{quote}
The table moca\_associations contains general information and comments on all the associations from MOCA. We then discarded groups when: their best estimated ages from the literature were above 100 Myr (age\_myr $>$ 100), their average distances to the Sun were greater than 200 pc (avg\_dist $>$ 200), or there existed another better definition of the group (suboptimal\_grouping $\neq$ 0), or the group could have been rejected for some reason (comments contains the strings \texttt{Rejected}, \texttt{Not in BANYAN} or \texttt{Not included in BANYAN}).

We then selected the individual sources of the identified NYMGs from the \texttt{moca\_associations} table by querying \texttt{calc\_banyan\_sigma} using:
\begin{quote}
\texttt{SELECT * FROM calc\_banyan\_sigma WHERE moca\_bsmdid = 23 AND max\_observables = 1 AND moca\_aid IN (\{groupToSearch\})}
\end{quote}
The \texttt{calc\_banyan\_sigma} table contains, for each source from MOCA, the results from running BANYAN, including membership probabilities. The condition \texttt{moca\_bsmdid = 23} ensures selection of the latest results from BANYAN (ran in April 2025), and \texttt{groupToSearch} refers to the list of association names selected from \texttt{moca\_associations}.

We discarded sources with membership probabilities lower than 0.95 (\texttt{ya\_prob < 95}) and with distances to the Sun greater than 200 pc (\texttt{d\_opt > 200}). We used unique source identifiers from different external surveys provided by MOCA to match the resulting sample to the following catalogues: Gaia DR1, DR2 and DR3, WISE and 2MASS, prioritizing in the case of multiple matches Gaia DR3 and DR2. Then, using Gaia proper motions, we propagated the sky positions of the Gaia sources from $S_{\text{NYMG}}$ back in time to the epochs of the other external surveys and sky-matched the catalogues. This allowed us to find sources from our query that had no GDR3 identifier but did have an identifier from another important survey. Only $\sim4\%$ of the sources initially queried were not found in $S_{\text{NYMG}}$.

\section{Kinematic biases and relevance of the LSR}
\label{appendix_section_lsr}

Because the typical size of many NYMGs is similar to or even larger 
than their distance from the Sun, their proper motion 
distributions are expected to be affected by three phenomena: $(i)$ NYMGs may 
cover very large regions of the sky, which means that even if all 
members of a NYMG share the exact same velocity vector, their proper 
motions may be different due to projection effects, $(ii)$ two 
members of the same NYMG may be at different distances from the 
Sun, which means that even if they have the same sky position and 
Cartesian velocity vector, their proper motion vectors may differ,
and $(iii)$ the proper motions are also affected by the Sun's 
motion in both the proper motion space and the tangential velocity 
space. Although it is not possible to correct the first bias without 
RV, the second bias can be addressed by computing the tangential 
velocity of each star using the parallax. To correct 
the third bias, it is necessary to work within the LSR.

\begin{figure*}
\sidecaption
  \includegraphics[width=12cm]{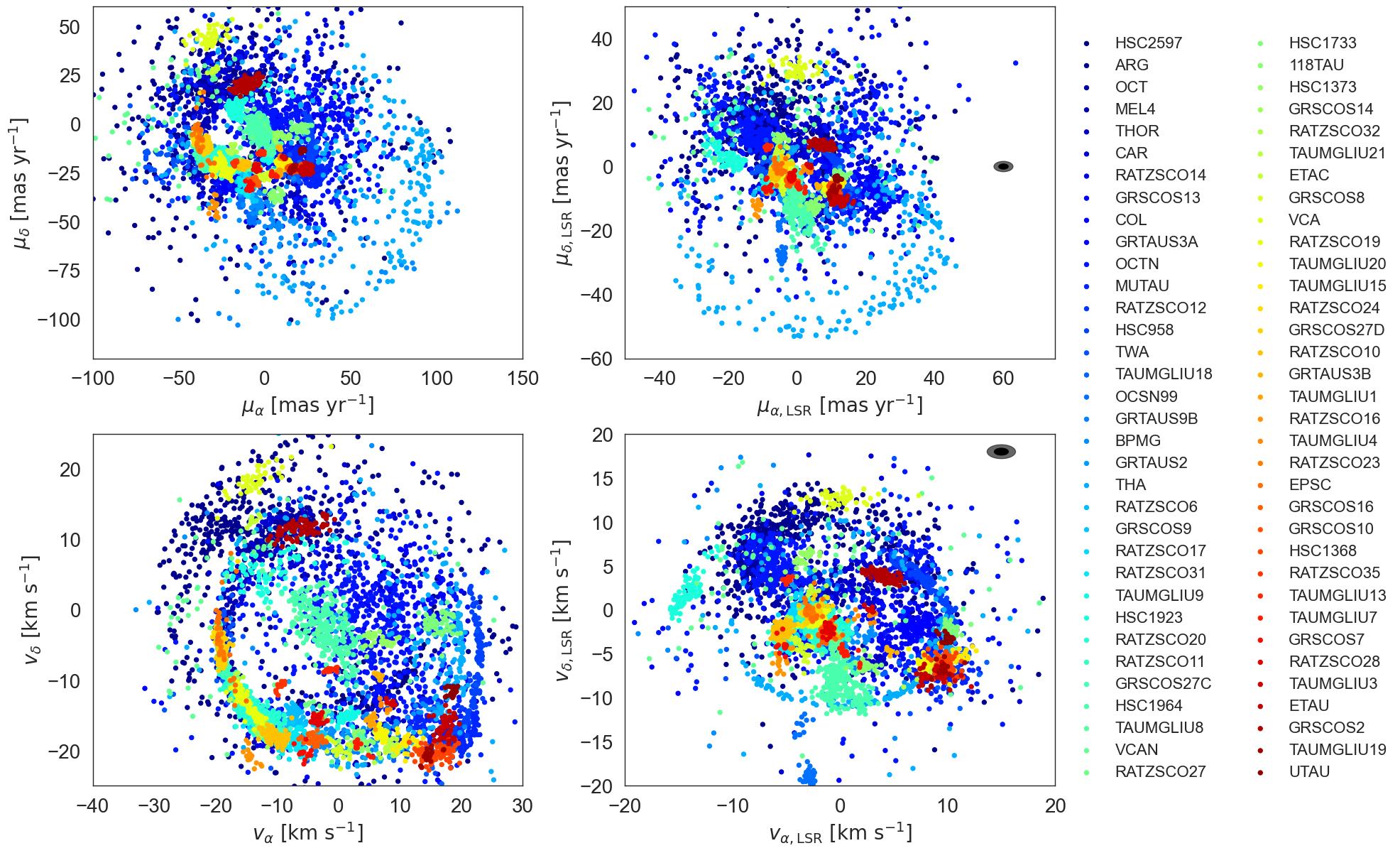}
     \caption{Distributions of proper motions (top panel) and tangential velocities 
(bottom panel) of known NYMGs in the heliocentric equatorial coordinate 
system (left panel) and the LSR equatorial coordinate (right panel). The black and grey filled ellipses on the right panels show how much the whole sample can be shifted due to respectively 1-$\sigma$ and 2-$\sigma$ uncertainties on the measurement of the LSR motion relative to the Sun, i.e. errors from \cite{Schonrich_2010}.}
     \label{fig:gagne_nymgs_kinematics}
\end{figure*}

Figure \ref{fig:gagne_nymgs_kinematics} shows the distribution 
of proper motions and tangential velocities of known NYMGs from $S_{\text{lit}}$ in both the heliocentric frame 
and the LSR, obtained using the Astropy library \citep{astropy_2022}, which uses $(U,V,W)_{\odot} = (11.1^{+0.69}_{-0.75}, 12.24^{+0.47}_{-0.47}, 7.25^{+0.37}_{-0.36})$ km s$^{-1}$ for the Sun's motion from \cite{Schonrich_2010}. The kinematics in the 
heliocentric frame are more scattered than in the LSR, and 
the distribution of the NYMGs changes radically when 
going from proper motions to tangential velocities. 
Finally, we notice that the NYMGs tangential velocities in the LSR do not go beyond $\sim20$ km s$^{-1}$, which is consistent with the fact that these groups are made of very young stars deep within the disc kinematics. It is based on this plot that we decided in Section \ref{subsec:GDR3} to impose the condition $[v_{\alpha_{LSR}}^2+v_{\delta_{LSR}}^2]^{1/2} < 20$ km s$^{-1}$.

\section[Setting the epsilon hyper-parameter of the DBSCAN]{Setting the $\varepsilon$ hyper-parameter in DBSCAN}
\label{appendix_dbscan_hyper_parameters}

Given that we are searching for NYMGs in both the position and the 
LSR tangential velocity space, we considered two different 
$\varepsilon$ parameters: $\varepsilon_r$ for the position space, 
and $\varepsilon_v$ for the velocity space. Then, a possible approach, that we shall call approach A, is to apply the DBSCAN separately in the velocity 
space and then in the position space. In this approach, the two conditions 
that two stars with positions $\vec{r}_0$ and $\vec{r}_1$ and LSR tangential velocities $\vec{v}_0$ and $\vec{v}_1$ should meet to be considered neighbours are $||\vec{v}_1-\vec{v}_0||^2\leqslant\varepsilon_v^2$ and $||\vec{r}_1-\vec{r}_0||^2\leqslant\varepsilon_r^2$. This approach is useful because it helps to reduce the field density by exploiting the fact that stellar groups in general are better defined in the velocity space than in the position space. Additionally, it allows us to estimate the minimum density of the detected group in both position ($N_{min}/(\frac{4}{3}\pi\varepsilon_r^3)$) and velocity ($N_{min}/(\pi\varepsilon_v^2)$) spaces separately.

However, applying the DBSCAN separately is computationally more expensive than applying it only once to a five-dimensional space that combines positions and LSR tangential velocities. Fortunately, the conditions given by approach A can be approximated to a single condition by observing that both equations imply:
\begin{equation}\label{eq_1_DBSCAN}
    \sqrt{\left(\frac{\Delta r_{0,1}}{\varepsilon_r}\right)^2+\left(\frac{\Delta v_{0,1}}{\varepsilon_v}\right)^2} \leqslant \sqrt{2}
\end{equation}
From now on, we shall call the approach that uses the DBSCAN in the SRPS through Equation \ref{eq_1_DBSCAN} approach B. 

We now introduce the similarity factor $\alpha$ that we use to define new scaling parameters $(\varepsilon_r^*,\varepsilon_v^*)$ based on $(\varepsilon_r,\varepsilon_v)$ through the equation $(\varepsilon_r^*,\varepsilon_v^*)=\alpha(\varepsilon_r,\varepsilon_v)$. With these new parameters, we can now find the value of $\alpha$ that best approximates approach B using $(\varepsilon_r^*,\varepsilon_v^*)$ to the approach A with $(\varepsilon_r,\varepsilon_v)$. We do so by finding the value of $\alpha$ that maximizes the Jaccard similarity index defined as $\mathbf{J}=V_i/V_u$. Here, $V_i$ and $V_u$ correspond to the volumes of the intersection and the union, respectively, between the SRPS region defined by the equations of approach A using $(\varepsilon_r,\varepsilon_v)$ and the SRPS region defined by Equation \ref{eq_1_DBSCAN} through approach B using $(\varepsilon_r^*,\varepsilon_v^*)$. 

Using a Monte Carlo method to estimate the volumes $V_i$ and $V_u$, we find that the value of $\alpha$ that maximizes $\mathbf{J}$ is $\alpha=0.847$ with $J=0.654$. With this change, we can re-write Equation \ref{eq_1_DBSCAN} for approach B as:
\begin{equation}\label{eq_2_DBSCAN}
    \sqrt{\left(\frac{\Delta r_{0,1}}{\varepsilon_r}\right)^2+\left(\frac{\Delta v_{0,1}}{\varepsilon_v}\right)^2} \leqslant \alpha\sqrt{2}=0.847\sqrt{2}
\end{equation}

\section{Over-density candidates rejection}
\label{appendix_overdensities_rejections}

To estimate which of the over-densities detected by our algorithm are more likely to be contaminants, we follow the approach described in section 3.5.1 (see their figure 5) of \cite{ratzenbock_2022} and fitted a double-Gaussian to the uni-variate distribution of the over-density sizes. The Gaussian of lower mean value is likely to be dominated by contaminant detections while the second Gaussian is more likely to be dominated by real detections. 

\begin{figure}
\includegraphics[width=7cm]{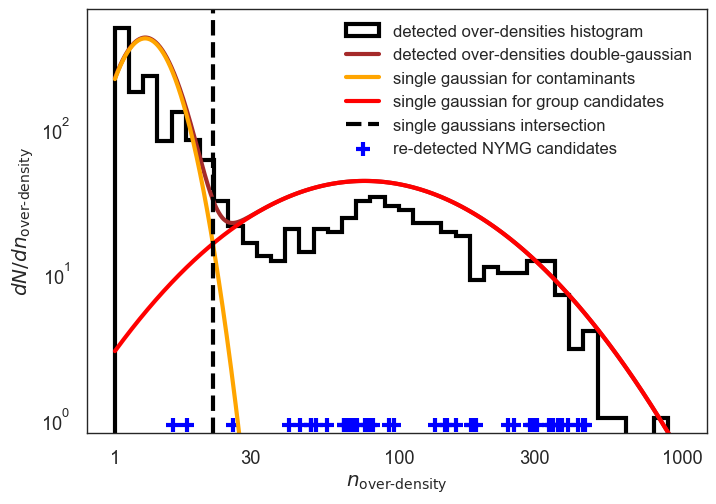}
  \caption{Uni-variate distribution of the over-density sizes (black histogram) with the double Gaussian fit (brown curve). The normal distributions of contaminants candidates and real over-density candidates are shown in orange and red respectively. The blue crosses indicate the number of members of the 33 detected group candidates associated with known NYMGs. The black dashed line shows where the two normal distributions intersect ($N_{\text{over-density}}\sim23$).}
  \label{fig:appendix_overdensities_rejections}
\end{figure}

Figure \ref{fig:appendix_overdensities_rejections} shows the uni-variate distribution and the fitted normal distributions. As we can see, 1232 over-densities are rejected when using the intersection of the two normal distributions as criterion while 473 are not rejected. We also notice that only 5 of the over-densities associated with known NYMGs are rejected.

\section{The IMFs of the detected NYMGs}
\label{appendix_individual_imfs}

\clearpage

\begin{figure*}[t]

\nointerlineskip
\includegraphics[width=35mm]{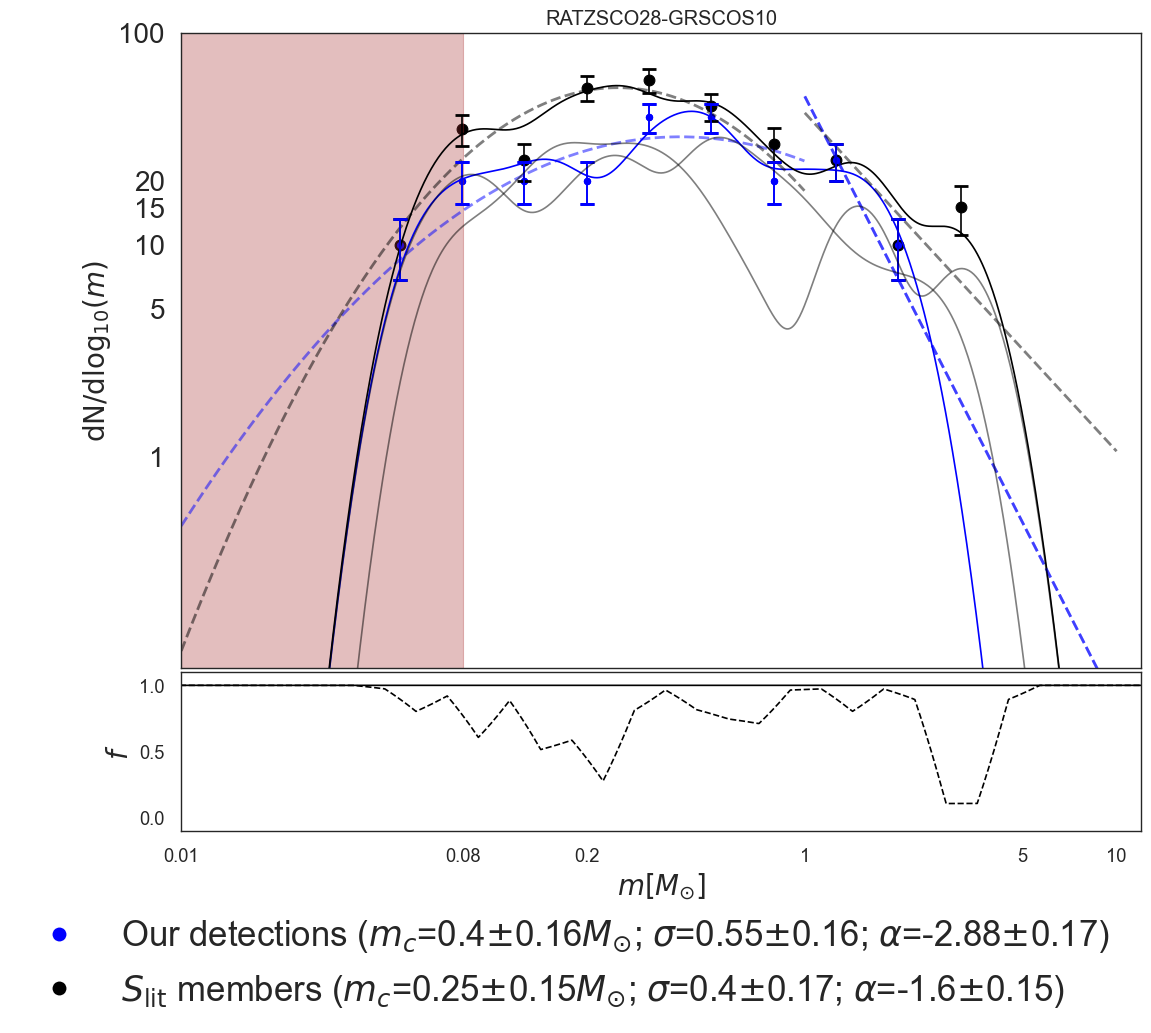}
\includegraphics[width=35mm]{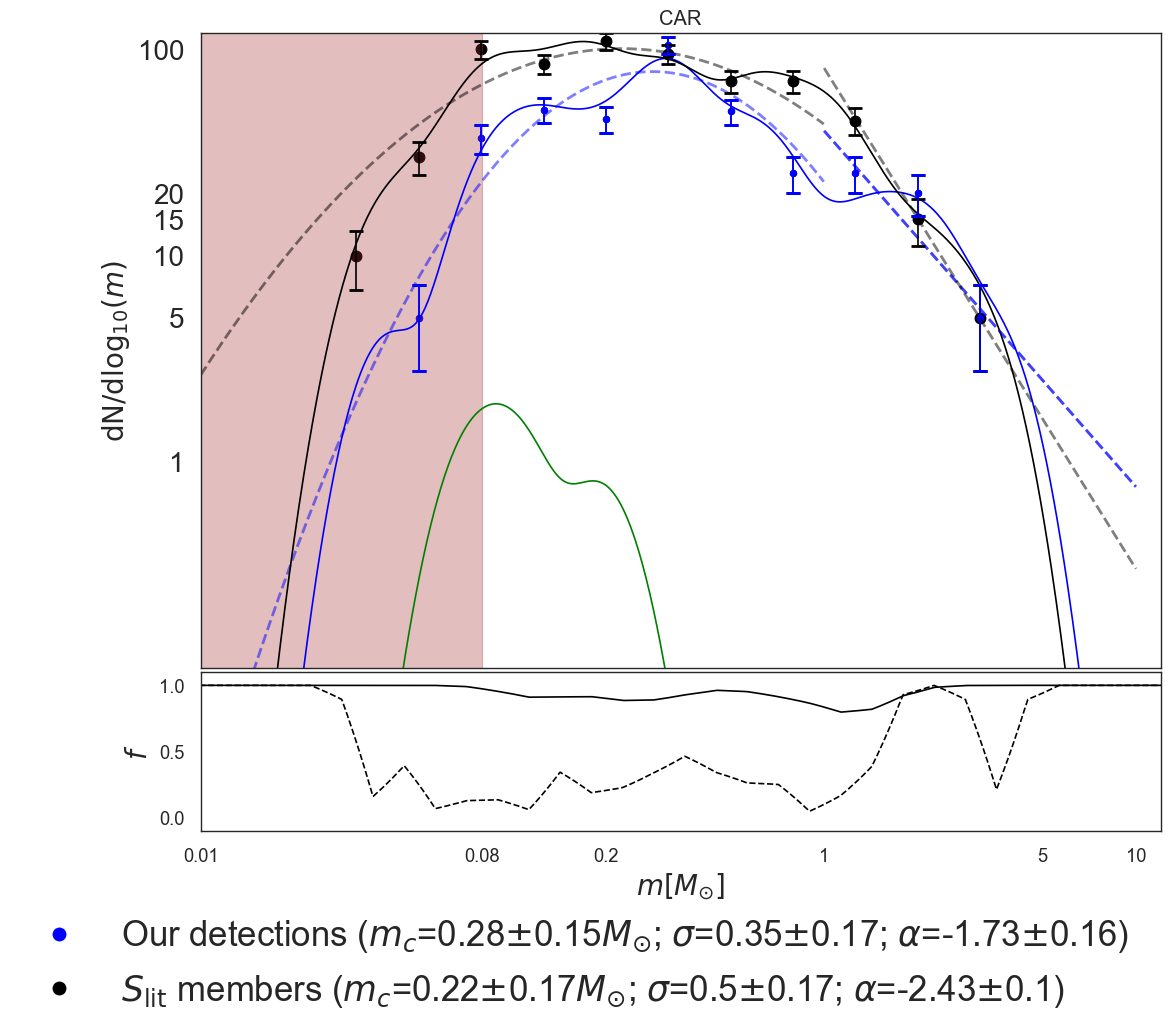}
\includegraphics[width=35mm]{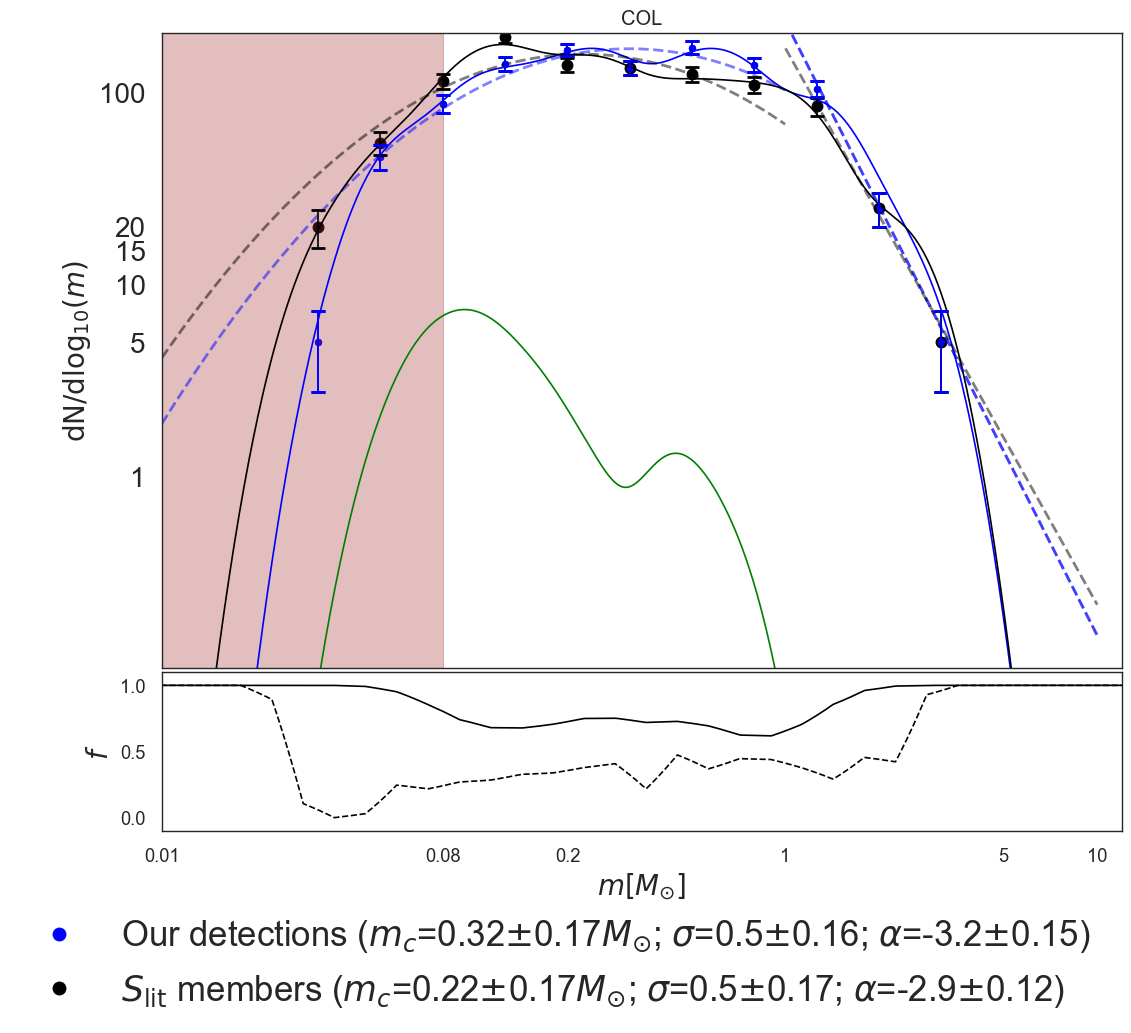}
\includegraphics[width=35mm]{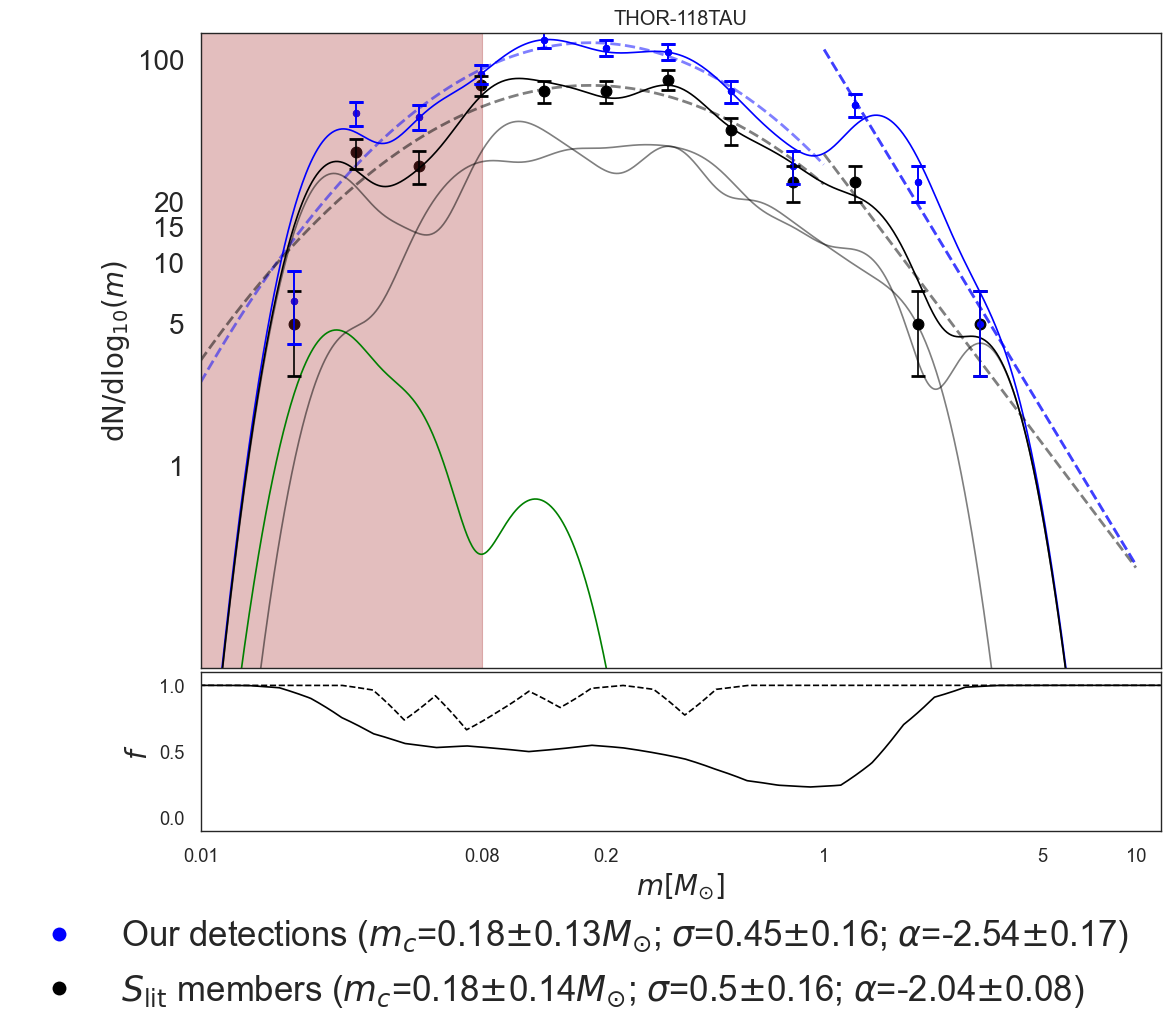}
\includegraphics[width=35mm]{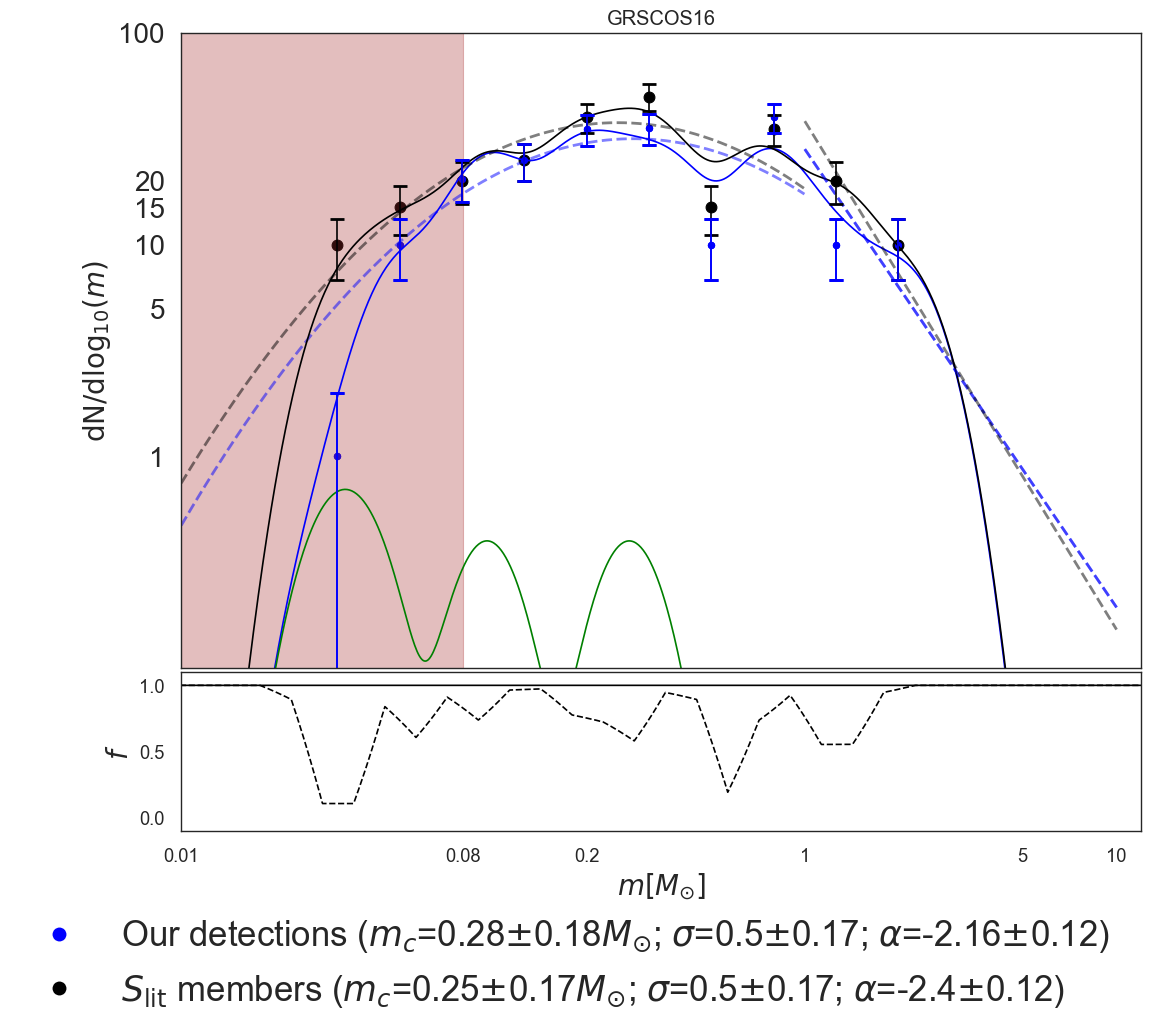}
\includegraphics[width=35mm]{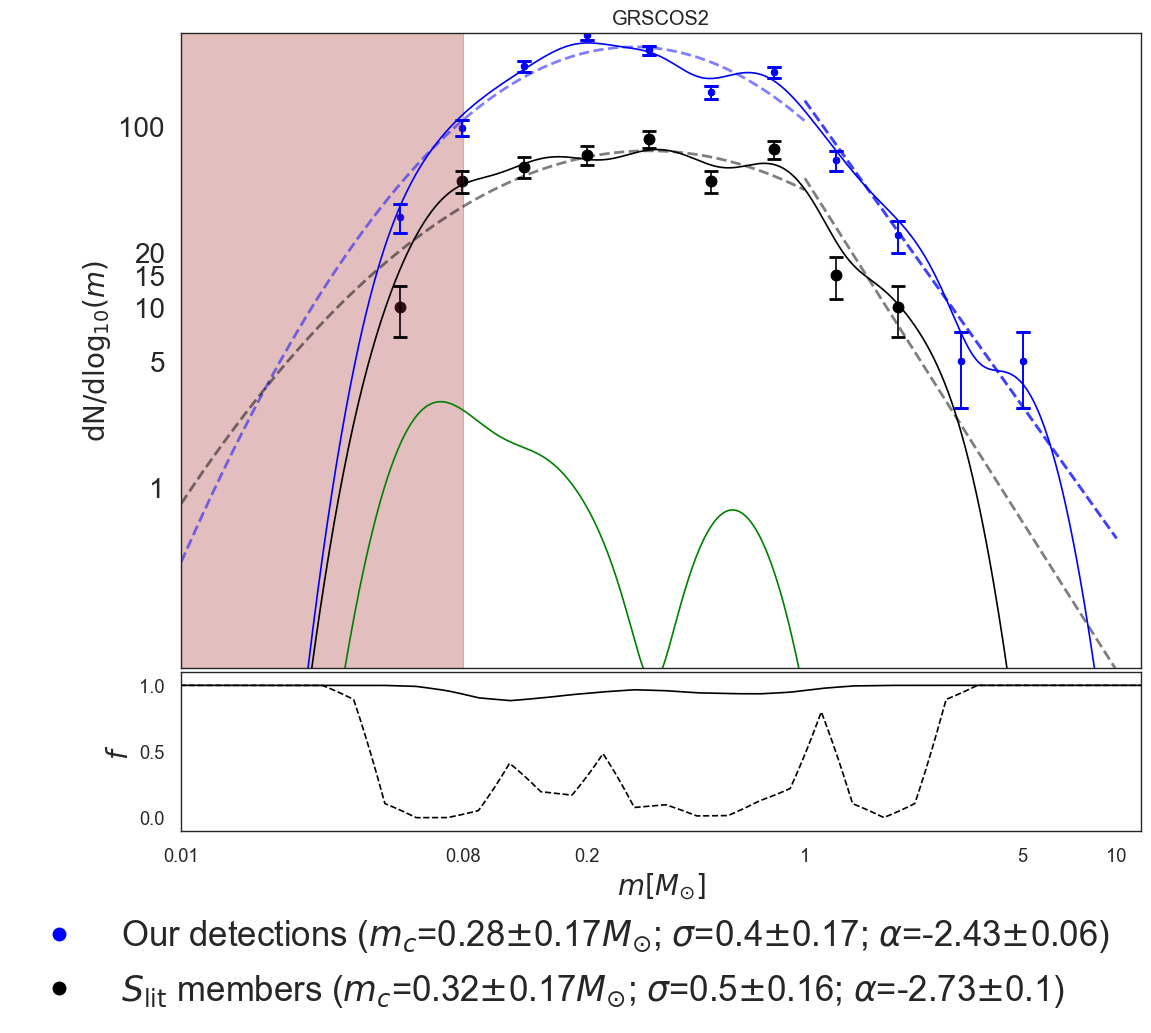}
\includegraphics[width=35mm]{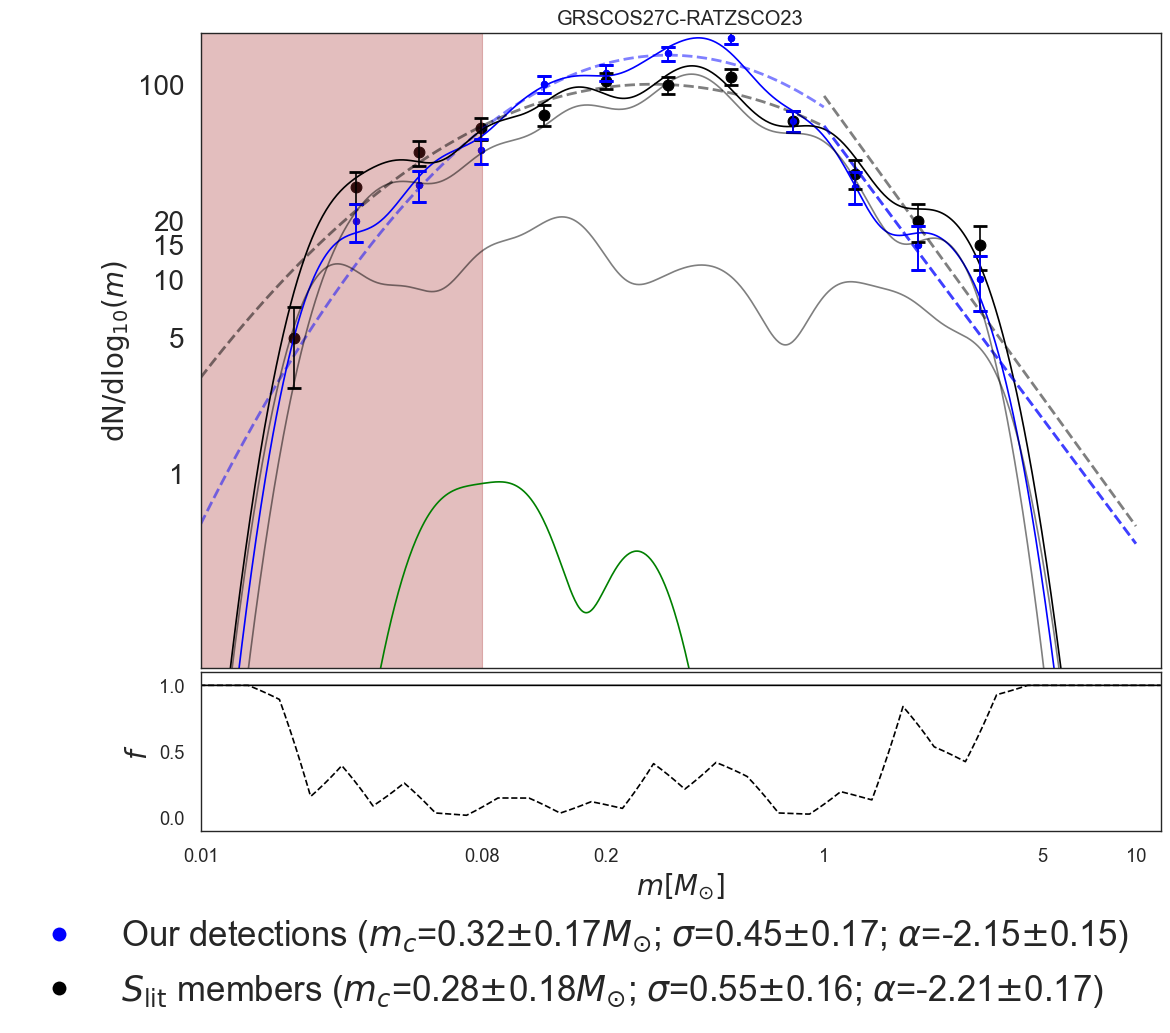}
\includegraphics[width=35mm]{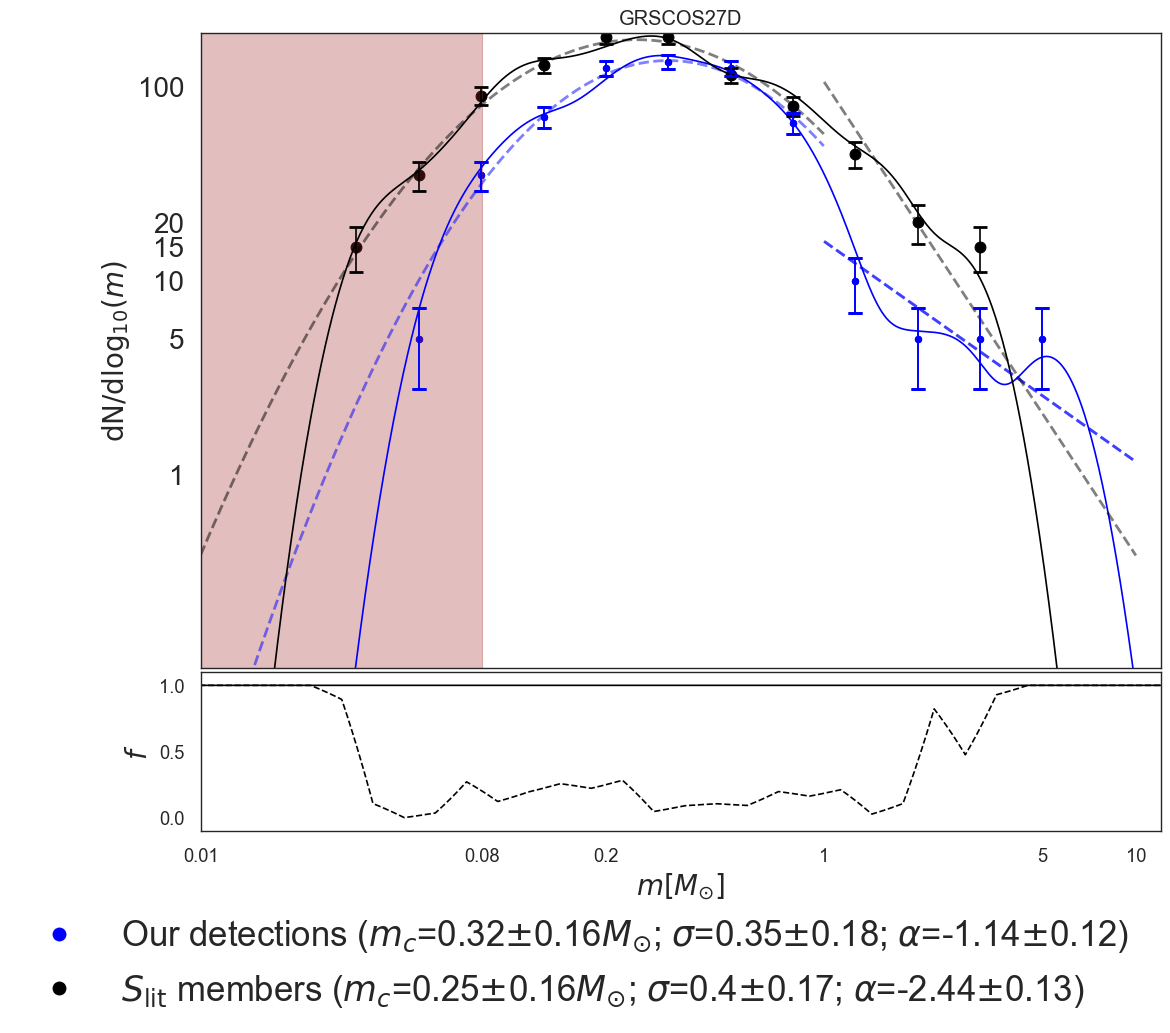}
\includegraphics[width=35mm]{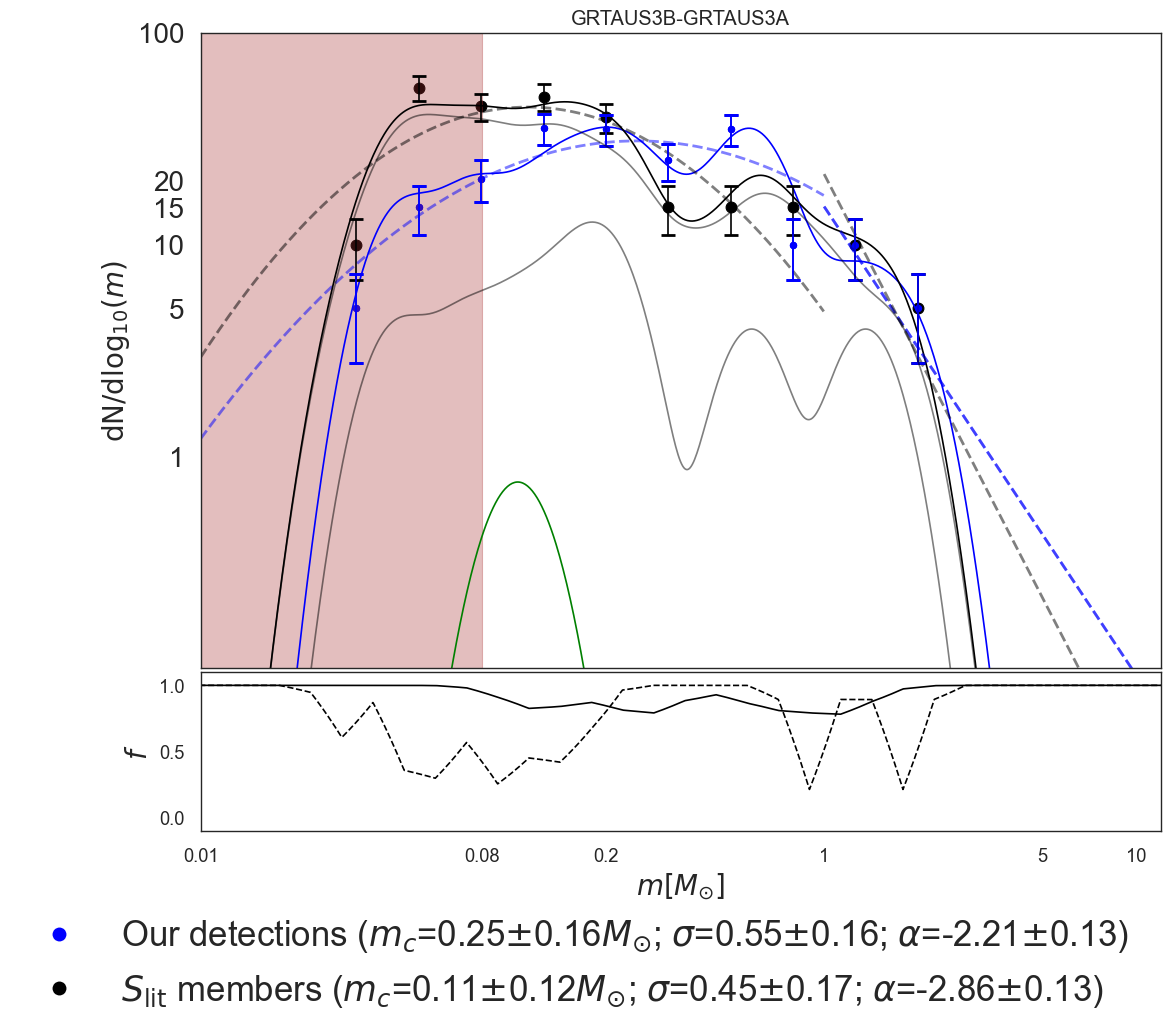}
\includegraphics[width=35mm]{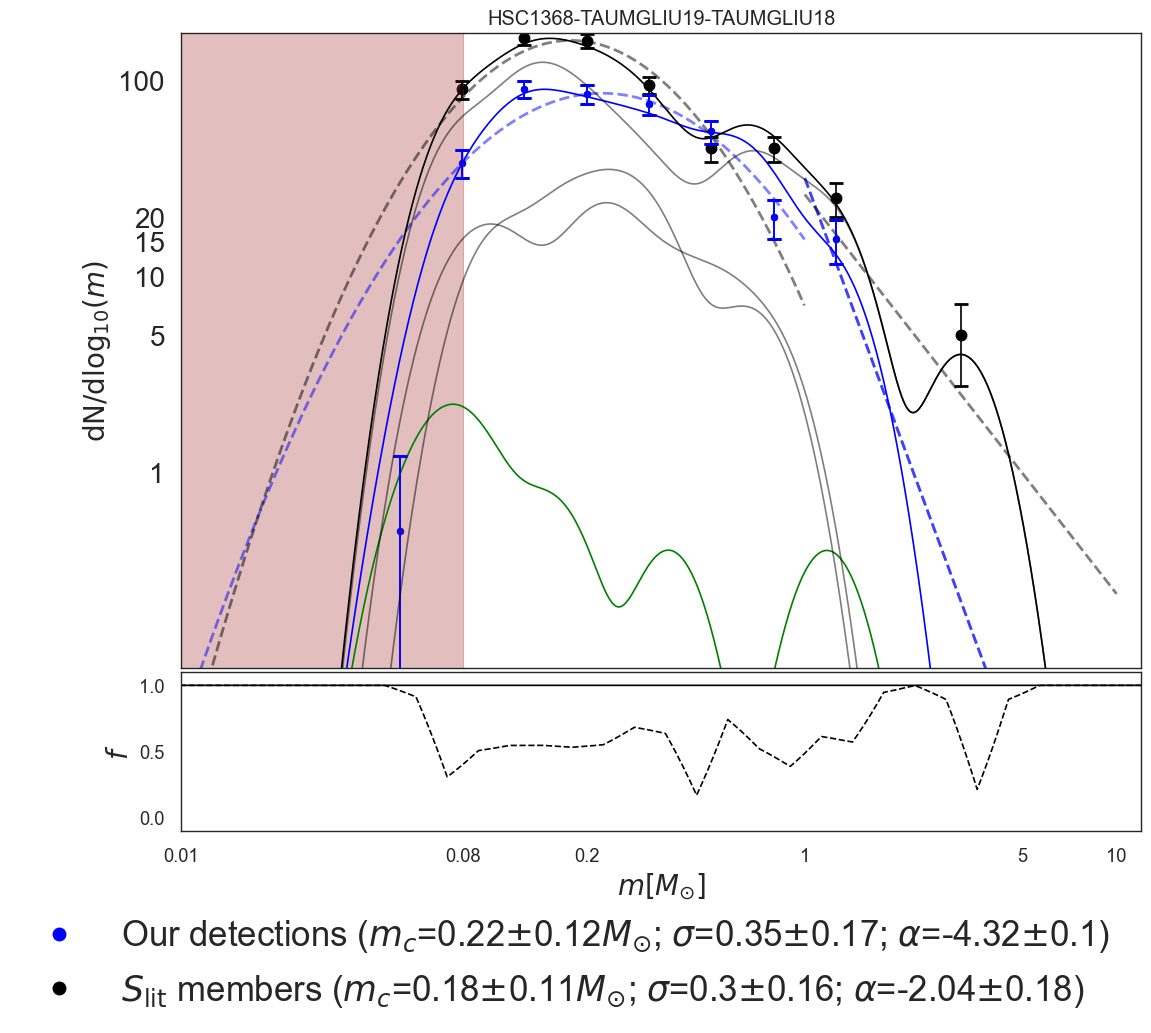}
\includegraphics[width=35mm]{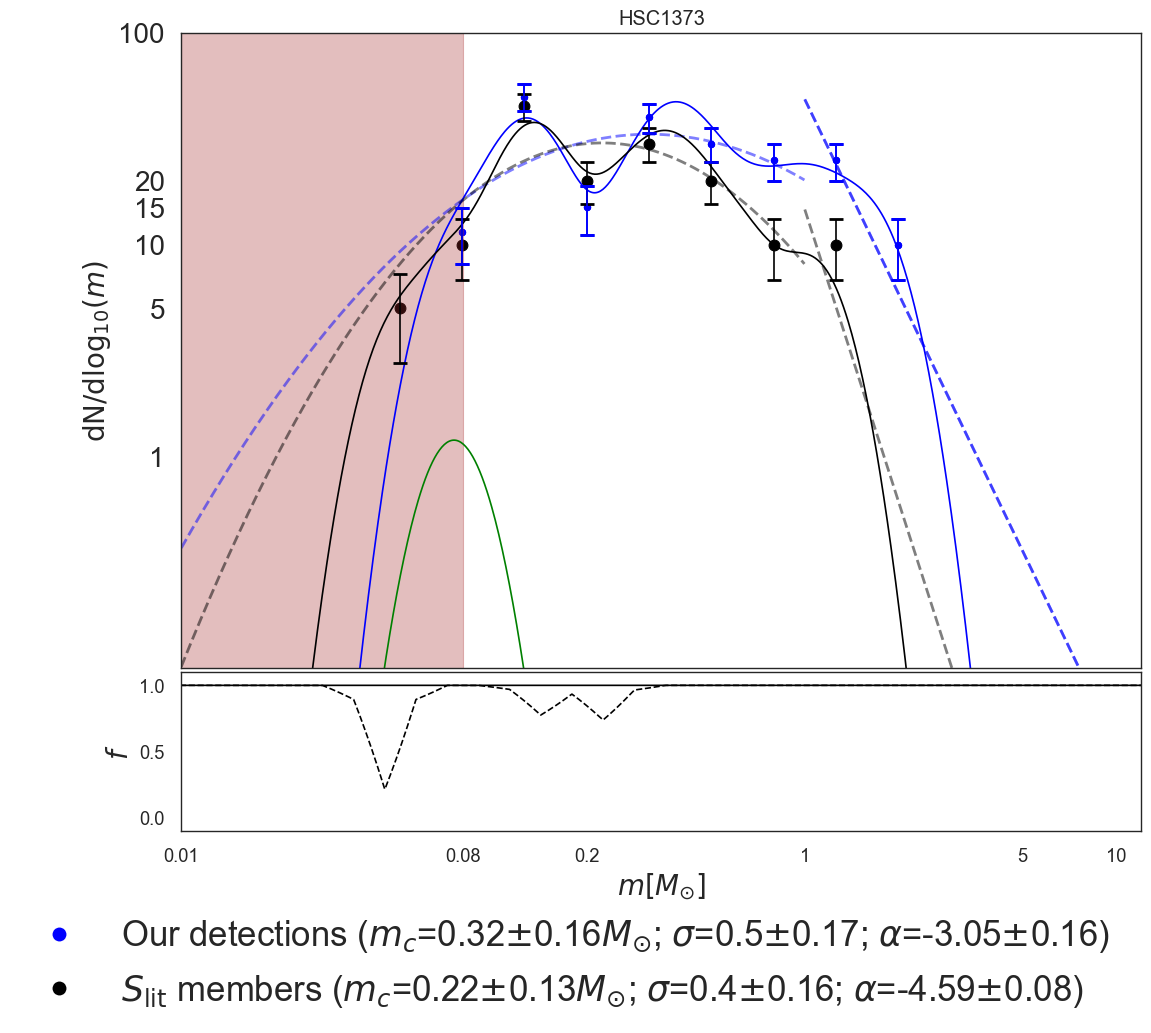}
\includegraphics[width=35mm]{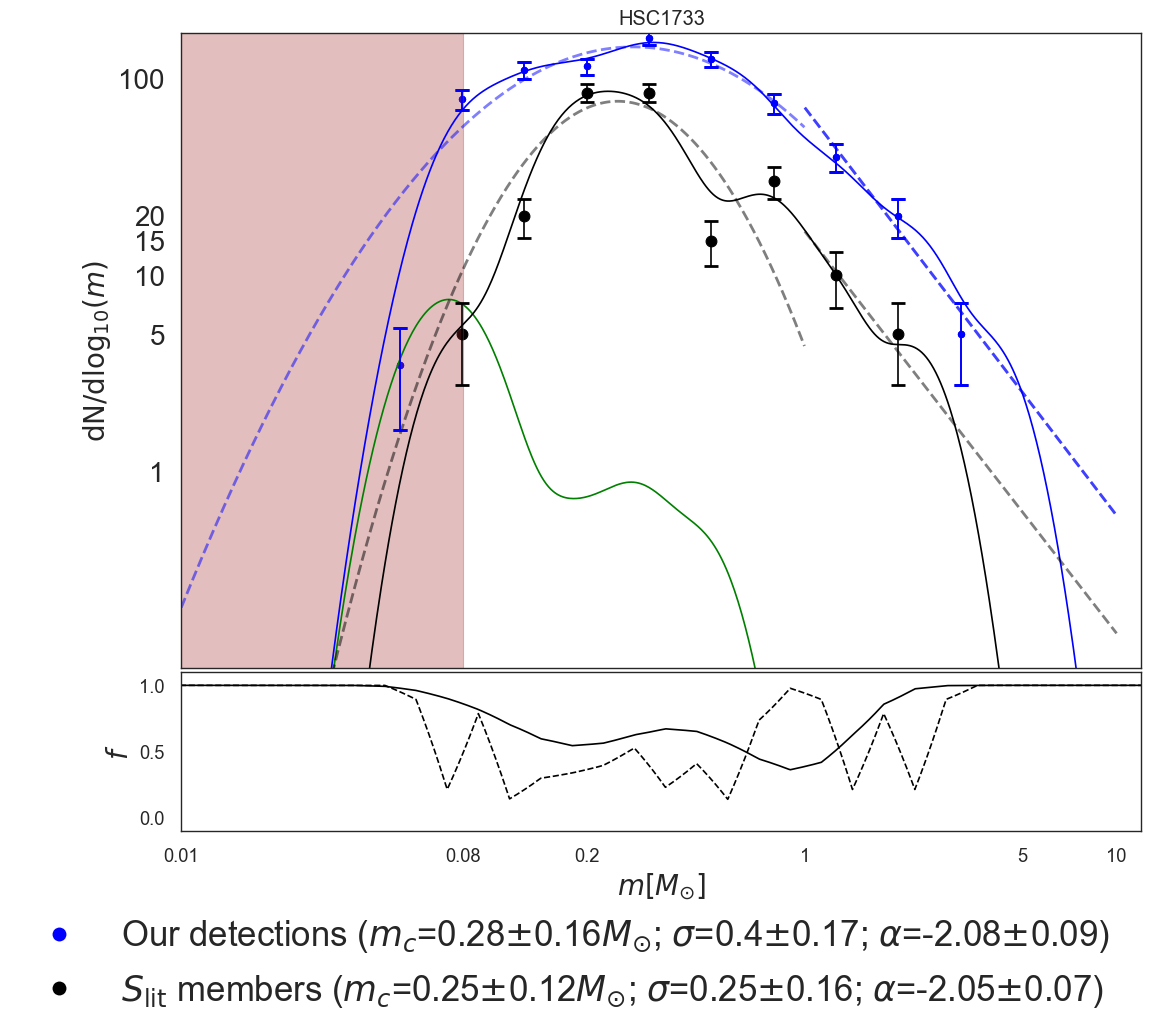}
\includegraphics[width=35mm]{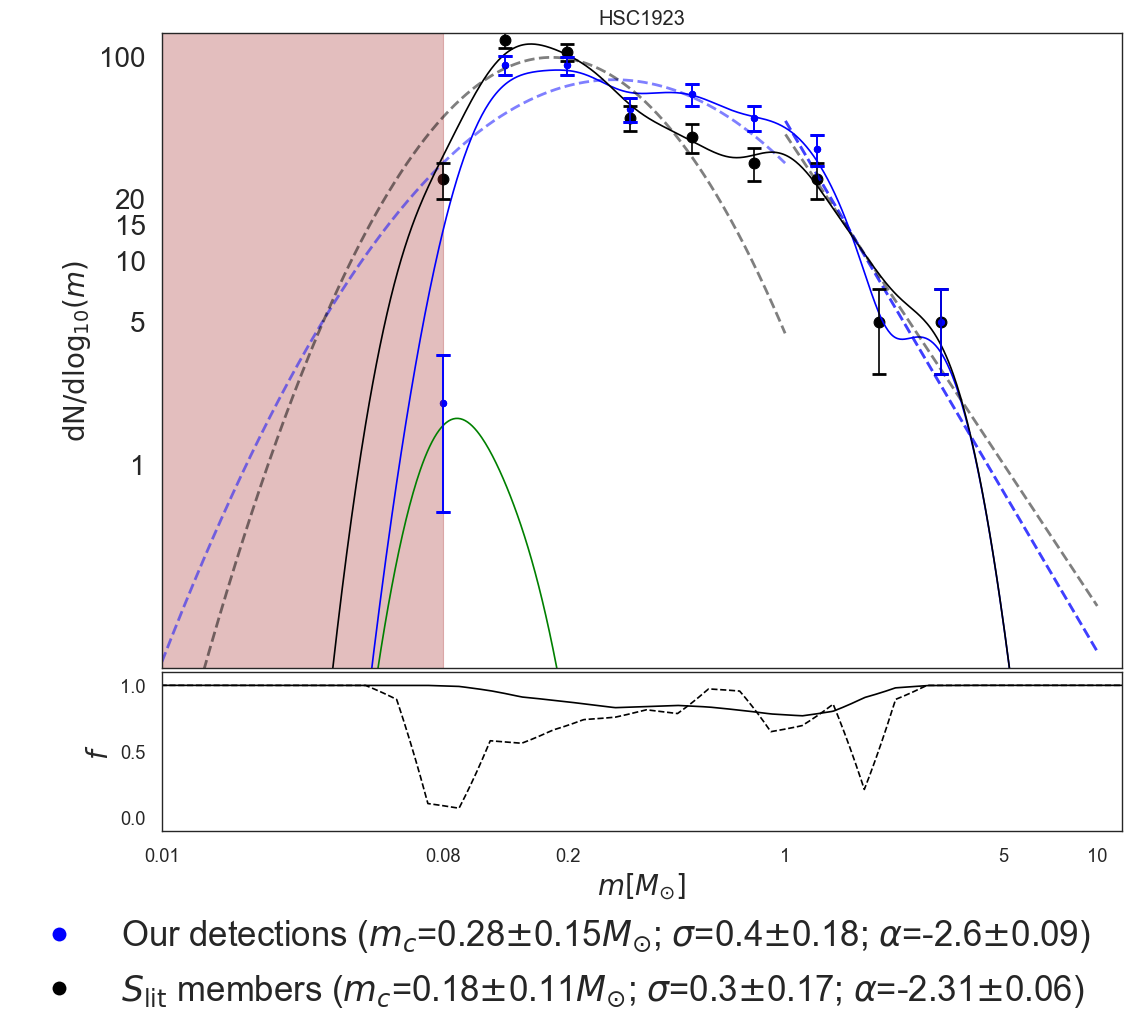}
\includegraphics[width=35mm]{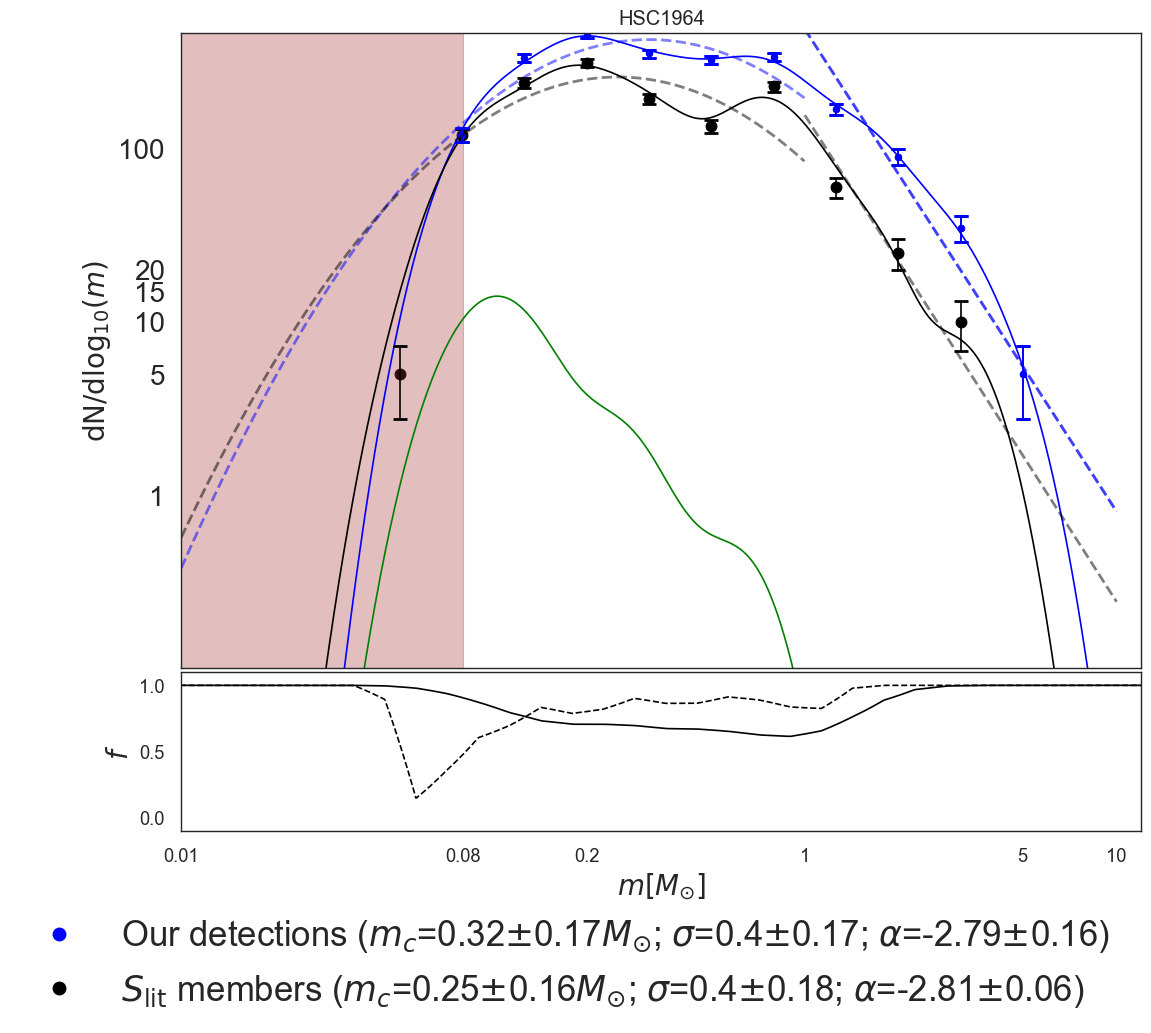}
\includegraphics[width=35mm]{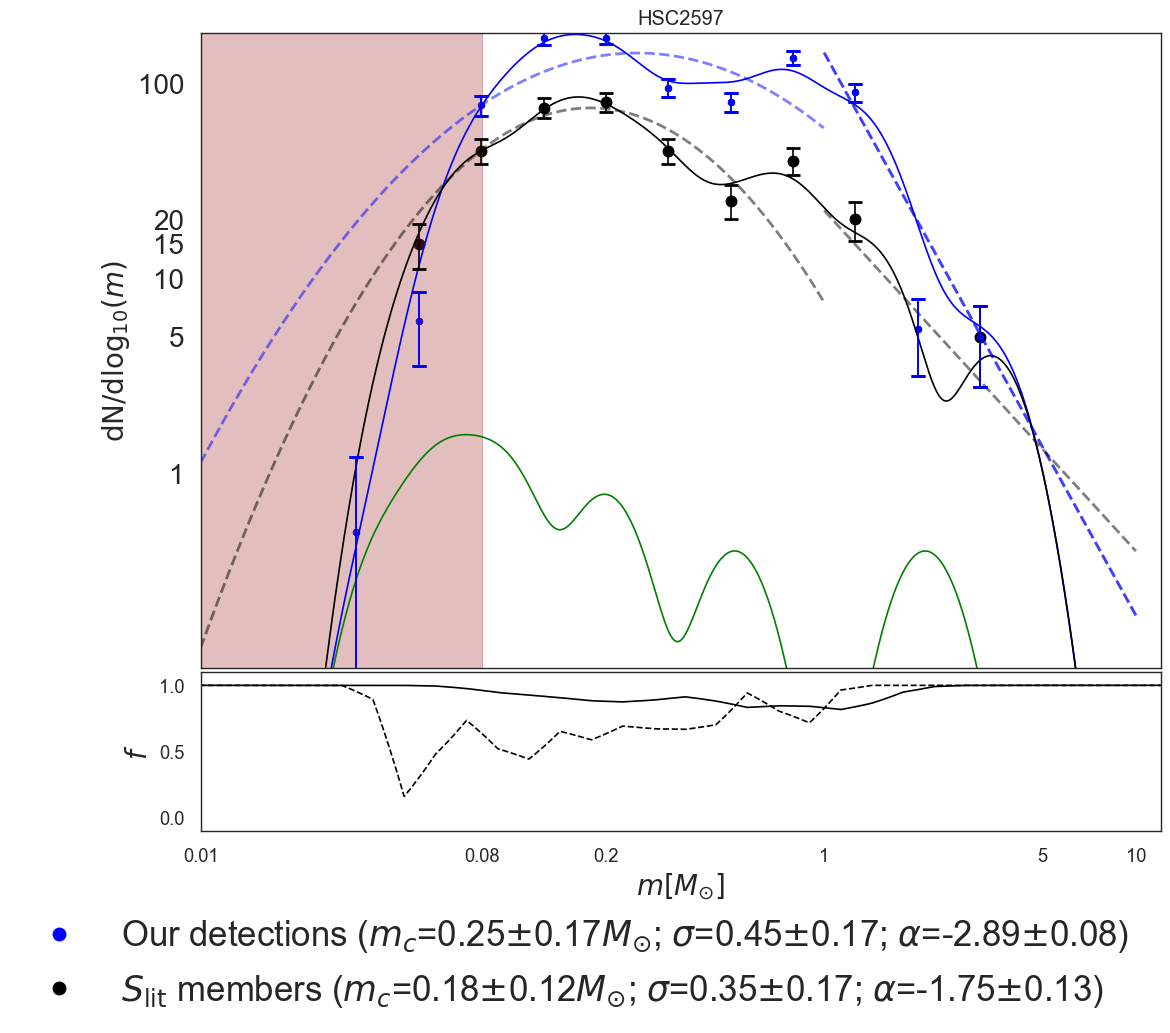}
\includegraphics[width=35mm]{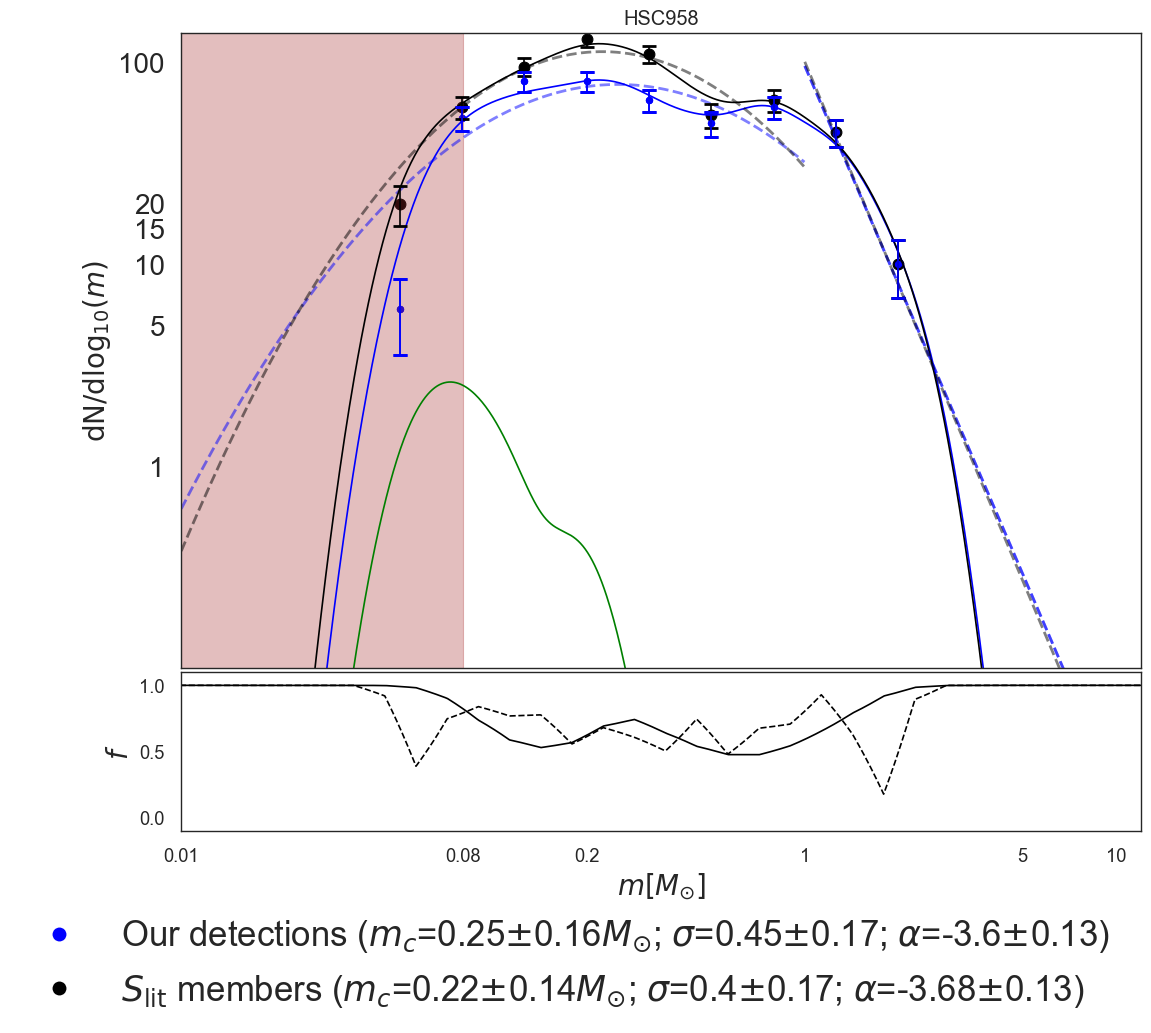}
\includegraphics[width=35mm]{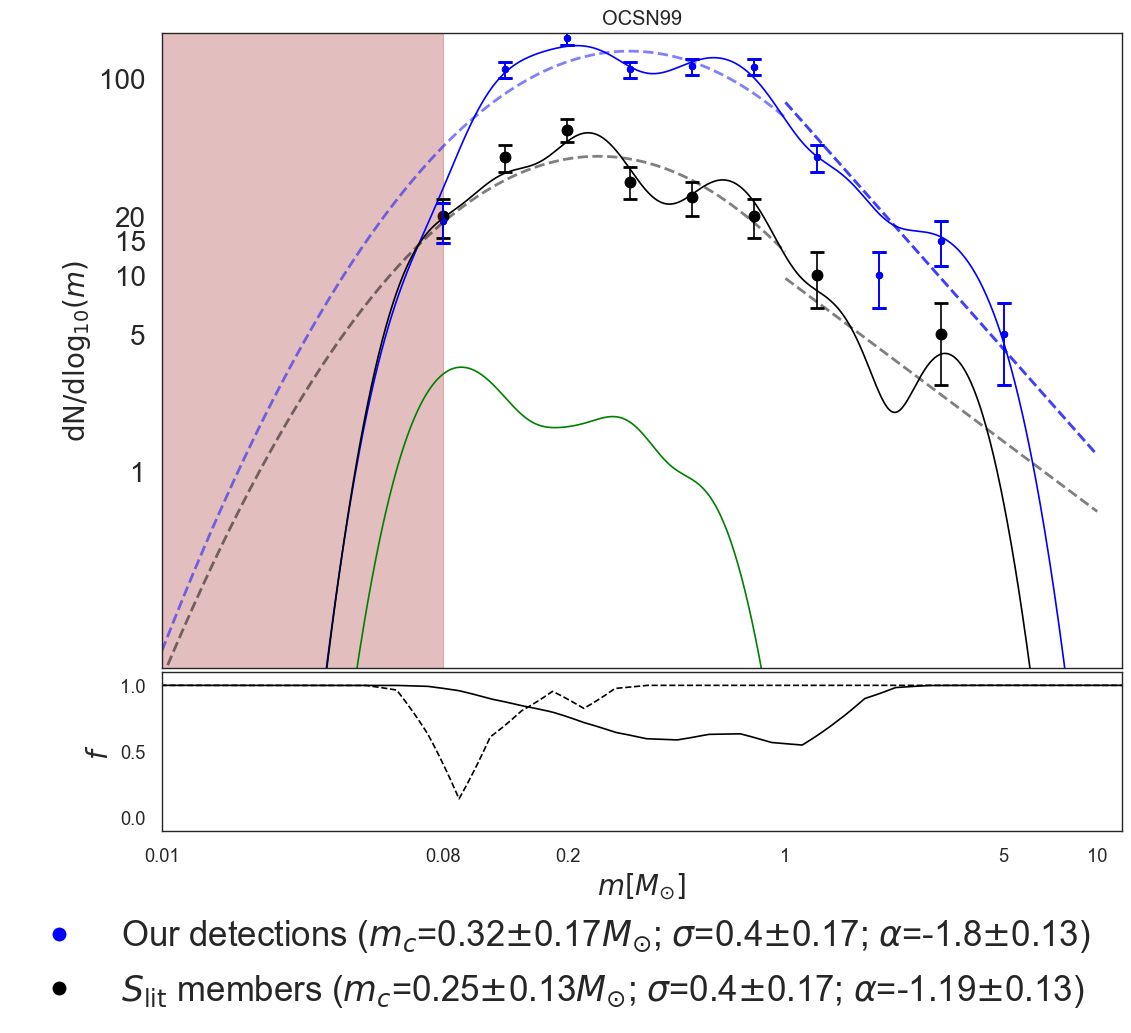}
\includegraphics[width=35mm]{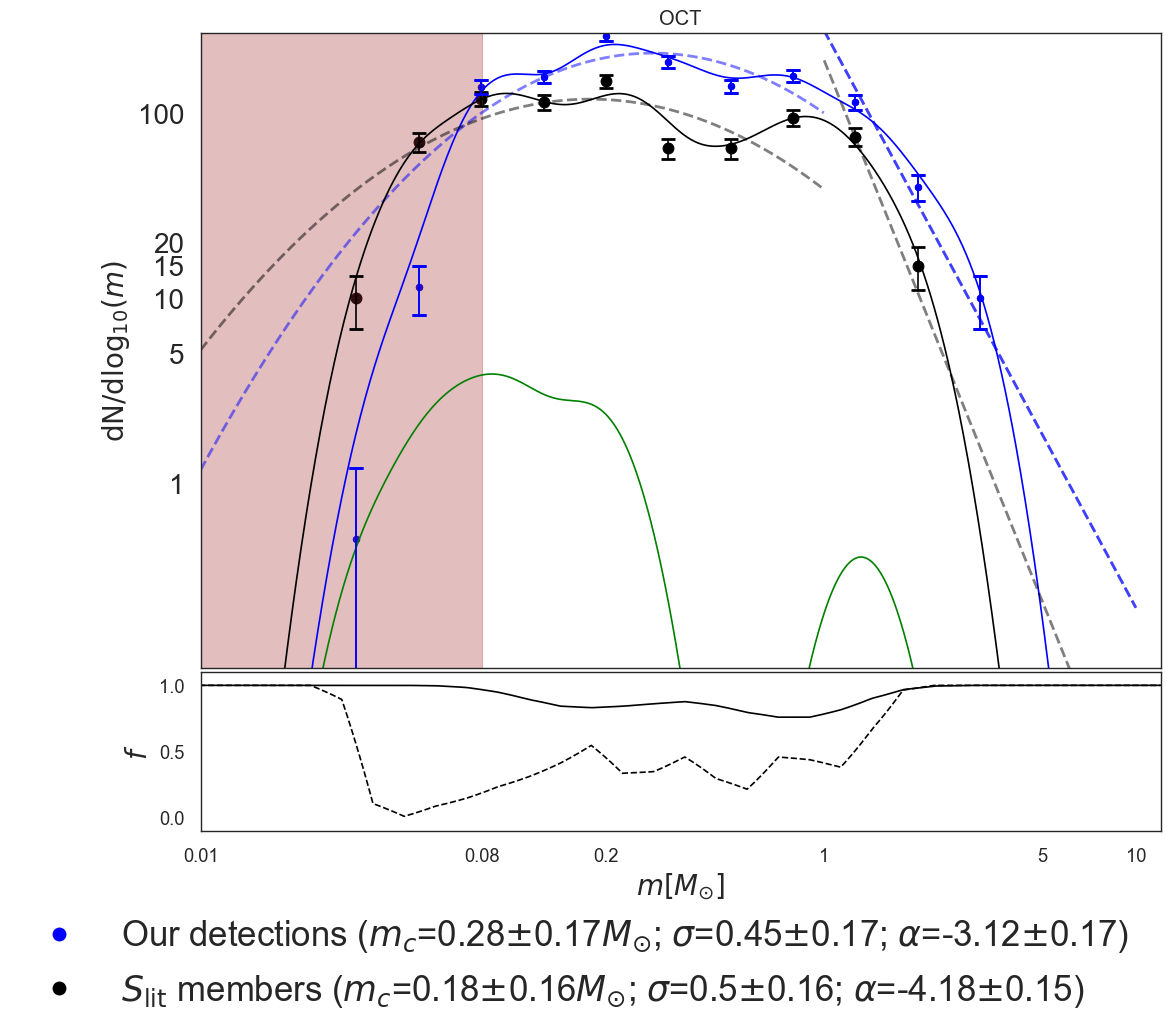}
\includegraphics[width=35mm]{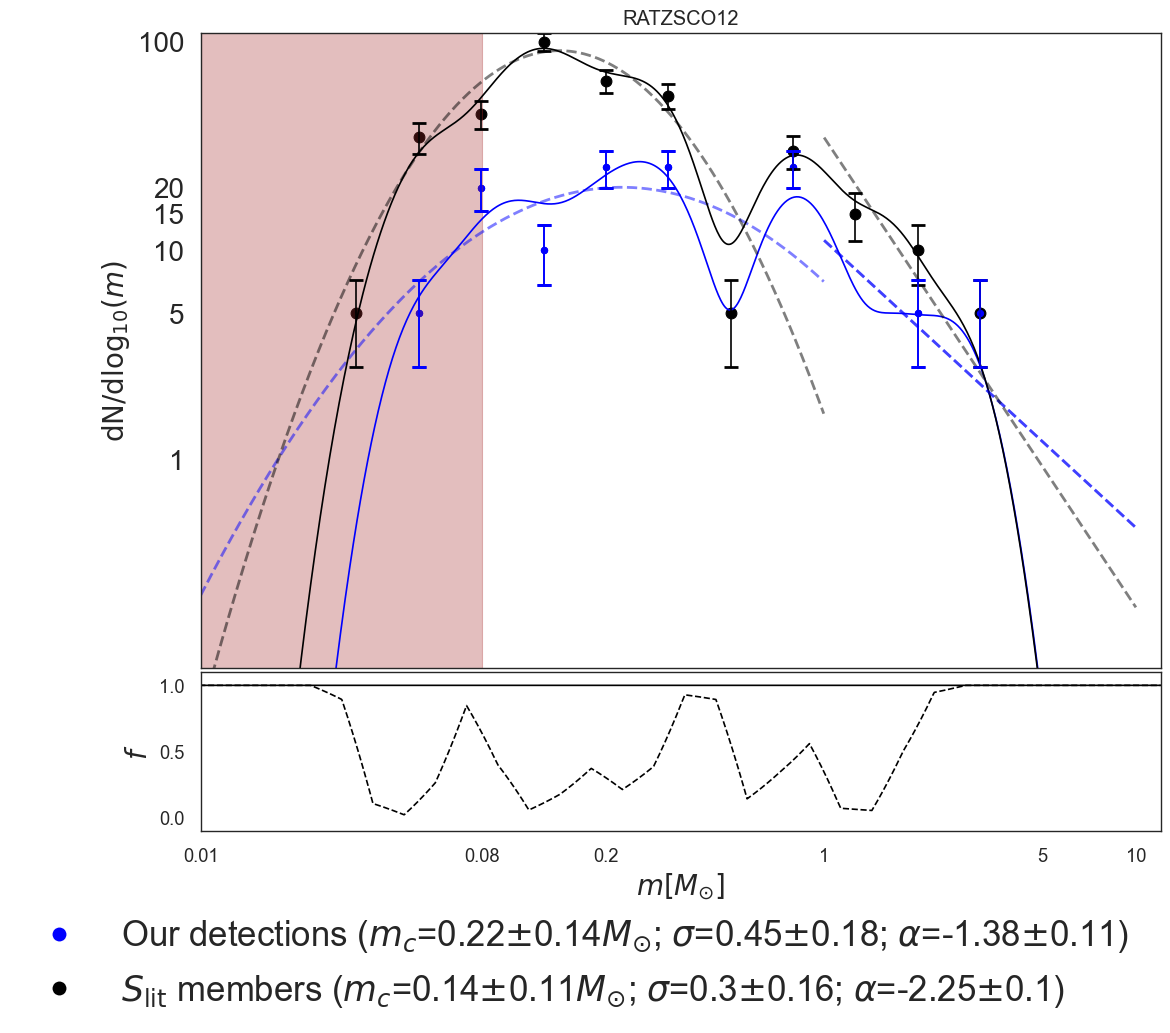}
\includegraphics[width=35mm]{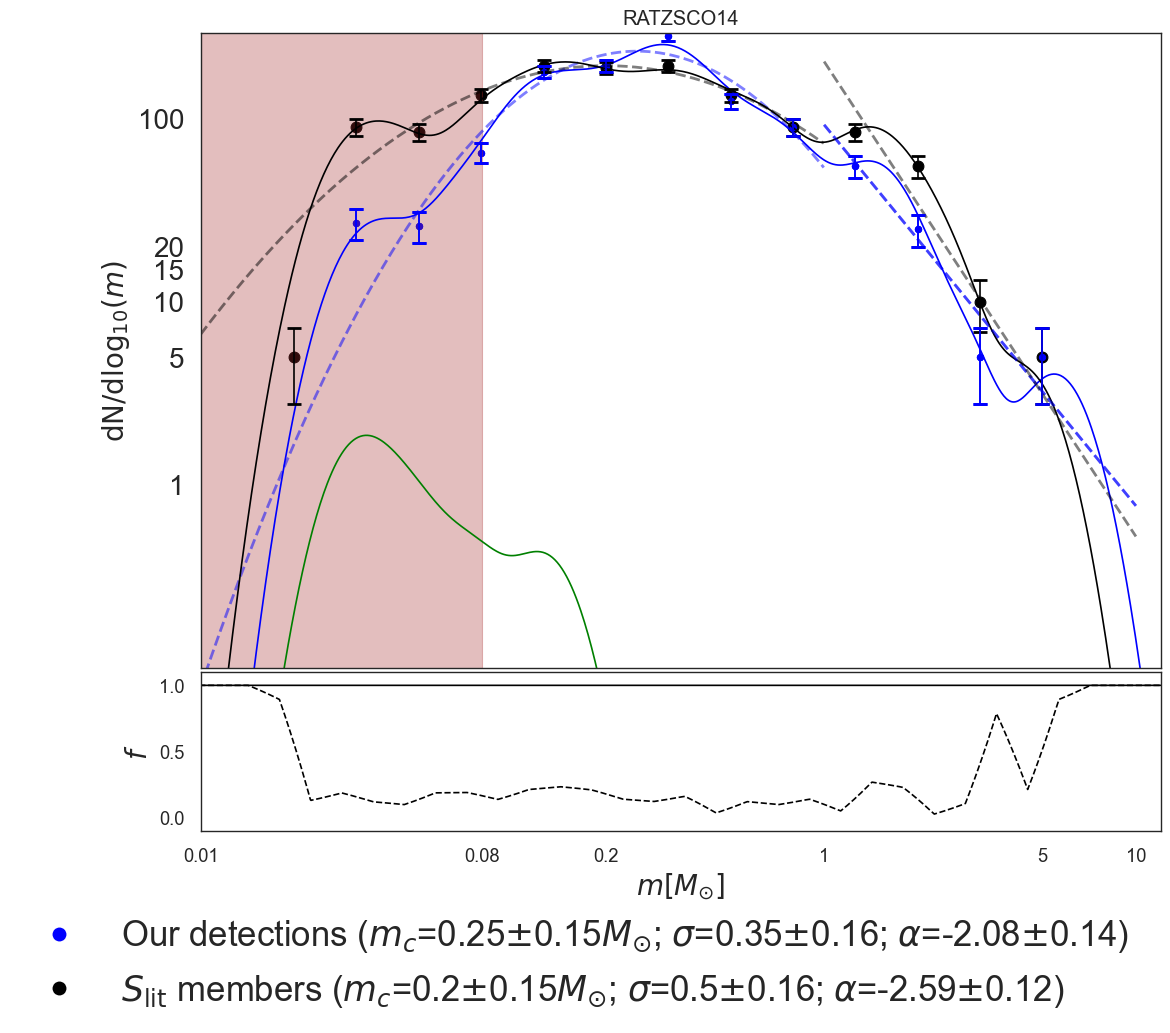}
\includegraphics[width=35mm]{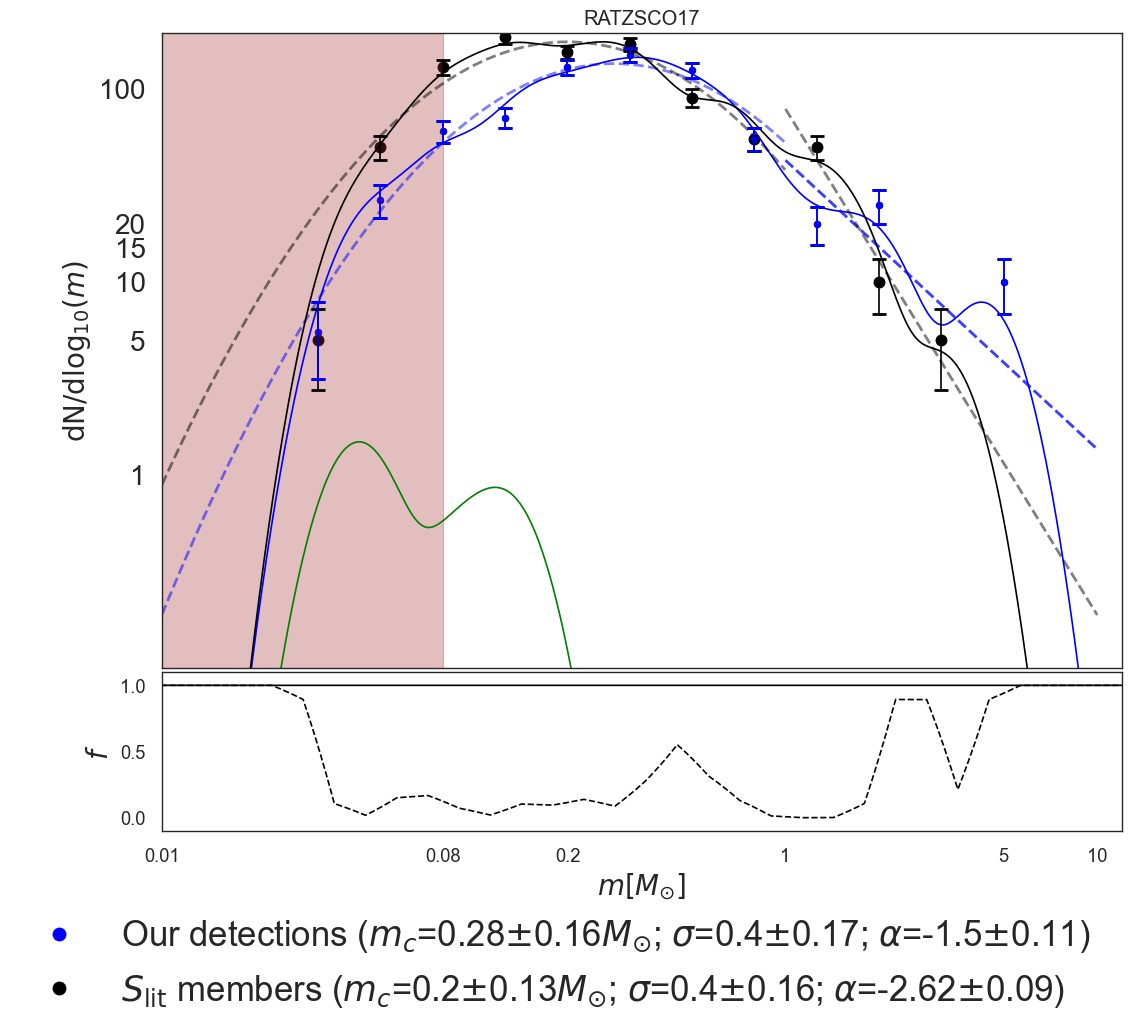}
\includegraphics[width=35mm]{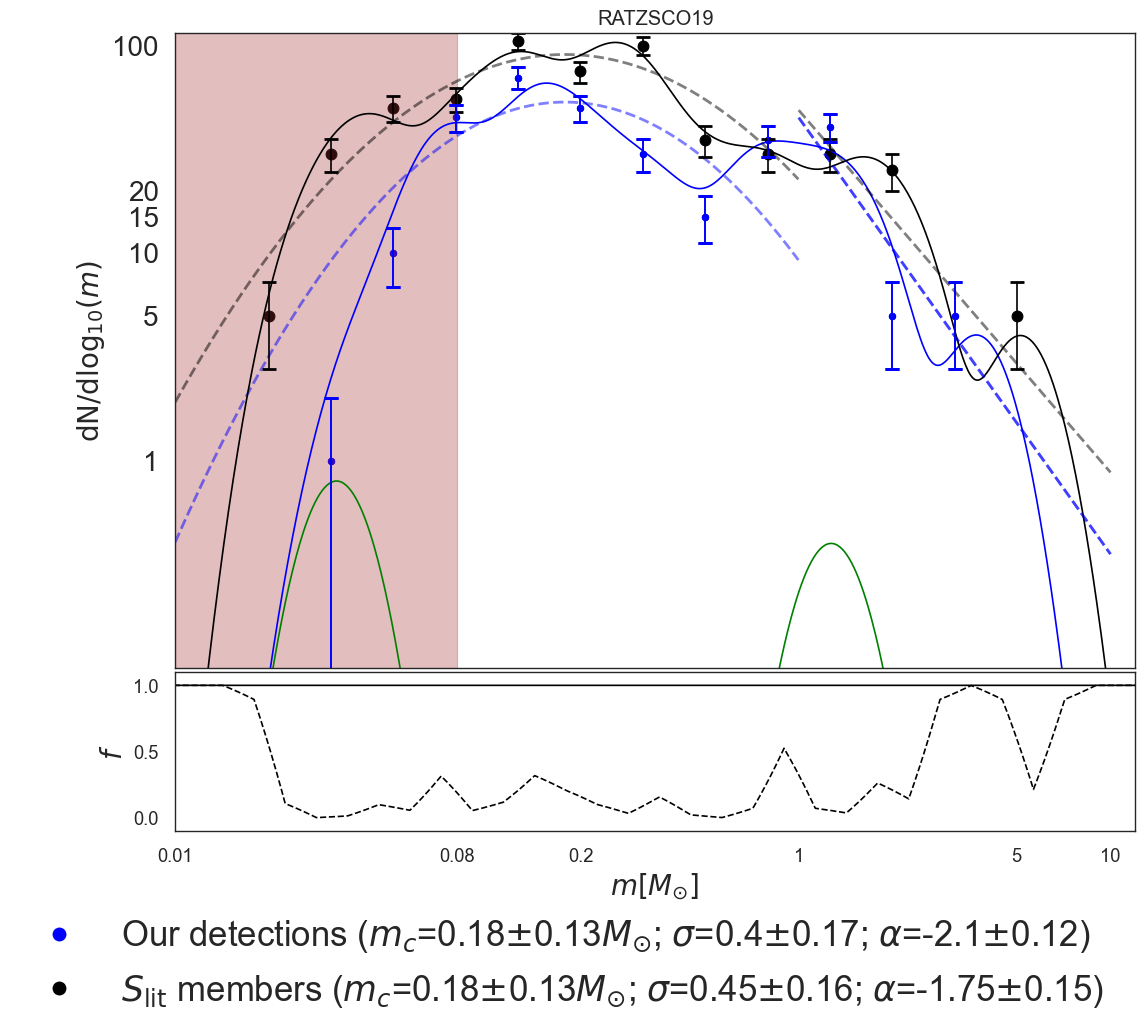}
\includegraphics[width=35mm]{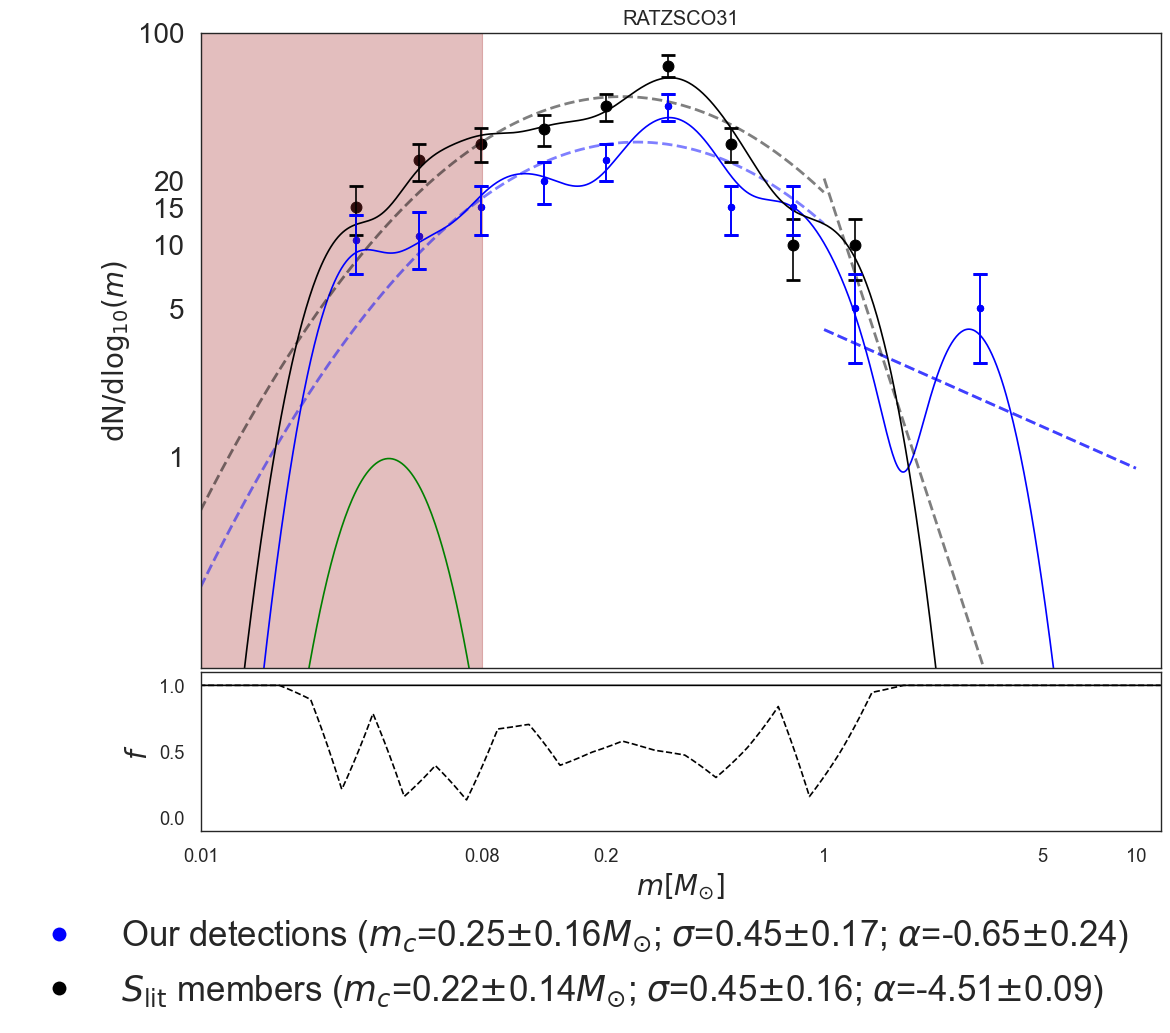}
\includegraphics[width=35mm]{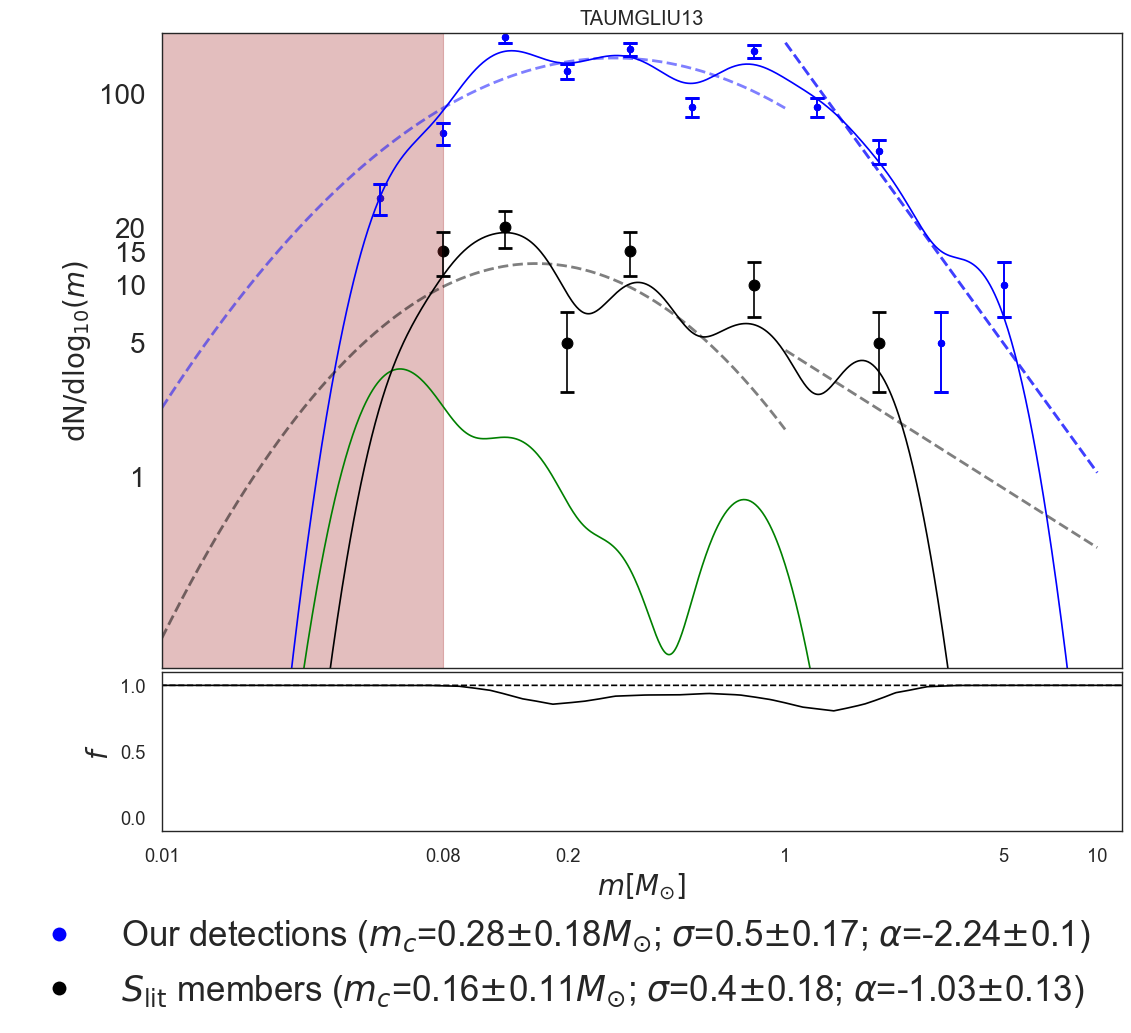}
\includegraphics[width=35mm]{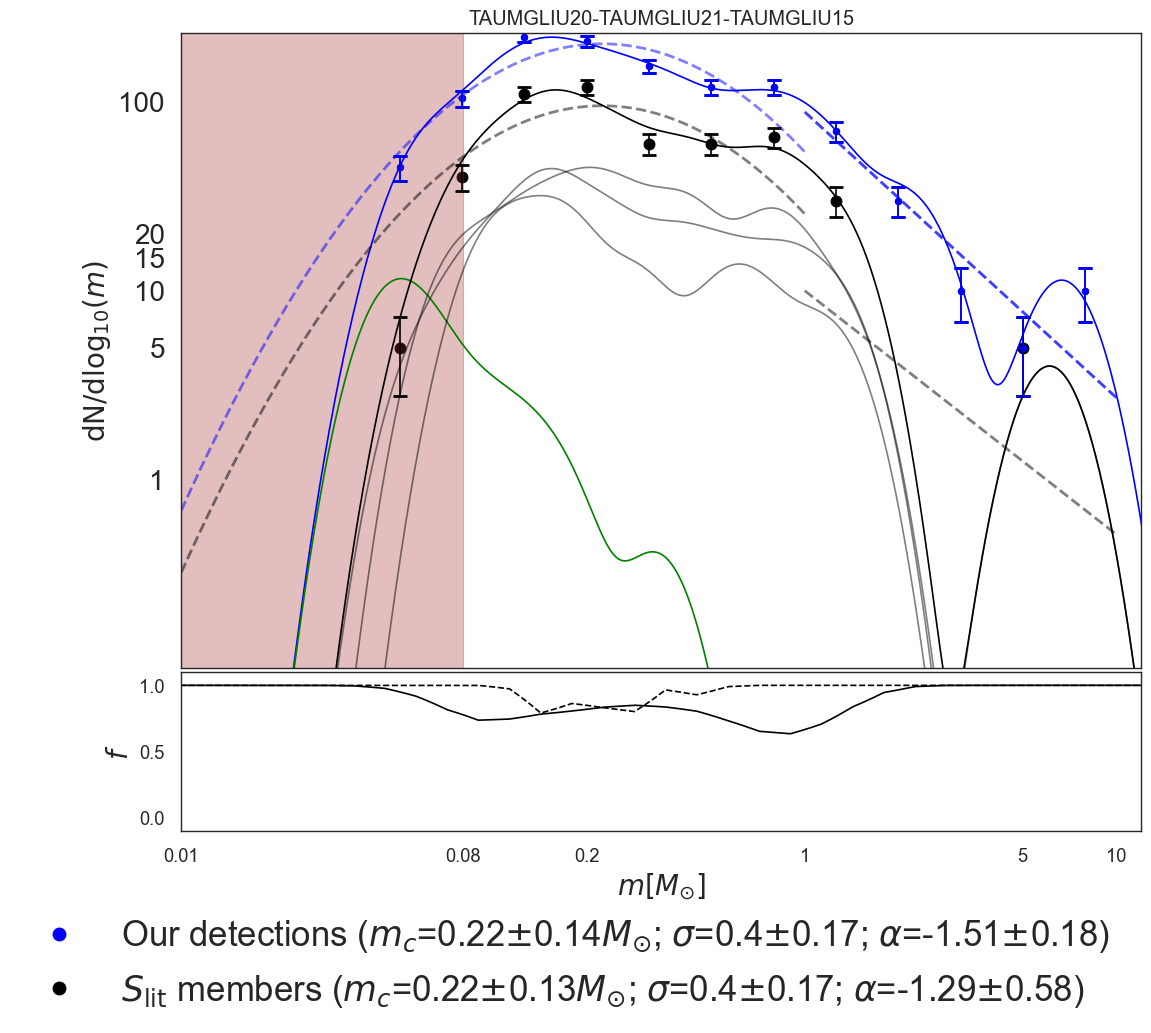}
\includegraphics[width=35mm]{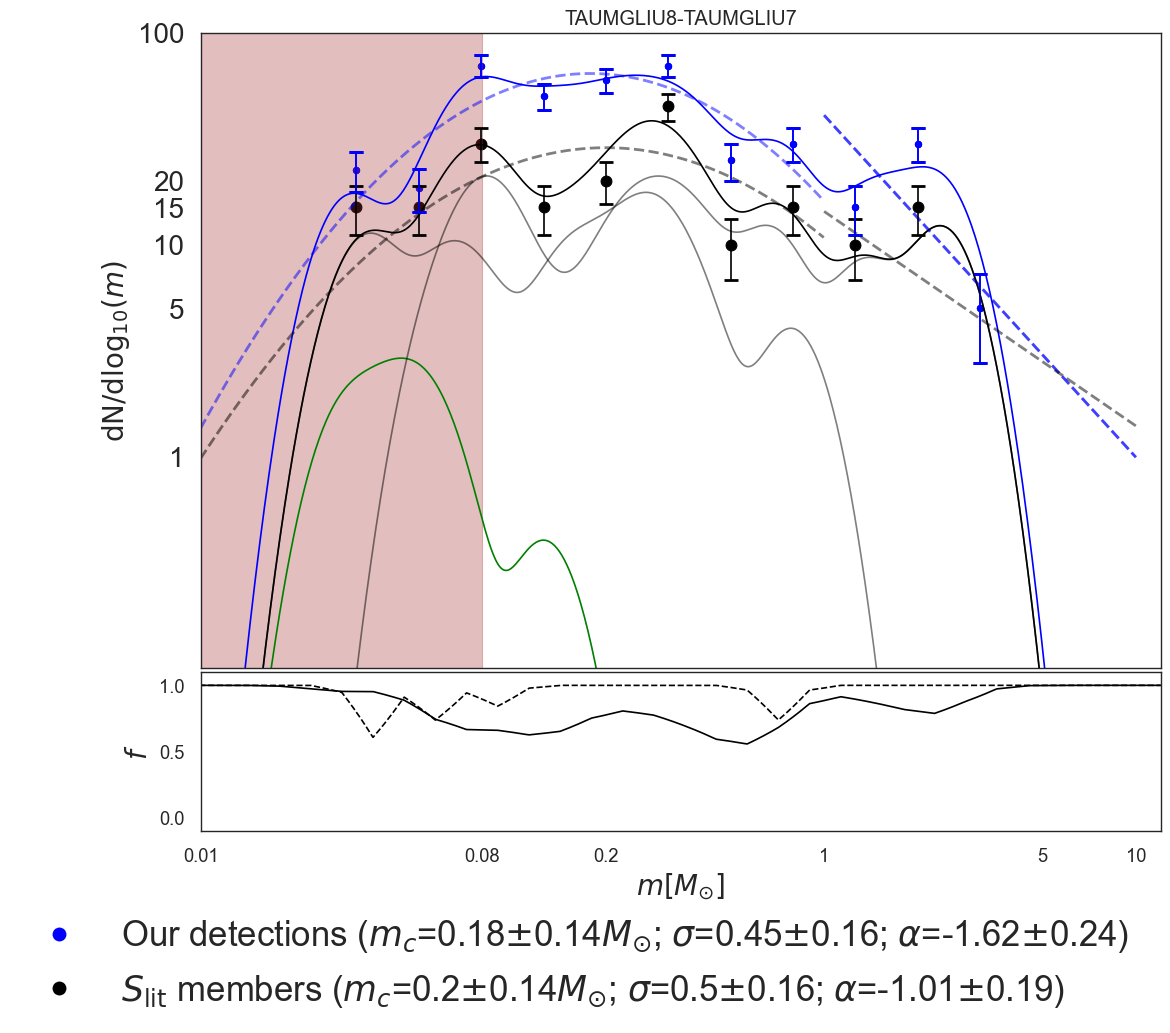}
\includegraphics[width=35mm]{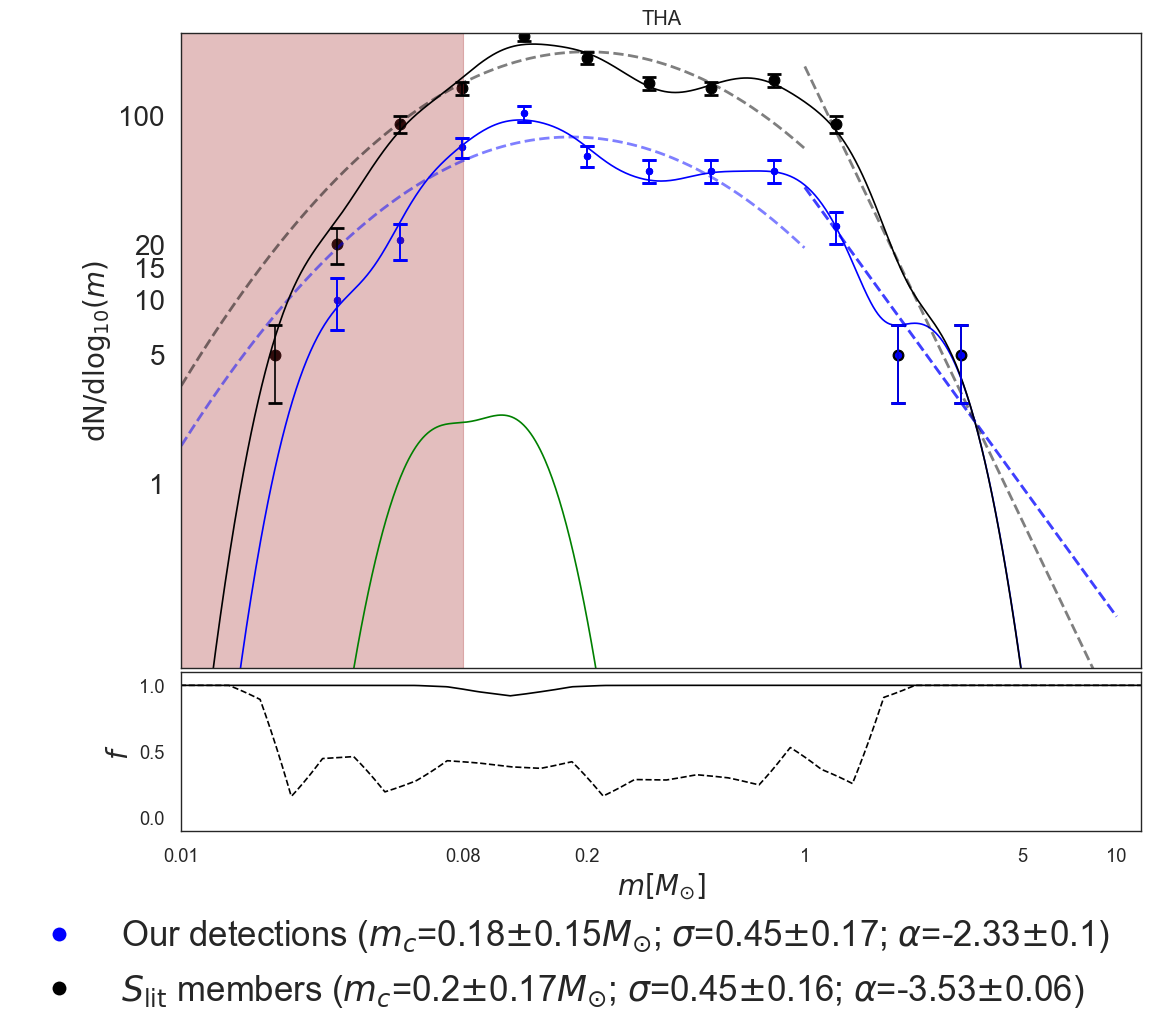}
\includegraphics[width=35mm]{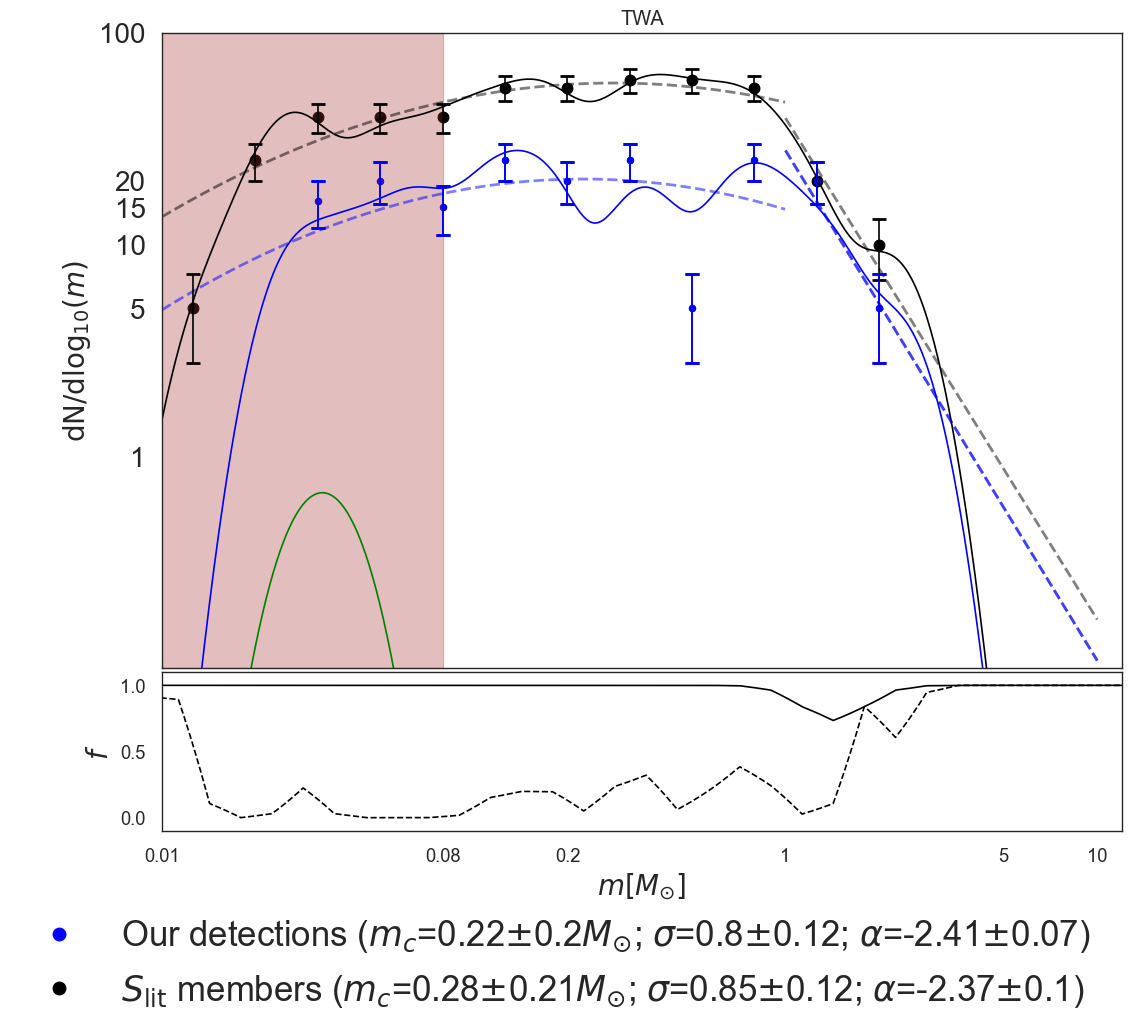}
\includegraphics[width=35mm]{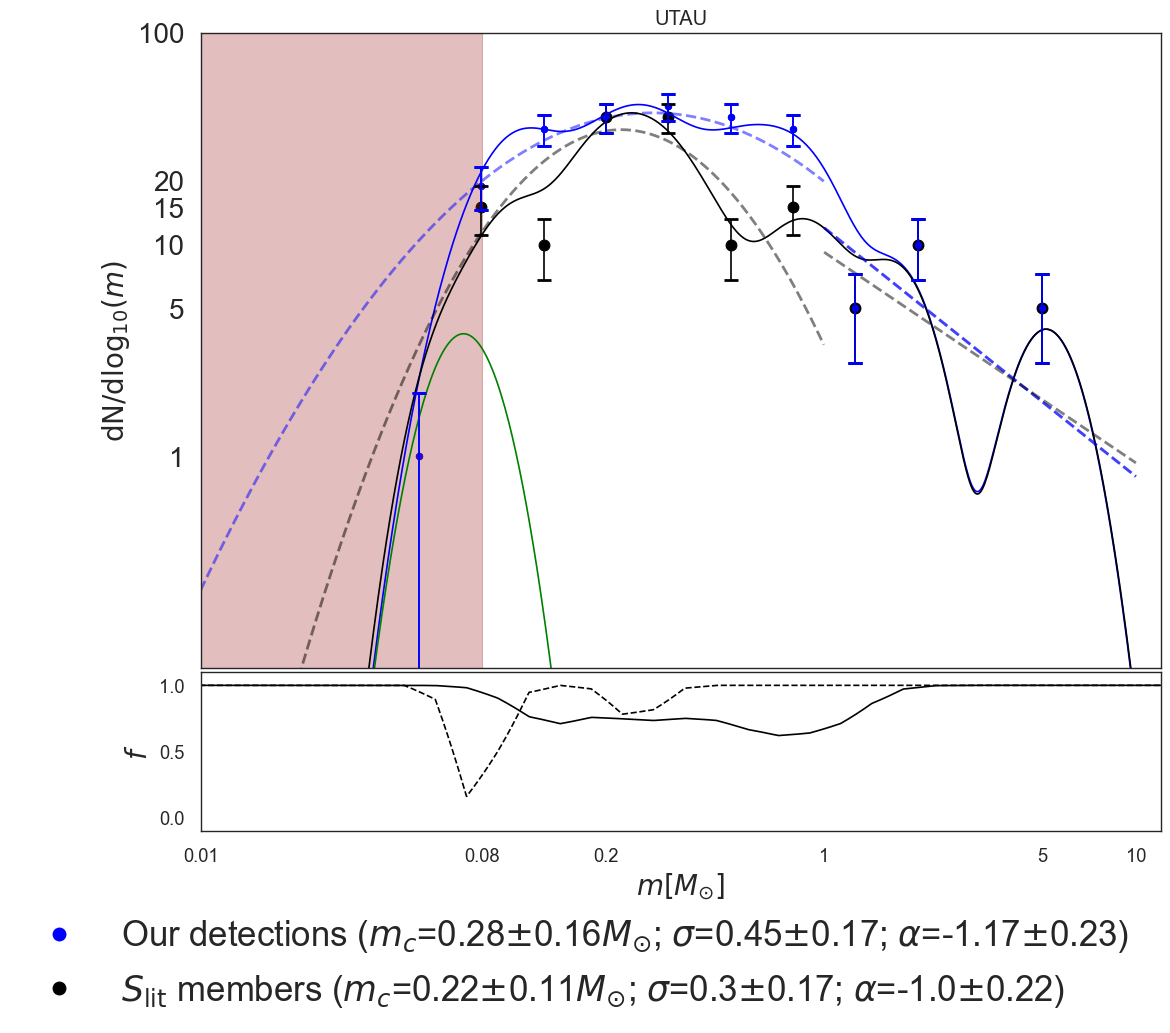}
\includegraphics[width=35mm]{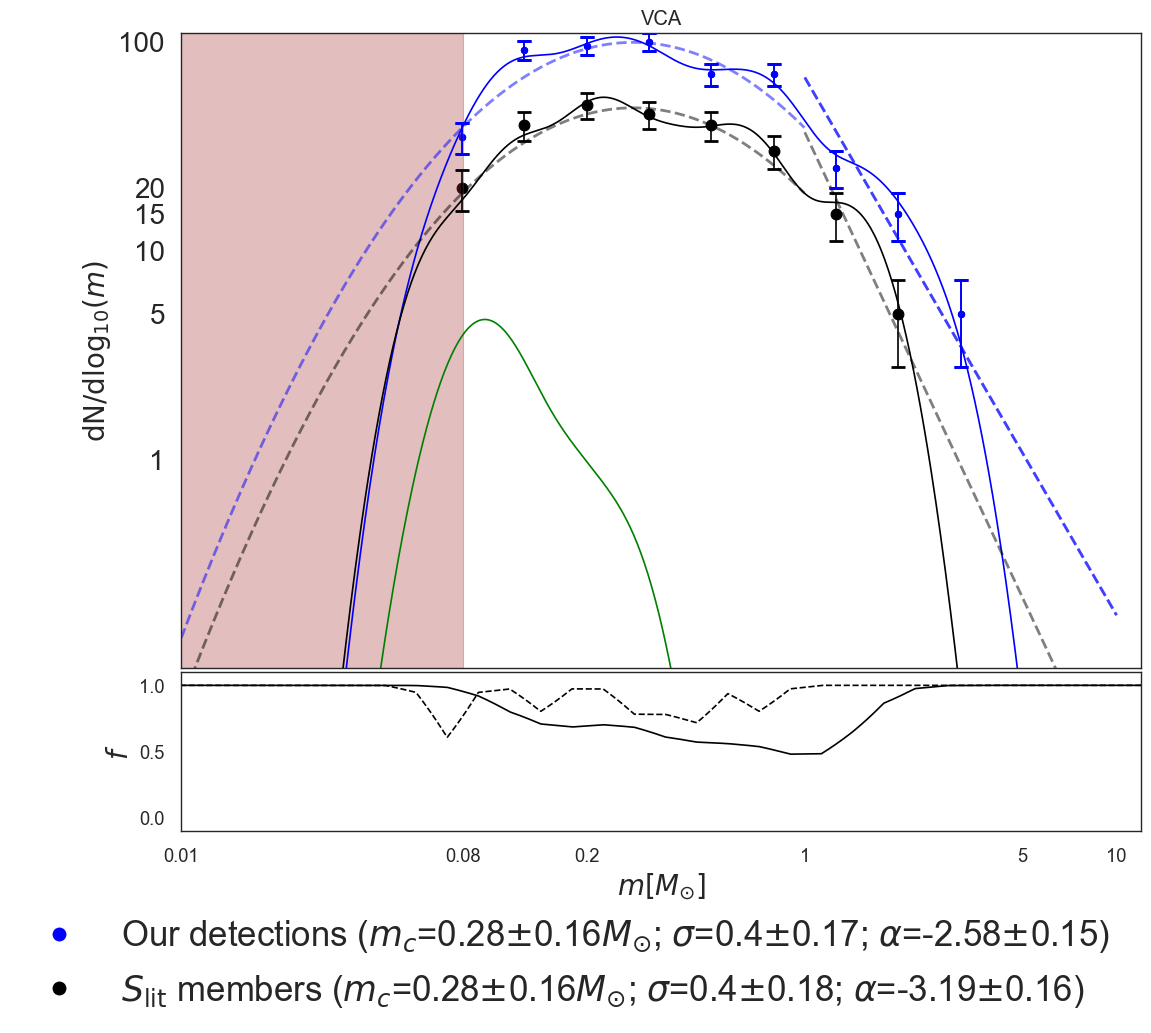}
\caption{Same plots as Figure \ref{fig:single_imf} for the remaining 30 detected groups.}
\label{fig:appendix_imfs}
\end{figure*}

\end{document}